\documentclass[review]{elsarticle}

\usepackage[colorlinks,citecolor=blue,linktoc=all,linkcolor=cyan]{hyperref}
\usepackage{graphicx}

\usepackage[T1]{fontenc}
\usepackage{dsfont}               
\usepackage{mathrsfs}             
\usepackage{slashed}              
\usepackage{amsmath}
\usepackage{amssymb}
\usepackage{amsbsy}
\usepackage{amsfonts}
\usepackage{multirow}
\usepackage{subcaption}
\usepackage{siunitx}
\usepackage{soul}
\usepackage{threeparttable}
\usepackage{CJKutf8}
\soulregister\cite7
\soulregister\ref7 

\numberwithin{equation}{section}
\numberwithin{table}{section}
\numberwithin{figure}{section}

\journal{Progress in Particle and Nuclear Physics}

\usepackage{titlesec}
\usepackage{sectsty}
\titleformat{\section}{\normalfont\Large\bfseries}{\thesection}{1em}{}
\titleformat{\subsection}{\normalfont\large\bfseries}{\thesubsection}{1em}{}
\titleformat{\subsubsection}{\normalfont\normalsize\bfseries}{\thesubsubsection}{1em}{}

\begin{document}
\begin{CJK*}{UTF8}{gbsn}
	
	\begin{frontmatter}
		
		\title{Nuclear Physics at BRIF}

		\author[ciae]{Wei~Nan~(南巍)}
		
		\author[ciae]{Bing~Guo~(郭冰)\corref{cor1}}
        \cortext[cor1]{Corresponding author.}
        \ead{guobing@ciae.ac.cn}

        \author[sustech]{Jie~Chen~(陈洁)}
        \author[ciae]{Baoqun~Cui~(崔保群)} 
        \author[ciae]{Wei~Fu~(付伟)}
        \author[ciae]{Xianlu~Jia~(贾先禄)}
        \author[ciae]{Chaoxin~Kan~(阚朝新)}
        \author[ciae]{Jiayinghao~Li~(李家英豪)}
        \author[ciae]{Yunju~Li~(李云居)}
        \author[ciae]{Chengjian~Lin~(林承键)}
        \author[ciae]{Yihui~Liu~(刘亦晖)}
        \author[ciae]{Nanru~Ma~(马南茹)}
        \author[ciae]{Zhaohua~Peng~(彭朝华)}
        \author[ciae]{Yangping~Shen~(谌阳平)}
        \author[ciae]{Guofang~Song~(宋国芳)}
        \author[bnu]{Jun~Su~(苏俊)}
        \author[ciae]{Bing~Tang~(唐兵)}
        \author[ciae]{Haorui~Wang~(王浩睿)}
        \author[ciae]{Youbao~Wang~(王友宝)}
        \author[ciae]{Lei~Yang~(杨磊)}
        \author[pku]{Xiaofei~Yang~(杨晓菲)}
        \author[ciae]{Zhiguo~Yin~(殷治国)}
        \author[ciae]{Yun~Zheng~(郑云)}
        
		\author[ciae]{Tianjue~Zhang~(张天爵)\corref{cor1}}
        \ead{13641305756@139.com}
        
  	\author[ciae]{Weiping~Liu~(柳卫平)\corref{cor1}}
        \ead{liuwp@sustech.edu.cn}    

		\address[ciae]{China Institute of Atomic Energy, P.O. Box 275(10), Beijing, 102413, China}
		\address[sustech]{College of Science, Southern University of Science and Technology, Shenzhen, 518055, China}
        \address[bnu]{Key Laboratory of Beam Technology of Ministry of Education, \\School of Physics and Astronomy, Beijing Normal University, Beijing 100875, China}        
        \address[pku]{School of Physics and State Key Laboratory of Nuclear Physics and Technology, Peking University, Beijing 100871, China}
		
		\begin{abstract}
			The Beijing Radioactive Ion-beam Facility (BRIF), based on the Isotope Separation On-Line (ISOL) technique, consists of a 100 MeV proton cyclotron as the driving accelerator, a two-stage ISOL system for ion separation, a 13-MV tandem accelerator for post-acceleration, a superconducting linac for further boosting beam energies. It is capable of providing ISOL beams in the energy range from 60 to 300 keV, and post-accelerated beams in the energy range from 3 to 10 MeV/u for nuclei with mass numbers of $A$ < 80. For nuclei with $A$ up to 170, energies are still able to reach 3 MeV/u. This facility offers opportunities to address key questions of current interest in nuclear astrophysics, nuclear structure and reactions of unstable nuclei. In this review we present a comprehensive introduction to the BRIF and the typical experimental instruments installed on it, and then summarize current experimental results on unstable Na and Rb isotopes and future plan for development of the BRIF to improve its performance.
		\end{abstract}
		
		\begin{keyword}
			Radioactive ion beams\sep Nuclear astrophysics\sep Isotope separation on-line\sep Proton cyclotron\sep Post-acceleration
			
		\end{keyword}
	\end{frontmatter}
\end{CJK*}		
	\newpage
	
	\thispagestyle{empty}
	\tableofcontents

	
	\newpage
	\section{Introduction}\label{intro}
    
Nuclear physics has been rapidly expanding its territory in the N-Z plane, due to the dramatic increase of available rare isotopes (equivalently radio-isotopes, denoted by RI) that have different combinations of neutrons and protons from those found in naturally occurring nuclei on the Earth~\cite{pfutzner12,nakamura17}. Present and planned rare isotope accelerators offer opportunities to explore the structure of unstable nuclei.


Since Becquerel's discovery of natural radioactivity in 1896 while studying phosphorescent materials, unstable nuclei have gradually become a frontier field in nuclear physics research. Out of over 3000 natural and artificially synthesized nuclides, only around 300 are stable. The study of unstable nuclei is one of the hot topics in current research of nuclear physics~\cite{keeley07,keeley09,bac14,canto15,ye2024}. The nuclei near the drip-line have low binding energy and can have exotic properties such as halo structure and new magic numbers, indicating that the structures and properties of nuclei far from the $\beta$-stability may have systematic evolution and reveal new nuclear physics with new reaction mechanisms and interactions. However, due to the limitations of the types and intensities of unstable beams provided by nuclear facilities, the nuclei being studied are still far from the possible boundaries, thus making the effort to generate unstable nuclear far from the $\beta$-stability still necessary.

Also, the investigation of nuclear reactions that involve unstable nuclei is one of the frontier and interesting field in nuclear astrophysics, including rapid proton capture (rp-) process~\cite{schatz98}, rapid neutron capture (r-) process~\cite{cowan91,arnould07,thielemann11,cowan21}, slow neutron capture (s-) process~\cite{kappeler11}, intermediate neutron capture (i-) process~\cite{jones15,hampel16,choplin21} and photo-dissociation (p-) process~\cite{arnould03,sauerwein11,rauscher13}. In high-temperature and high-density conditions such as in novae,  X-ray bursts and supernovae, these processes take place and directly affect the abundance of key nuclei through the production and elimination of unstable nuclei~\cite{kubono2022,diehl2018}. In order to perform measurements of these important nuclear processes in the laboratory, it is necessary to study the structures, reactions, and decay characteristics of a large number of unstable nuclides. These studies can help us to understand the cosmic evolution and the origin of matter.

\begin{figure}[!htbp]
    \centering
    \includegraphics[width=0.6\linewidth]{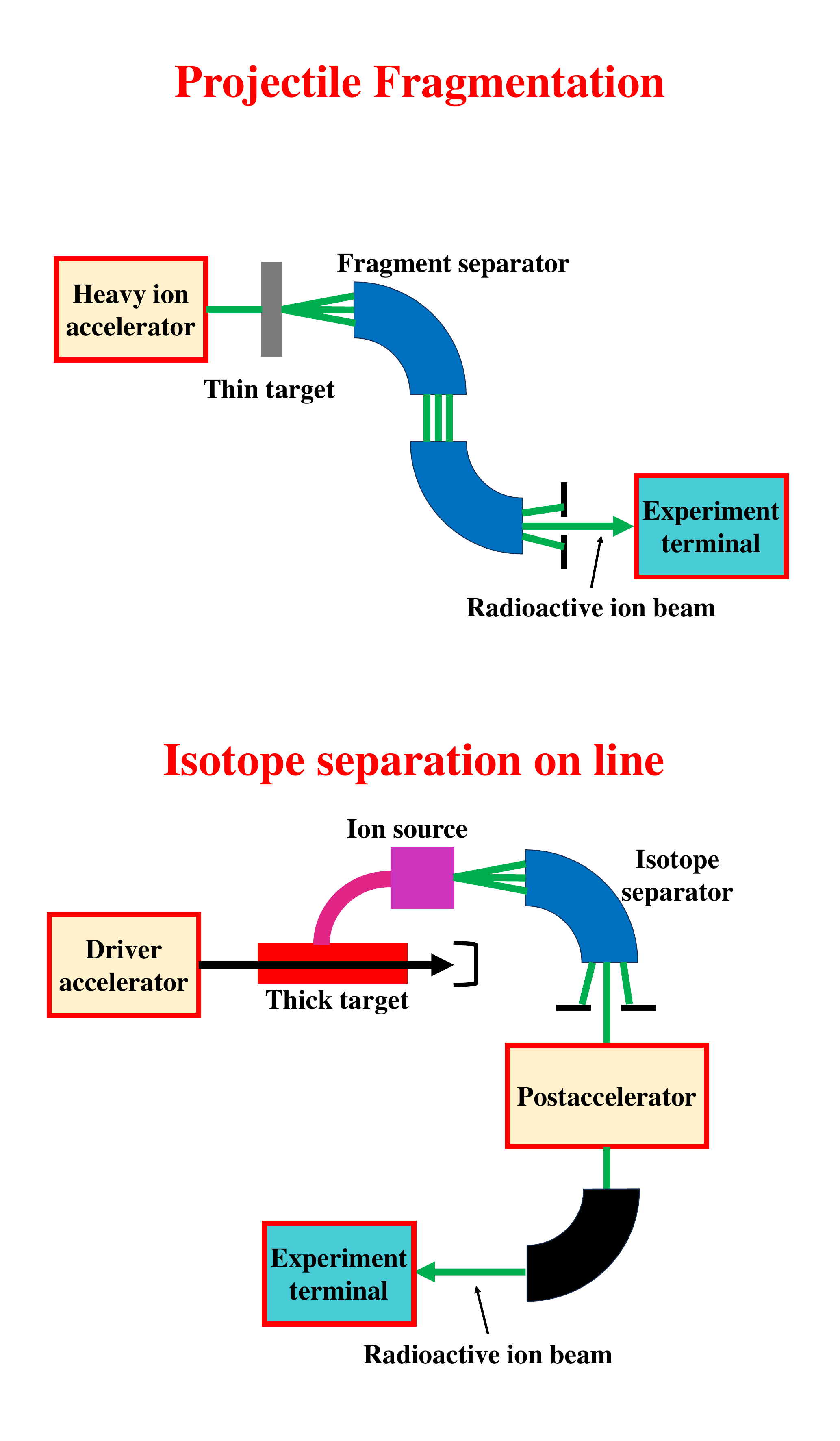}
    \caption{Schematic diagram of the generation of radioactive nuclear beams.}
    \label{fig:methods}
\end{figure}

	
Unstable nuclei are difficult to make as targets, so stable nuclei are usually used as targets in studies of unstable nuclear reactions, and unstable nuclei are used as beams. The unstable nuclear beam, also called the radioactive ion beam (RIB), was first introduced in 1967 by Hansen et al. at the Isotope mass Separator On-Line facility (ISOLDE) and several short-lived isotopes were studied~\cite{isolde69}. RIBs were first used for exotic nuclear study in 1985 by Tanihata et al.~\cite{tanihata85} at the Lawrence Berkeley National Laboratory to measure the interaction cross section and nuclear radius of the light p-shell region. The generation of unstable nuclear beam opened radioactive nuclear physics the frontier of nuclear physics~\cite{ma21}. Radioactive nuclear beam facility was developed based on this technique and has been one of the main focuses in nuclear physics for decades, which provides a powerful tool to explore the nuclides in the domain of incognita and extend our knowledge accumulated with stable beams. About 270 stable and 50 long-lived nuclides are naturally occurring, which come out as remnants of various stellar nucleosynthesis processes in different astrophysical scenarios including the recently observed merger of neutron stars. The methods for generating unstable nuclear beams include projectile fragmentation (PF, also called in-flight isotope separation), Isotope Separation On-Line (ISOL) and two-step method (ISOL+PF). The principle of PF and ISOL methods is shown in~\ref{fig:methods}.

The PF method uses a heavy ion accelerator to generate a heavy ion beam with an energy ranging of tens of MeV/u$\sim$1~GeV/u, which is used to bombard a thin target. The products are separated, collected, and purified by an electromagnetic separation device to obtain the unstable beam required for the experiment. The half-life of the beam generated by this method can be very short, and it is easy to produce unstable beams close to the drip-line. However, the energy dispersion and spot of the beam are relatively large, and the energy of the beam is also close to the energy of the primary beam with limited purity. The ISOL method uses a driver accelerator to generate light particle beams with an energy ranging of tens of MeV/u-1~GeV/u, such as proton and $\alpha$, and uses them to bombard thick targets. The products are collected by ion sources, and then separated by online isotope separators. Finally, the separated beams can be post-accelerated by traditional accelerators and transferred to experimental terminal. The purity of the beam generated by the ISOL method is high, and the energy of the beam can also be adjusted over a large range, but only some long-lived beams can be produced. The PF type facilities in operation include the FRagment Separator (FRS)~\cite{geissel92} at Gesellschaft f{\"u}r Schwerionenforschung (GSI), the RI Beam Factory (RIBF)~\cite{sakurai08} at RIkagaku KENkyusho/Institute of Physical and Chemical Research (RIKEN), the Radioactive Ion Beam Line in Lanzhou (RIBLL)~\cite{sun03} at Institute of Modern Physics (IMP) of Chinese Academy of Sciences (CAS), and the Facility for Rare Isotope Beams (FRIB)~\cite{portillo23} at Michigan State University (MSU), while the ISOL type facilities in operation include ISOLDE~\cite{kugler92} at European Organization for Nuclear Research (CERN), the Production System of Radioactive Ion and Acceleration On-Line (SPIRAL)~\cite{gales11} at Grand Acc{\'e}l{\'e}rateur National d$'$Ions Lourds (GANIL), the Isotope Separator and Accelerator (ISAC)~\cite{bricault97} at Tri-University Meson Facility (TRIUMF) and the Beijing Radioactive Ion-beam Facility (BRIF)~\cite{LIU03,cui08} at China Institute of Atomic Energy (CIAE). Some other facilities under construction include Facility for Antiproton and Ion Research in Europe (FAIR) at GSI~\cite{fair}, Heavy Ion Accelerator Facility (HIAF)~\cite{HIAF} in China and Rare isotope Accelerator complex for ON-line experiments (RAON)~\cite{kim20,lee23} in Korea.

The quality of the beam generated by the PF method is relatively poor, and it is difficult for the ISOL method to produce a short-lived beam, so one of the future development directions for the generation of radioactive nuclear beams is the two-step method (ISOL+PF). The ISOL+PF method is a combination of the PF method and the ISOL method, neutron-rich unstable nuclei are produced by a light particle beam or by neutron-induced fission and then separated by the ISOL method and then post-accelerated for further PF. The PF method is used to generate the very neutron-rich nuclear beam further away from the stability line. In addition, the PF+ISOL method can also be used in inverse way, starting from the PF method, the required radioactive nuclei are selected by electromagnetic separation and stopped in gas stopper, and then accelerated by the post-acceleration method, which can greatly improve the beam quality, e.g., the ReA project at FRIB~\cite{reafrib}.

In this paper, we introduce the BRIF facility and review the progress of nuclear physics at BRIF.  This paper is structured as follows. Section~\ref{second} introduces the composition of BRIF and the principle of generating unstable beams. Section~\ref{third} introduces the experimental instruments on BRIF. Section~\ref{fourth} introduces some physical results on BRIF. Section~\ref{fifth} introduces future developments in multiple aspects of BRIF, including the cyclotron, the ion source, the new post-accelerator and new instruments. Section~\ref{sec:sum} summarizes the main points of this paper and gives an outlook of this facility and unstable nuclear physics.

	\newpage
	\section{The ISOL-type facility BRIF}\label{second}

\begin{figure}[!htb]
    \centering
    \includegraphics[width=1\linewidth]{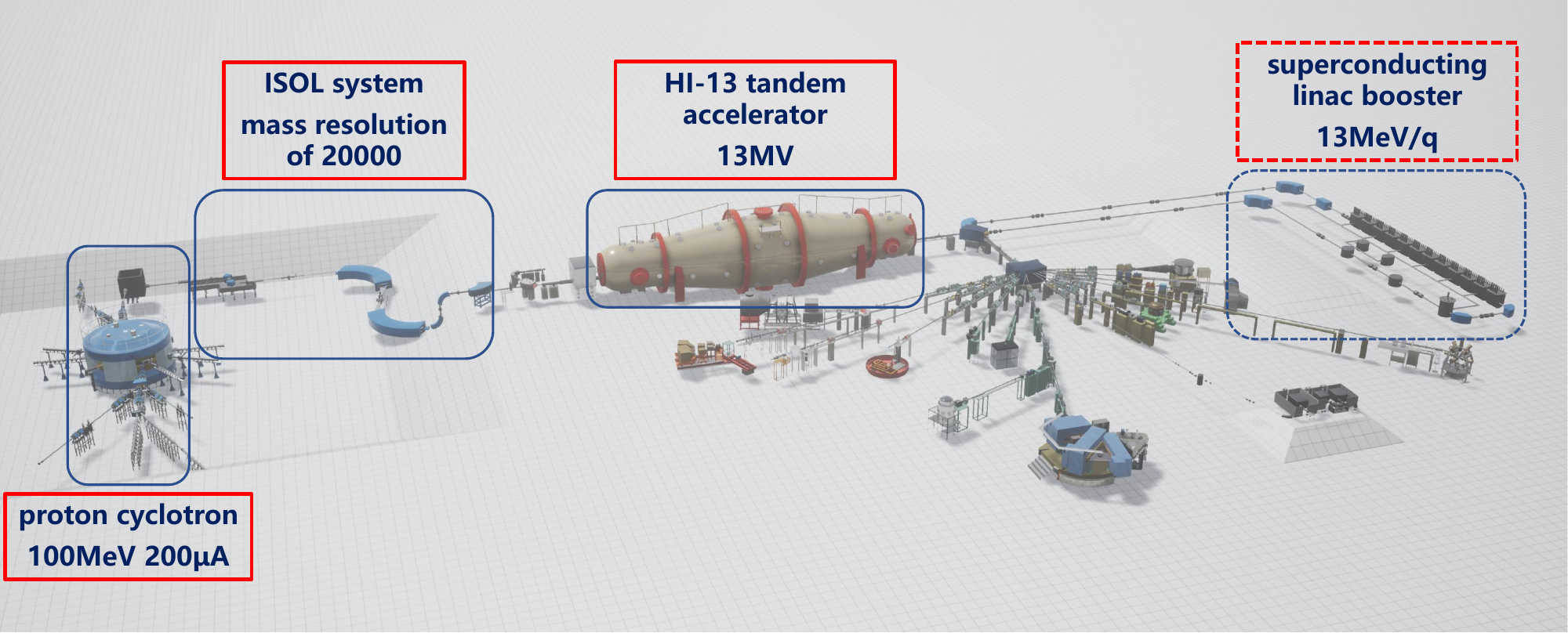}
    \caption{The overall layout of BRIF project.}
    \label{fig:2.2.1}
\end{figure}

	%
The Beijing Radioactive Ion-beam Facility (BRIF) is the first ISOL-type RIB facility in Asia and provides high-intensity proton beams and radioactive ion beams, making it an important experimental facility for nuclear physics research and nuclear industry applications in various fields, e.g. the irradiation facility for space science~\cite{zhangtianjue2020nimb}. BRIF is an ISOL-type facility which consists of a 100-MeV 200-$\mu$A compact proton cyclotron (CYCIAE-100) serving as the driving accelerator~\cite{zhangtianjue2007nimb}, a two-stage ISOL system with mass resolution of 20,000 for ion separation~\cite{cui14}, a 13-MV (HI-13) tandem accelerator for post-acceleration~\cite{tian1986}, a superconducting linac for further boosting beam energies and various fundamental research terminals. BRIF aims to conduct physics research using unstable beams in the energy range of 60~keV to 10~MeV/u, focusing on topics such as nuclear astrophysics, nuclear reactions, and the study of neutron-rich nuclei. BRIF produces proton-rich and neutron-rich beams for nuclear physics research. The project was approved in 2009 after a feasibility review and budget revisions. The first beam was obtained in 2014 for proton, and in 2015 for RIB. The overall layout of BRIF project is shown in Fig.~\ref{fig:2.2.1}.

As shown in Fig.~\ref{fig:2.2.1}, a two-stage isotope separation on-line system (ISOL) with a mass resolution of 20,000 was installed between the cyclotron and the tandem accelerator. The superconducting linear accelerator booster (SCB), added after the tandem, can increase the energy of heavy ions from the tandem by 13~MeV/q. The facility's focus is on stable operation, increasing beam intensity, and improving mass resolution for RIB production. Notable achievements include the acceleration of proton beams to 520~$\mu$A, production of neutron and RIBs, and advancement in beam stability and diagnostics. The facility is designed for diverse applications, including isotope production and nuclear physics research.

	\subsection{100-MeV proton cyclotron}\label{sub:2-2}

\begin{figure}[!htbp]
    \centering
    \includegraphics[width=1\linewidth]{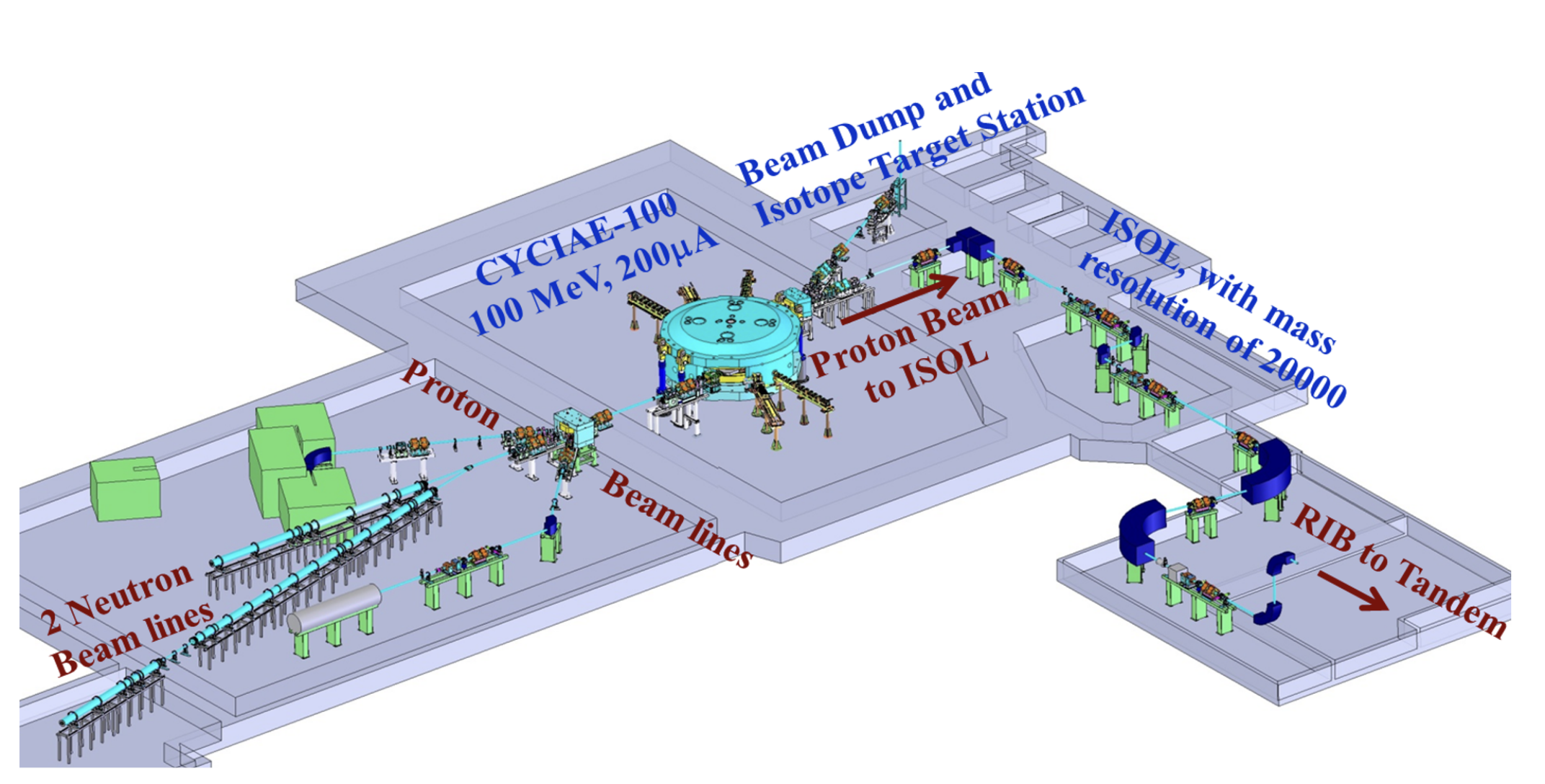}
    \caption{The overall layout of the proton cyclotron and ISOL system.}
    \label{fig:2.2.1a}
\end{figure}

\begin{figure}[!htb]
    \centering
    \includegraphics[width=0.7\linewidth]{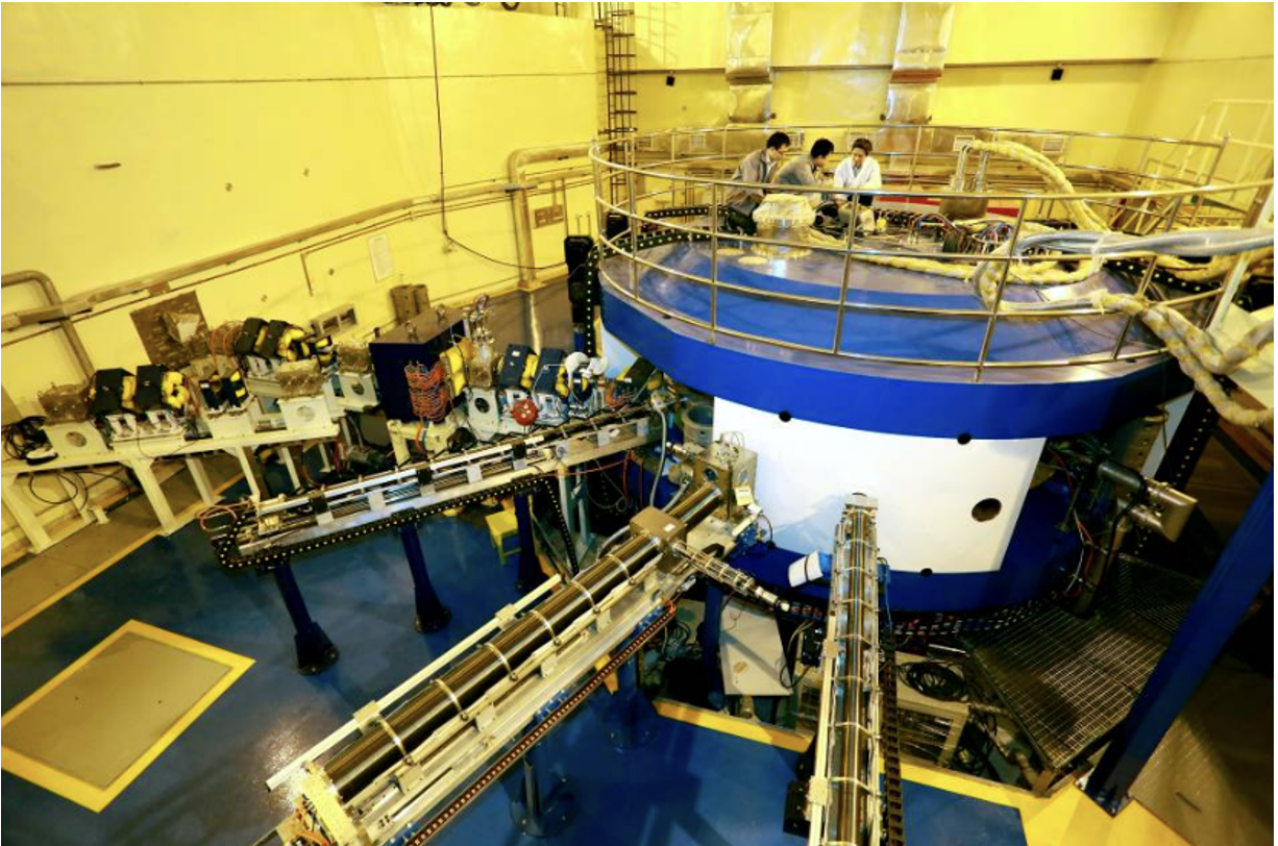}
    \caption{The 100 MeV compact cyclotron, CYCIAE-100.}
    \label{fig:2.2.2}
\end{figure}

As the front-end driver accelerator for the BRIF facility, the 100~MeV cyclotron can provide a proton beam with energy continuously adjustable between 70 and 100~MeV (from 10 to 100~MeV with energy degrader) and a current from 0.5~pA to 520~$\mu$A. A proton beam extracted from the 100~MeV cyclotron is used directly for physics experiments, while the proton beam extracted in the opposite direction is used to generate radioactive ion beams, as shown in Fig~\ref{fig:2.2.1a}, providing around 40 proton-rich beams and 80 neutron-rich beams with beam intensities exceeding 10$^{6}$~pps. CYCIAE-100 is a high-intensity compact cyclotron in this energy range. Due to its continuously adjustable energy and high beam intensity, the conventional cyclotron scheme using electrostatic deflection extraction is not employed. Instead, the compact high-intensity H$^{-}$ cyclotron, extracting proton beam by stripping, designed by the China Institute of Atomic Energy (CIAE) became the preferred solution. By choosing a compact magnet and adopting the acceleration of negative hydrogen ions with stripping extraction technology, the accelerator's structure is more compact and cost-effective. This front-end accelerator is a four-sector fixed-field cyclotron with a magnet height of 2.57~meters and a diameter of 6.16~meters, as shown in Fig.~\ref{fig:2.2.2}. Two RF cavities were installed in the valleys of the magnet independently, allowing ions to be accelerated four times, to get higher energy gain per turn. The main parameters of CYCIAE-100 are shown in Tab.~\ref{tab:2.2.1}~\cite{zhangtianjue2007nimb}.

\begin{table}[!htbp]
     \caption{Main parameters of 100 MeV H$^{-}$ Cyclotron CYCIAE-100.} 
     \centering
    \begin{tabular}{ll}
    \hline
         Parameter & Value \\
         \hline
Injection energy&$\approx$40~keV\\
Ion source current&$\geq$5~mA\\
Number of Dees&2\\
Dee angle&36$^\circ$\\
Dee Voltage&60$\sim$120~kV\\
Harmonic mode&4\\
Frequency&44.37~MHz\\
Number of sectors&4\\
Sector angle&$\approx$47$^\circ$\\
Field in hill&1.35~T\\
Field in valley&0.15~T\\
Radius of the pole&2000~mm\\
Outer radius of the yoke&3080~mm\\
Gap between the hills&50$\sim$60~mm\\
Total weight of iron&435~tons\\
\hline
    \end{tabular}
    \label{tab:2.2.1}
\end{table}

At the end of 2013, all subsystems of the cyclotron CYCIAE-100 were installed and assembled onsite. Through continuous efforts, the CYCIAE-100 ultimately integrated the compactness of AVF cyclotrons and the strong focusing of separated sector cyclotrons. The main magnet, weighing 416~tons, achieved a precision of 0.05~mm and the two RF resonators achieved a quality factor of 9500, the highest among existing compact cyclotrons. A high-speed cryo-panel system enhanced the vacuum to 6.7$\times$10$^{-8}$~mbar to minimize beam losses. 

After all systems were installed and operated normally, beam commissioning subsequently began. First, the ion source was tuned to provide a 5~mA, 35~keV hydrogen beam. On December 18, 2013, a 320~$\mu$A DC beam was measured on an internal target, with over 60$\%$ transmission efficiency. By June 16, 2014, the internal target was moved to the 1 MeV region, yielding a 109~$\mu$A beam current and over 10$\%$ injection efficiency. The first 100~MeV beam was observed on July 4, 2014, marking a project milestone. The beam current stabilized above 25~$\mu$A for over 8~hours, meeting acceptance criteria for the project's first phase~\cite{zhangtianjue2016nimb}.

\begin{figure}[!htbp]
    \centering
    \includegraphics[width=0.8\linewidth]{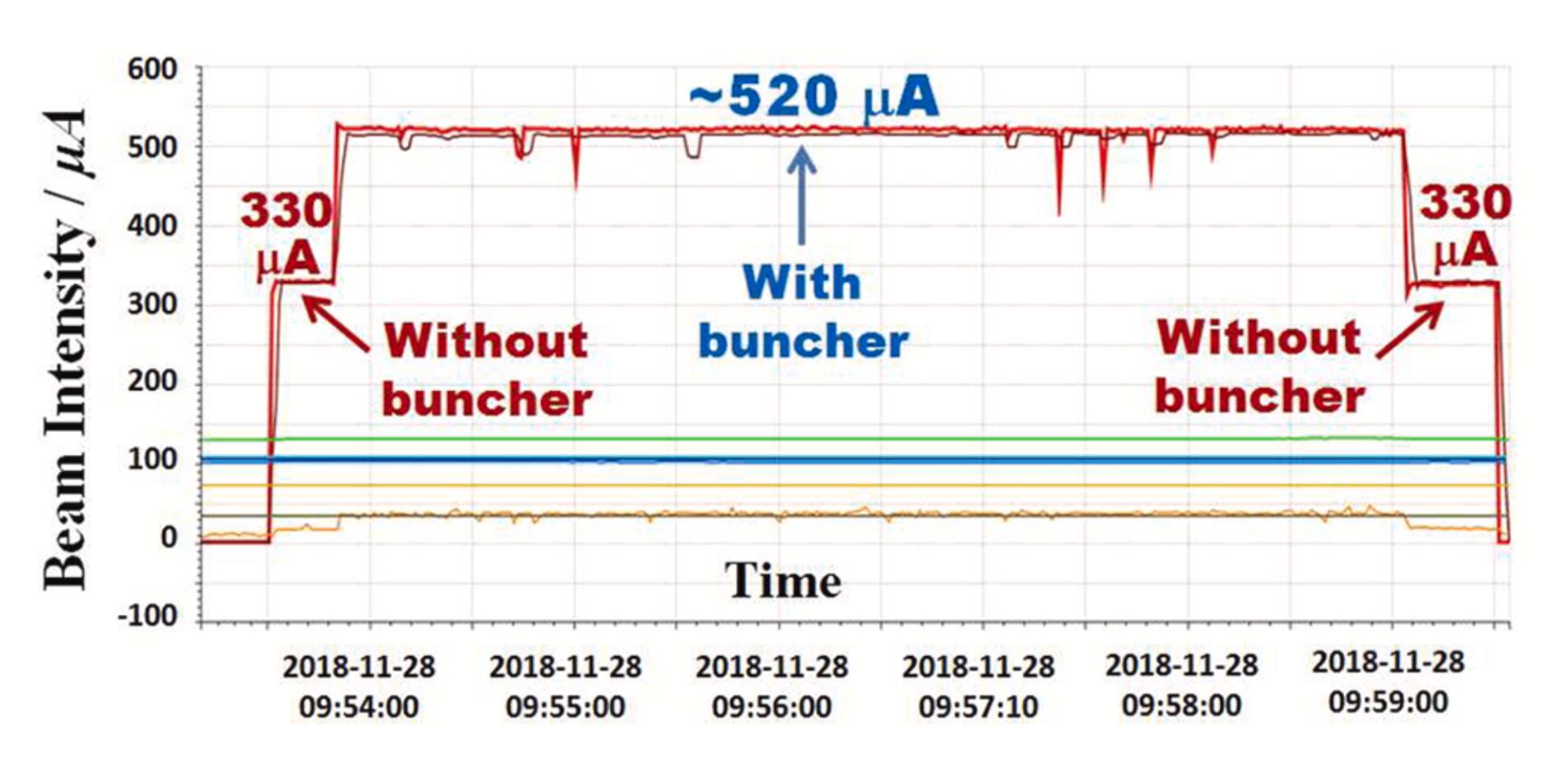}
    \caption{The 52 kW CW proton beam production.}
    \label{fig:2.2.18}
\end{figure}

In 2018, the 100~MeV extracted beam was first commissioned to 200~$\mu$A, and after using an internal target at~1 MeV to block the beam, the beam intensity in the central region was increased to over 500~$\mu$A. The internal target was then removed, resulting in an extracted beam of over 500~$\mu$A at the extraction beam line. During this process, only the RF cavity phase needed to be adjusted to limit the reflected power, ensuring the stability and acceleration efficiency of the RF system. As shown in Fig.~\ref{fig:2.2.18}, an average power of $\approx$52~kW~CW proton beam is achieved, and the bunching efficiency is up to 160$\%$. The time for such high power beam extracting was limited by 5$\sim$10~minutes. The overall efficiency, including acceleration from 1~MeV to 100~MeV, stripping extraction, and transmission of the beam line, is approximately 99.07$\%$~\cite{zhangtianjue2020nimb2}.

\subsubsection{Beam dynamics, beam loss and radiation field simulation}\label{sub:2-2-2}

In this subsection, we mainly introduce the beam dynamics studies conducted for high-intensity beams, including static orbit, acceleration path, phase space matching, resonance crossing, space charge effect, beam loss, and radiation field. 

\begin{figure}[!htbp]
    \centering
    \includegraphics[width=0.5\linewidth]{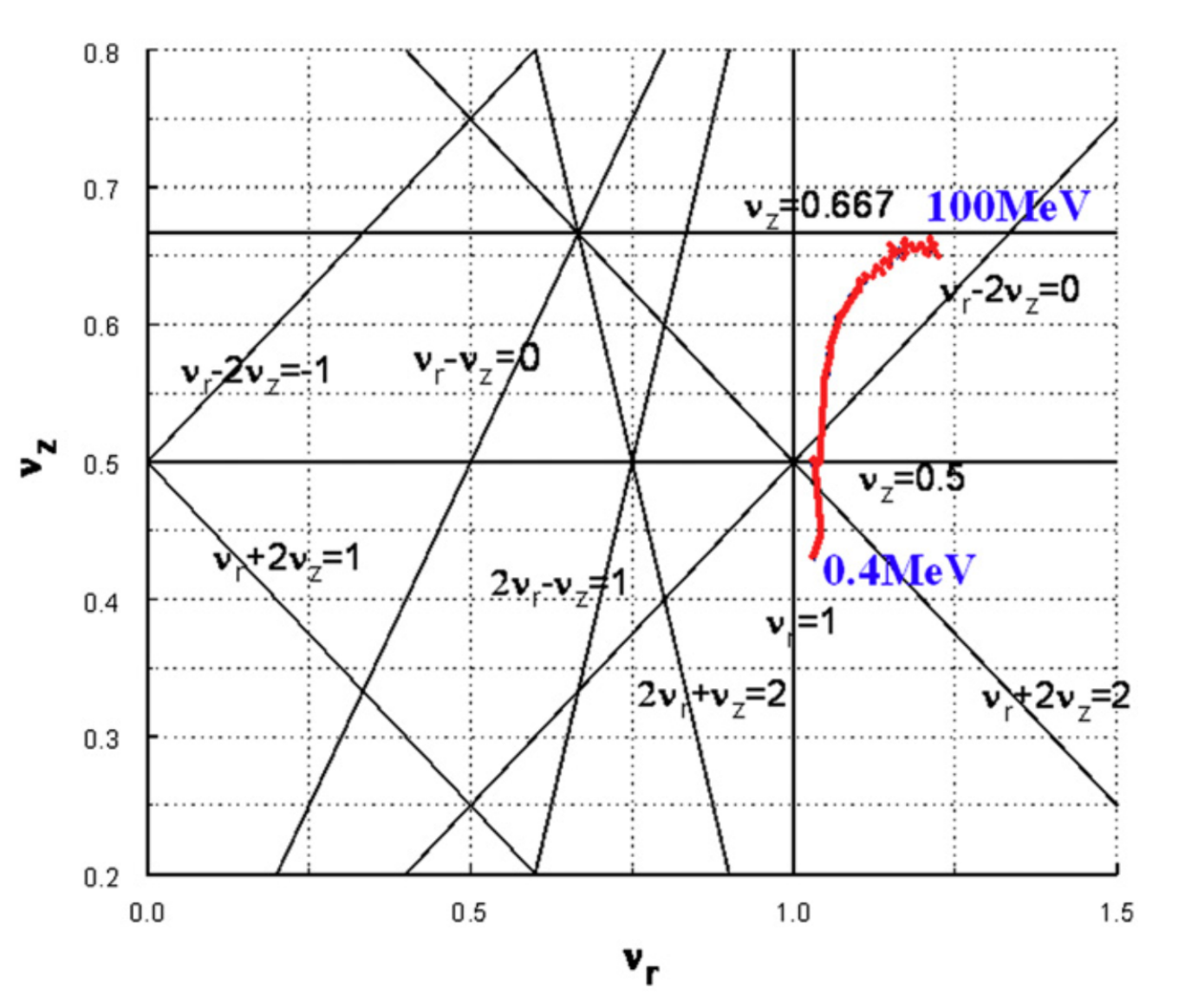}
    \caption{Tune diagram of CYCIAE-100.}
    \label{fig:2.2.3}
\end{figure}

\begin{figure}[!htbp]
    \centering
    \includegraphics[width=0.5\linewidth]{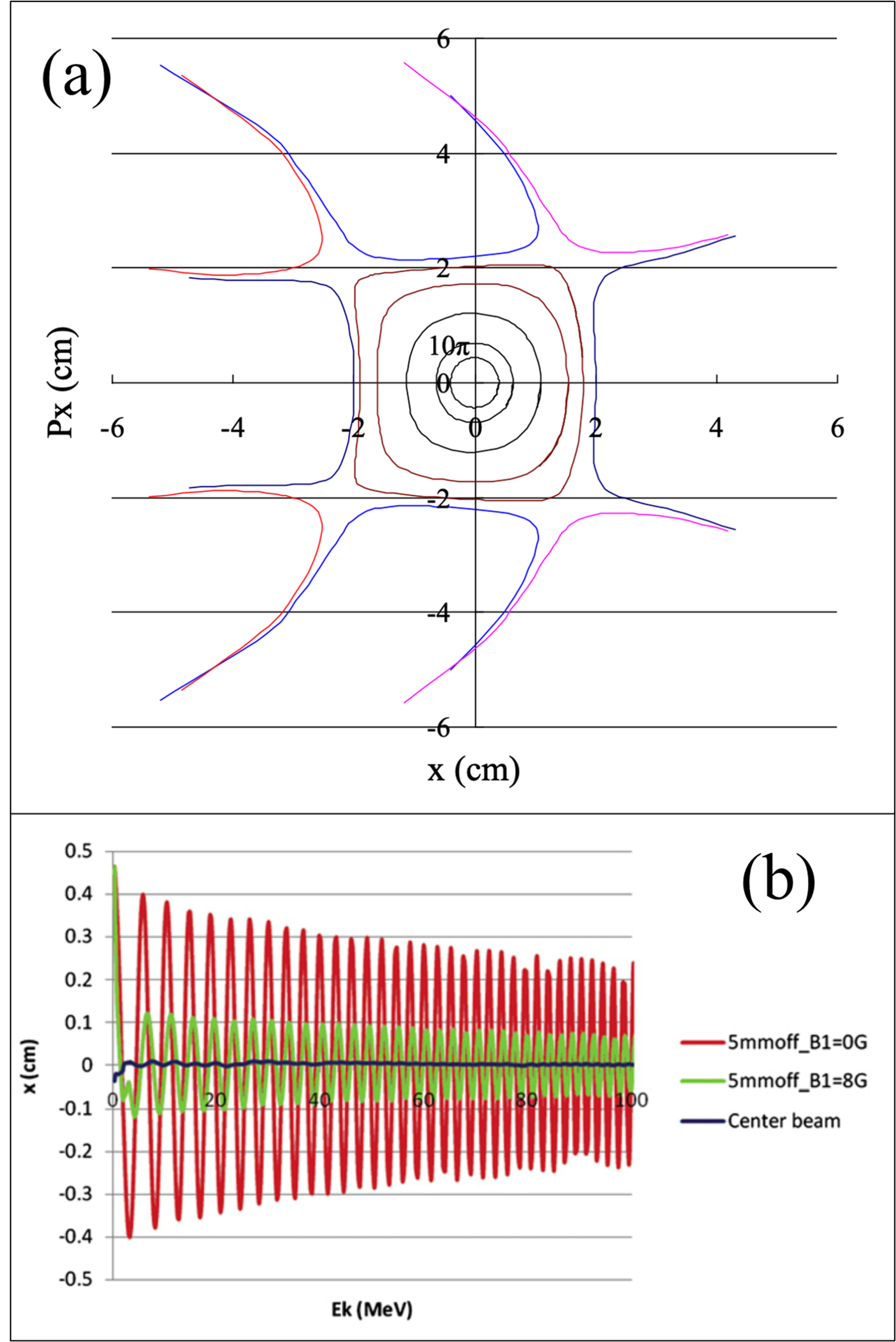}
    \caption{(a)Radial phase space diagram at 0.1~MeV. (b)Coherent oscillation of center beam and 5~mm off center beam.}
    \label{fig:2.2.4}
\end{figure}

Detailed simulation calculations confirmed that the focusing is enough and the stability region is sufficiently large, with good phase space matching in both radial and vertical directions~\cite{zhangtianjue2012nima}. The tune diagram of CYCIAE-100 calculated based on the designed magnetic field is shown in Fig.~\ref{fig:2.2.3}. As can be seen from the diagram, the axial oscillation frequency reached above 0.6, ensuring the stable acceleration of high-intensity beam, even mA level dominated by space charge effect. Additionally, numerical calculations revealed a large stable region of radial phase space with a linear acceptance area of 10$\pi$ mm-mrad (normalized emittance) at 0.1~MeV where vertical focusing was most difficult to handle. Results indicated that the linear stable region remained sufficiently large at low energy. The accelerated orbits were also well matched with the static equilibrium orbit and it can be concluded that the radial centering error was less than $\pm$0.03 cm during the accelerating process. Considering the off-centering correction, if the beam is off center of 5~mm, 8~Gs first harmonic Gaussian bump is enough to compensate, as shown in Fig.~\ref{fig:2.2.4}(b). The coherent oscillation of the off-center beam was effectively reduced by the first harmonic bump.

\begin{figure}[!htbp]
    \centering
    \includegraphics[width=1\linewidth]{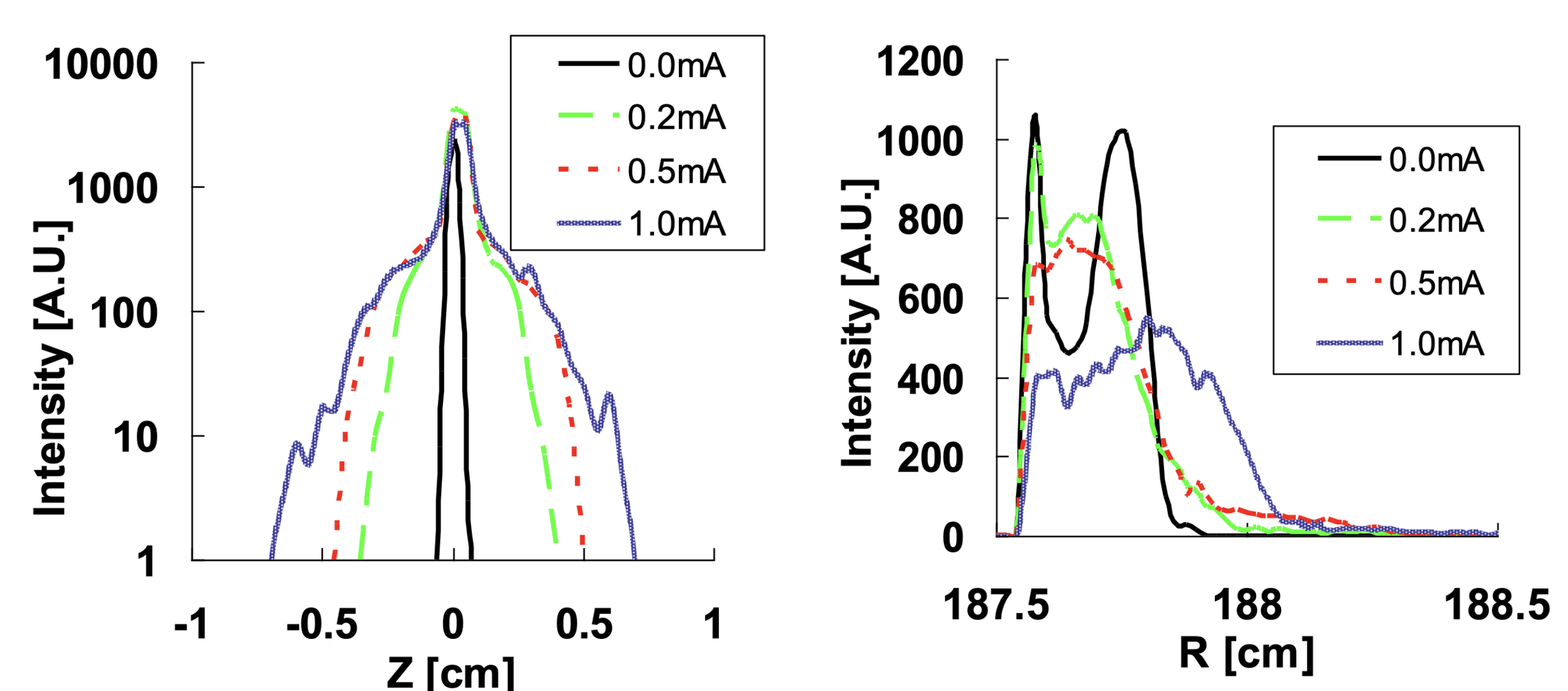}
    \caption{Comparisons of the histograms along the Z and R directions for the extracted beam for different beam currents under the steady state.}
    \label{fig:2.2.5}
\end{figure}

\begin{figure}[!htbp]
    \centering
    \includegraphics[width=0.6\linewidth]{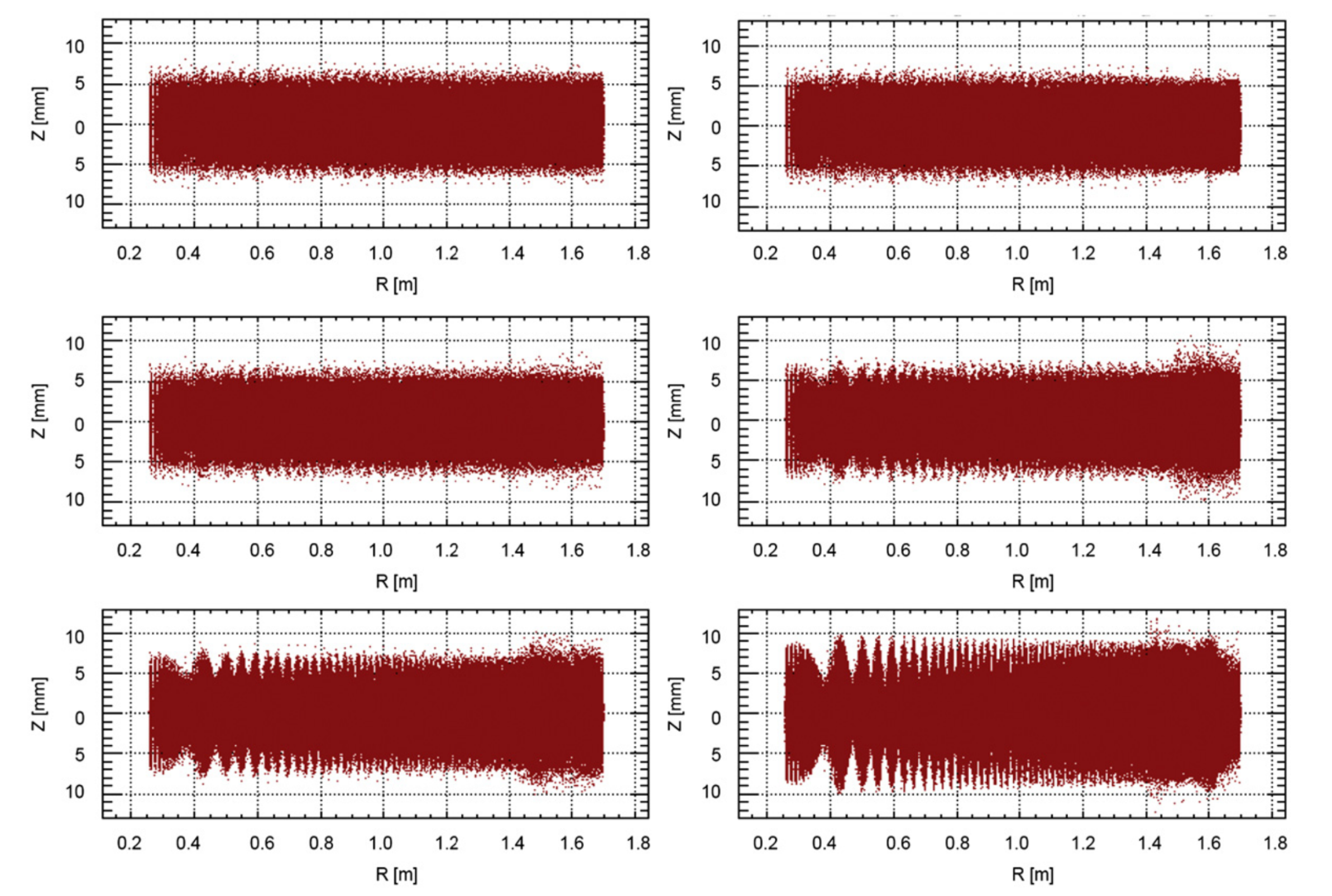}
    \caption{Projection of the particle distribution on the radial-vertical plane when bunches crossing 01 azimuth under steady state. The beam currents are 0 (top-left), 0.2 (top-right), 1 (middle-left), 5 (middle-right), 10 (bottom-left) and 20 (bottom-right) mA respectively.}
    \label{fig:2.2.6}
\end{figure}

The CYCIAE-100 aims for a beam current of 0.2$\sim$0.5~mA, with plans to increase it to 1~mA in next upgrades. In order to simulate the space charge effects of neighboring bunches in compact cyclotrons self-consistently, a precise and efficient model called "Start-to-Stop" was established~\cite{yangjianjun2011}. The model was implemented in the framework of the parallel Particle in Cell (PIC) program OPAL-CYCL~\cite{yangjianjun2010,adelmann2019} and first indicated a positive impact of the neighboring bunch effect in reducing transverse size and energy spread, which was verified in PSI Ring cyclotron~\cite{yangjianjun2010}. Subsequently, simulations were conducted using the new model to investigate, for the first time in the world, the effects of space charge from multiple bunches on the beam dynamics in the CYCIAE-100 at different intensity levels. The results indicated that space charge effects lead to an increased phase width during the acceleration process, causing the extracted beam size to expand in both axial and radial directions. In Fig.~\ref{fig:2.2.5} we show the histograms along the radial and axial directions respectively~\cite{zhangtianjue2009}. From the quantitative predictions by high performance PIC simulation with multi-bunches, it was predicted  that the vertical beam sizes at currents of 0, 0.2, and 1~mA remain nearly constant, each being less than 15~mm throughout the whole acceleration from the center region to the extraction region as shown in Fig.~\ref{fig:2.2.6}. When the beam current increased up to 5~mA, the oscillation of the beam envelope caused by the mismatch due to nonlinear space charge forces exceeded the acceptable range. This suggested that the space charge force was significantly controllable by vertical focusing force at a current around 1$\sim$2~mA and it did not result in significant beam loss.

Beam losses in high-intensity negative hydrogen cyclotrons mainly involved stripping losses caused by the Lorentz force and residual gases in the vacuum, beam losses during extraction, and losses due to space charge effects. For the last two types of beam loss, a precise model was specially developed and embedded into the OPAL-CYCL, which was able to simulate the beam dynamics in high-intensity machine more realistically~\cite{biyuanjie2011}. 

\begin{figure}[!htbp]
    \centering
    \includegraphics[width=0.8\linewidth]{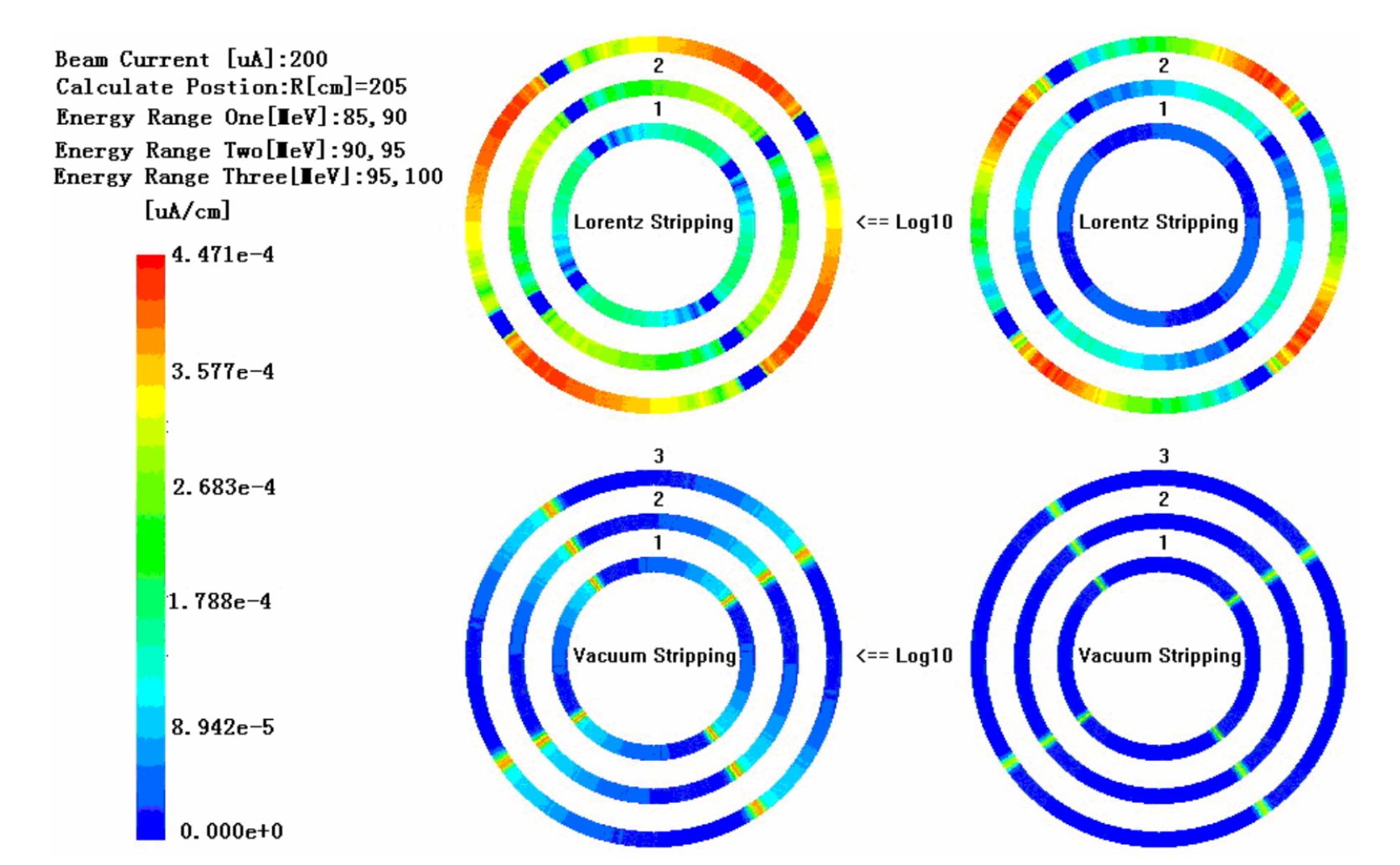}
    \caption{The distribution of beam loss on the inner side of vacuum tank.}
    \label{fig:2.2.7}
\end{figure}

For the first two types of beam loss, if one aims to accurately simulate beam losses caused by Lorentz stripping and dissociation due to residual gases, it is necessary to develop models that include particle-matter interactions. For every step of orbit tracking during beam acceleration, the beam losses were also calculated numerically. The results of beam losses by Lorentz stripping and vacuum dissociation then were used to determine the dimension of main magnet, and the vacuum system design~\cite{zhangtianjue2009pac}. To accurately simulate the radiation field of CYCIAE-100, the beam loss rate per unit length was calculated based on the equivalent lifetime of H$^{-}$ ions and integrated with OPAL. In Fig.~\ref{fig:2.2.7} we show the distribution of beam losses produced by Lorentz and gas stripping, mainly on the inner side of the vertical tank wall at the median plane. The Monte-Carlo program FLUKA~\cite{fluka} was used to estimate the radiation field produced by collisions between the stripped beam and cyclotron components. Simulated results were given for running and cooled-down cyclotrons to clarify the cooling time required before maintenance.

\subsubsection{Design, construction and shimming of the main magnet}\label{sub:2-2-4}

It is well known that increasing axial focusing will enhance the beam intensity limit of cyclotrons~\cite{baartman1995}. Most compact high-intensity cyclotrons with energies above 70~MeV adopt spiral sectors to provide sufficient vertical focusing at higher energies. However, the application of spiral sectors introduced technical challenges in practice, and higher construction costs. The CYCIAE-100 is the first high-intensity compact cyclotron in this energy range using straight-sectors. To realize strong focusing, the profile of hill gap was approximately elliptical to adjust the radial gradient of magnetic field in large radial range, with the gap decreasing from 60~mm at the center to 48~mm at the final radius, while the height of valley region reaching 1310~mm. Comprehensive studies on beam dynamics have confirmed that the deep-valley structure with adjustable radial gradient of magnetic field, provide sufficient axial focusing, especially at the high energies~\cite{zhangtianjue2016ieee}. 

The precise simulation of beam dynamics has enabled the successful implementation of quality control procedures during magnet design, mechanical machining, and assembly for the CYCIAE-100 main magnet. 

\begin{figure}[!htbp]
    \centering
    \includegraphics[width=0.6\linewidth]{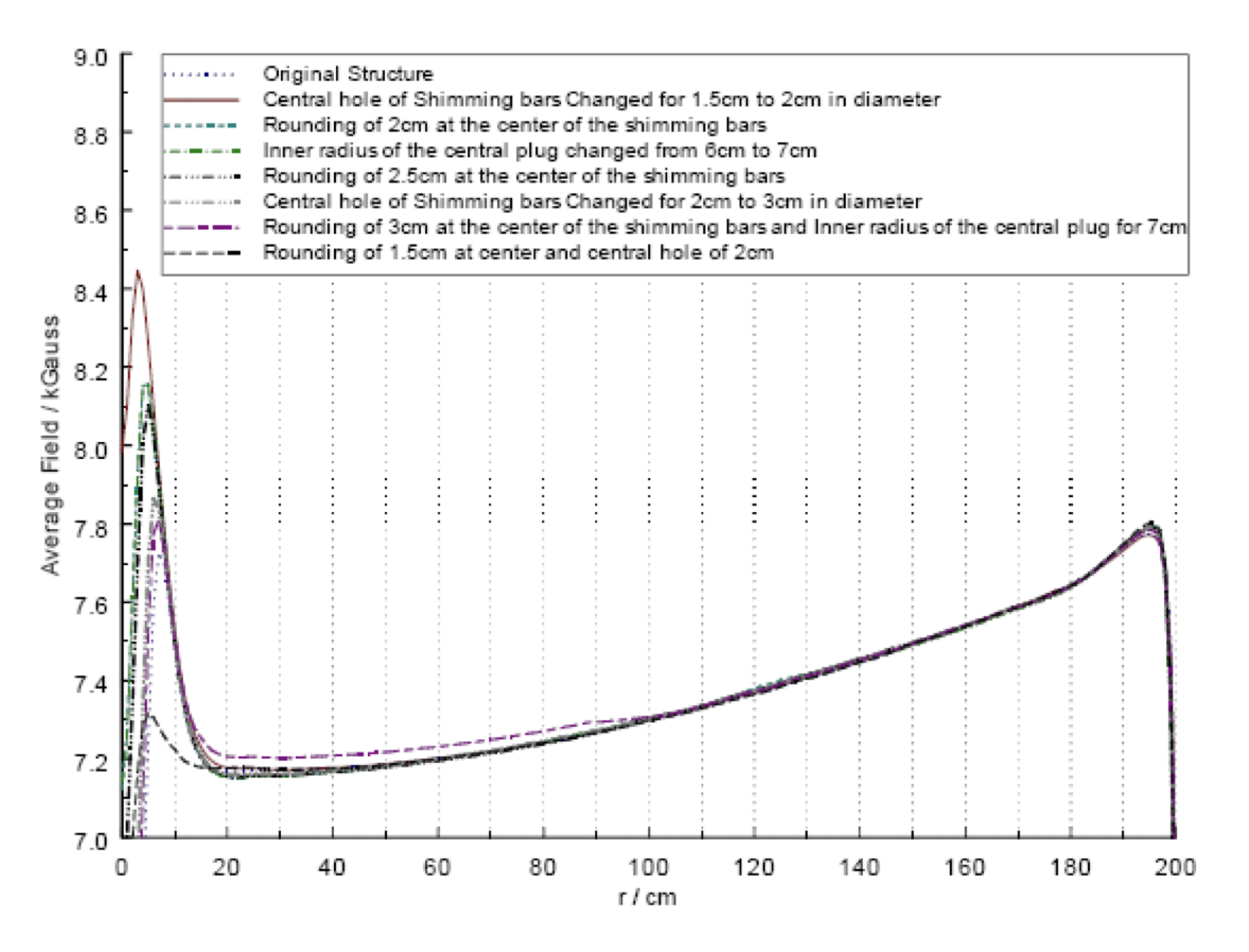}
    \caption{Comparison of the average field results after magnet structure adjustments.}
    \label{fig:2.2.8}
\end{figure}

For compact cyclotrons, the magnetic field must primarily satisfy isochronous conditions and provide sufficient axial focusing. Through three-dimensional numerical simulations, it was possible to design and study the magnetic performance of the main magnet structure and the functions of its key components, as well as to fine-tune parameters like the angle width of the shim to achieve precise adjustments of the average magnetic field. To meet the design requirements for the central region's orbits, including longitudinal acceptance and transverse focusing, the geometry of the central region's magnets has undergone multiple fine adjustments. The resulting average field outcomes are shown in Fig.~\ref{fig:2.2.8}, which plays a crucial role in beam dynamics for the central region.

The fabrication of the CYCIAE-100's top/bottom and return yokes, using Chinese standard $\#$8 low carbon steel weighing over 520 tons and requiring precise casting and thermal treatment to optimize magnetic properties, was a crucial step~\cite{wangchuan2010}. The machining precision of the elliptical hill gap is better than 0.05~mm, which is the most critical part of the main magnet. This presents a significant engineering challenge considering the magnet's weight of approximately 435~tons after finishing. The main magnet of the CYCIAE-100 cyclotron was installed 4~m underground, utilizing a 7~m wide by 6~m high hole in the west wall of the building for the installation of all magnet components, coils, and hydraulic systems, with a specially designed steel platform capable of bearing 200~tons to accommodate the height difference; precise positioning was achieved through the use of a total station and preset targets, resulting in installation tolerances of 0.10~mm in height and 0.20~mm horizontally, while other equipment, including the main coil and hydraulic elevating system, was also successfully installed with similar precision.

\begin{figure}[!htbp]
    \centering
    \includegraphics[width=0.6\linewidth]{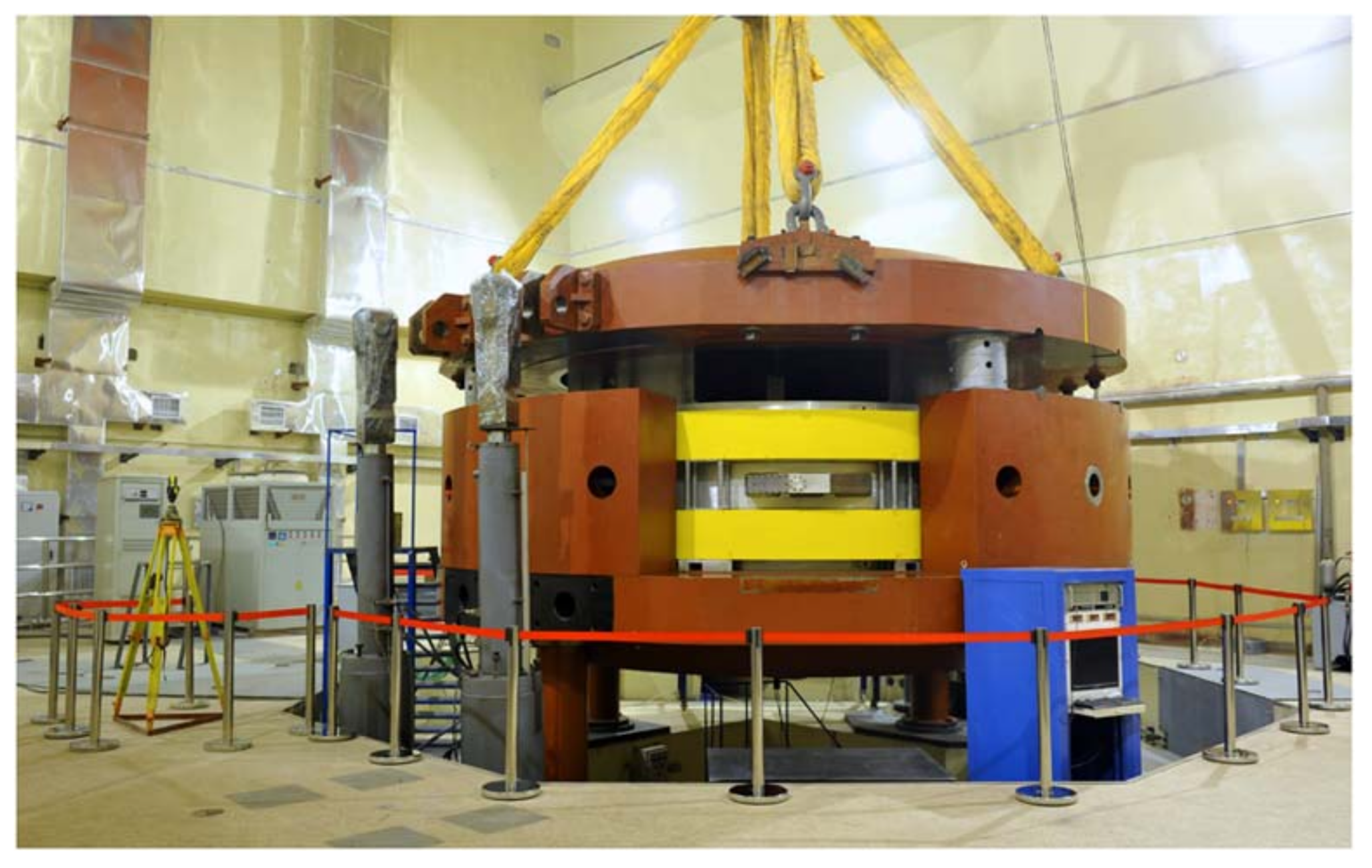}
    \caption{Main magnet, main coil, vacuum chamber, and hydraulic elevating equipment after installation.}
    \label{fig:2.2.9}
\end{figure}

\begin{figure}[!htbp]
    \centering
    \includegraphics[width=0.6\linewidth]{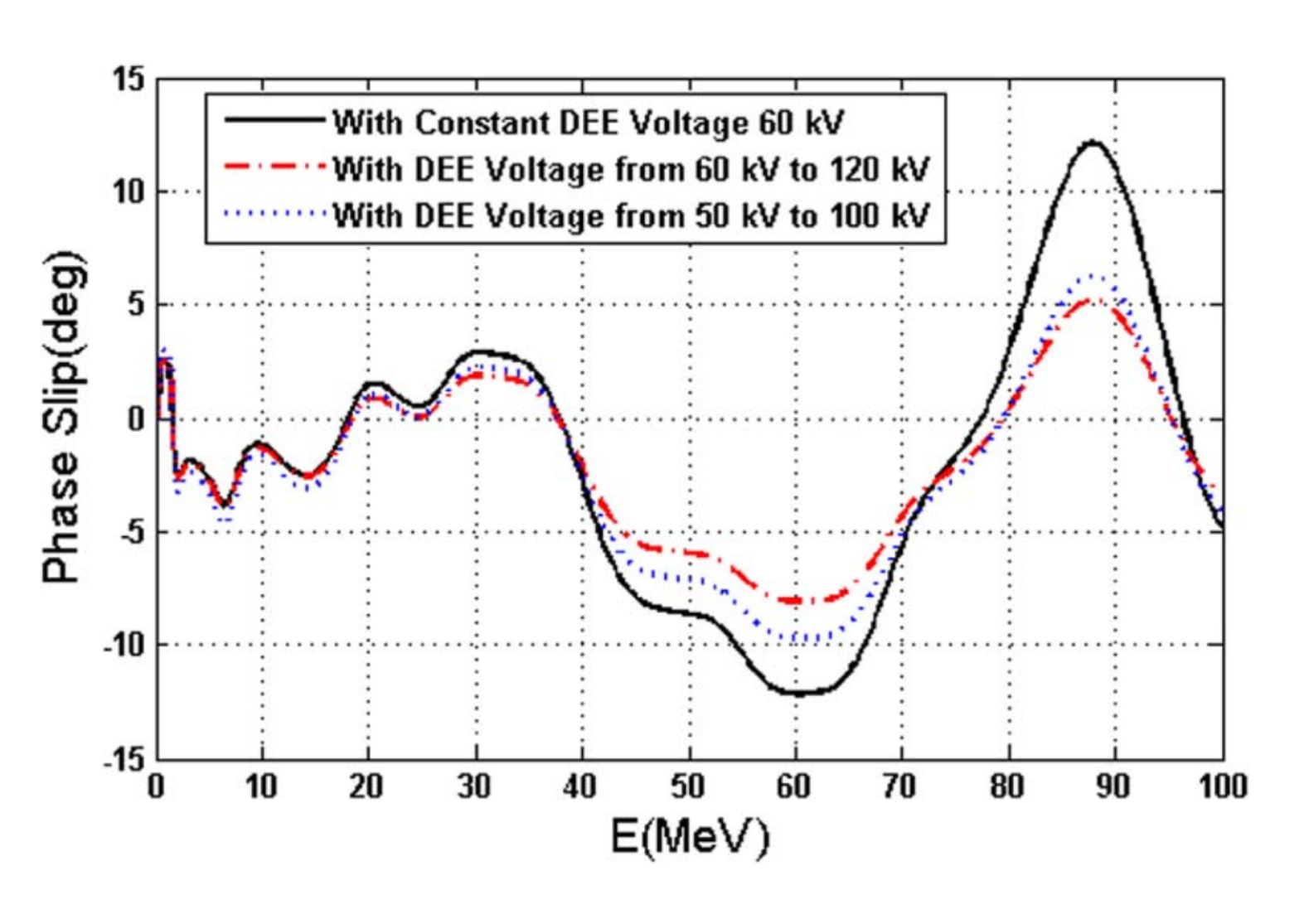}
    \caption{Integral phase slip with different cavity voltages after the final mapping field.}
    \label{fig:2.2.10}
\end{figure}

\begin{figure}[!htb]
    \centering
    \includegraphics[width=0.6\linewidth]{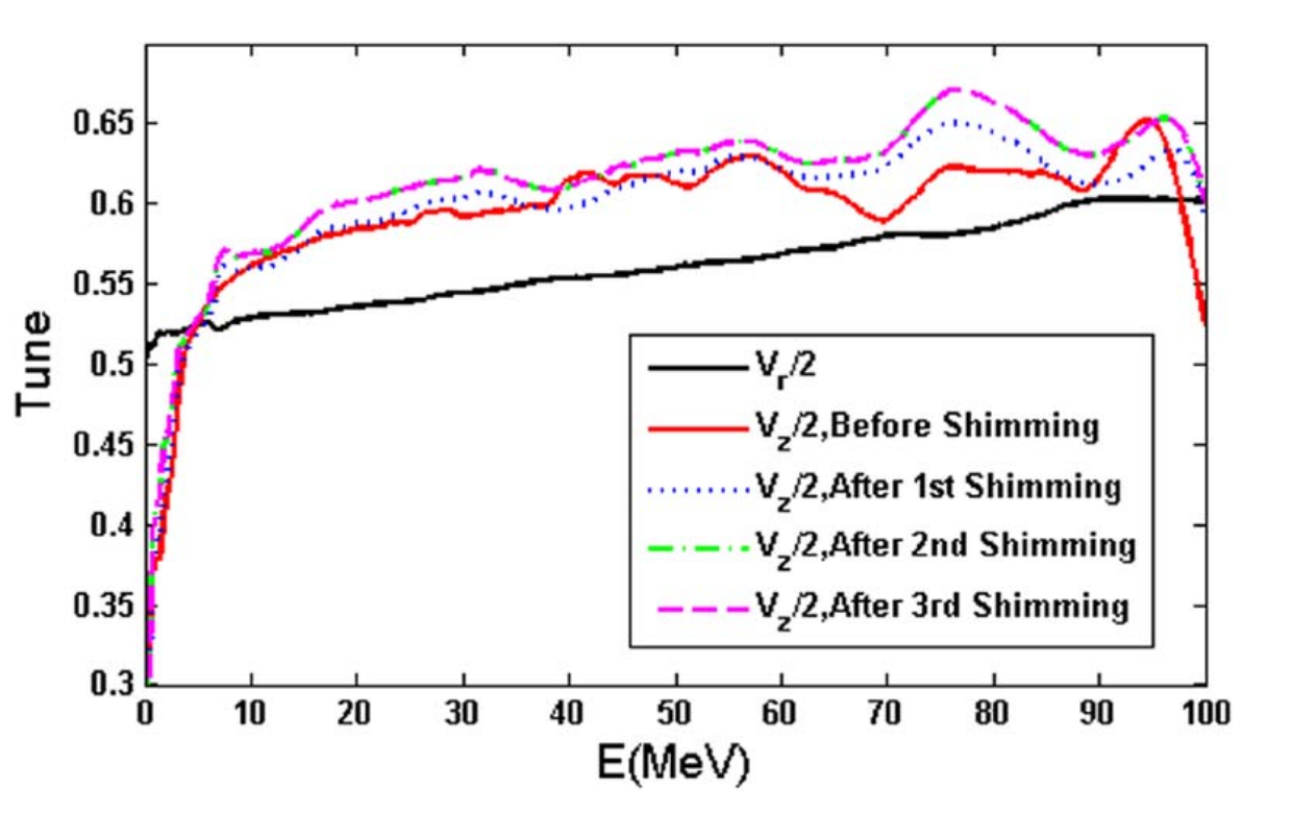}
    \caption{Vertical oscillation frequency change in the process of shimming.}
    \label{fig:2.2.11}
\end{figure}

For the first time globally, field mapping in a vacuum was performed on such a large machine. After nine rounds of magnetic mapping and shimming, the desired results were achieved. As shown in Fig.~\ref{fig:2.2.10}, the integral phase slip of the 100~MeV cyclotron after shimming ensures a high acceleration efficiency. In Fig.~\ref{fig:2.2.11} we indicate that the $\nu_{z}$ has been obviously improved at the large radius after the second time of shimming. The $\nu_{z}$ has reached 0.7 at the extraction~\cite{zhangtianjue2011nimb}. It is the highest $\nu_{z}$ in the compact cyclotrons. With the help of efficiency and highly precise mapping devices, field mapping and shimming of the main magnet to achieve the required isochronous field were completed in July 2013.

Additionally, the installation and assembly of other systems, including the RF system, injection, and extraction, were finalized right after field mapping and shimming.

\subsubsection{RF system}\label{sub:2-2-5}

\begin{figure}[!htb]
    \centering
    \includegraphics[width=0.6\linewidth]{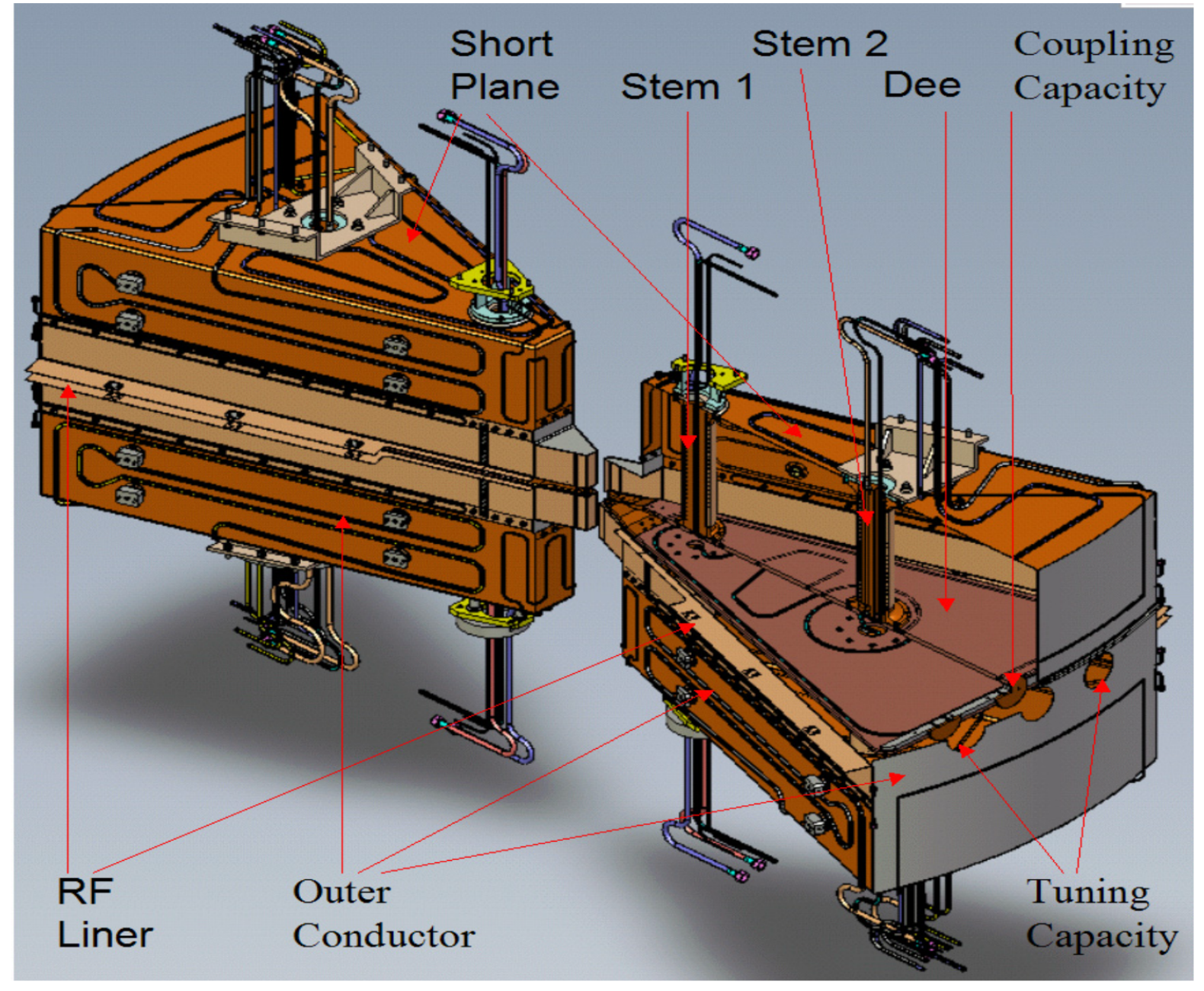}
    \caption{The structure of two stem cavity.}
    \label{fig:2.2.12}
\end{figure}

\begin{figure}[!htb]
    \centering
    \includegraphics[width=0.6\linewidth]{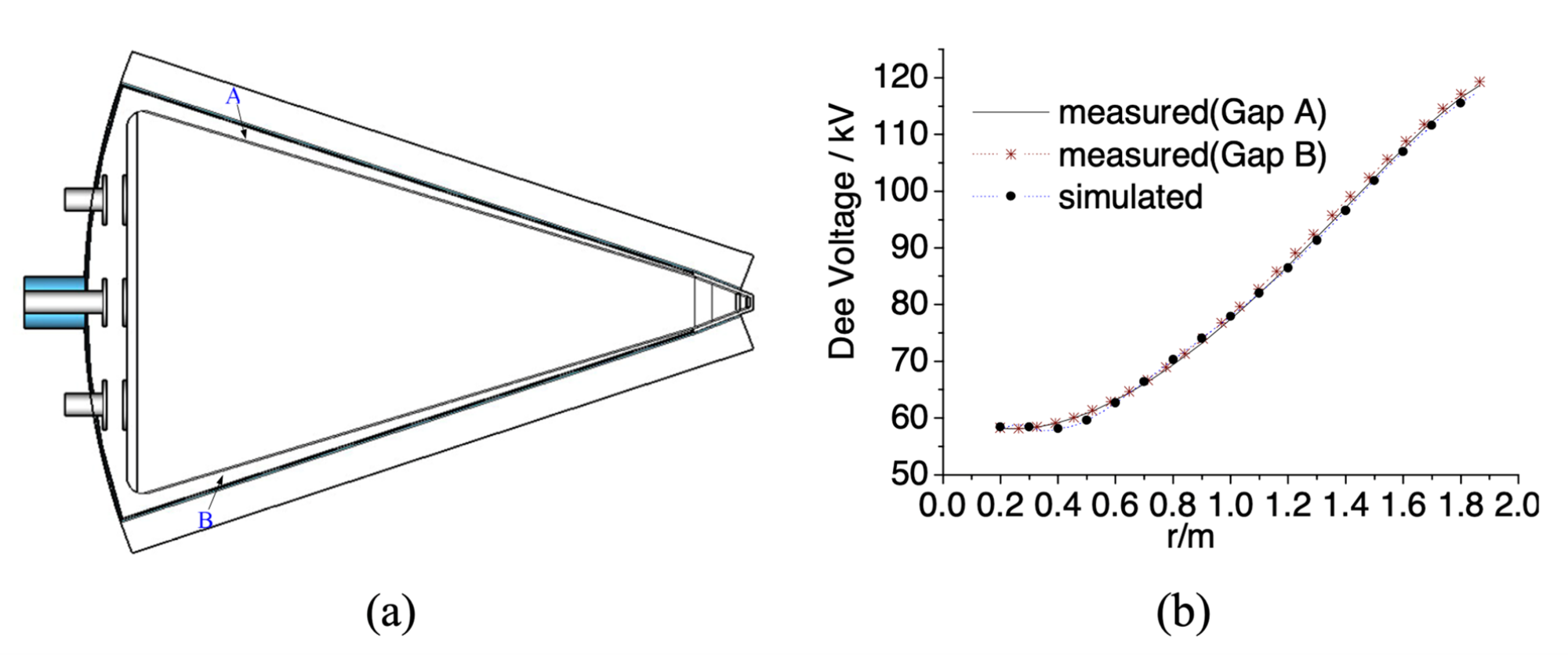}
    \caption{The accelerating gap (a) and dee voltage of the model vs radius (b).}
    \label{fig:2.2.13}
\end{figure}

The RF system of the 100~MeV cyclotron consists of two identical RF cavities as shown in Fig.~\ref{fig:2.2.12}. Two power transmission systems and two 100~kW tetrode RF amplifiers are also included. The high-power RF sections are completely independent, while the low-power RF sections share a common reference clock. As a result, the Low-Level Radio Frequency (LLRF) system can align the two cavities to the same phase for proton beam acceleration. A full-scale RF cavity was constructed following simulations using finite integral codes, with test results aligning well with the simulations, as shown in Fig.~\ref{fig:2.2.13}. Based on initial tests of the cavity, improvements were made to the design and fabrication process.

The LLRF control hardware design for CYCIAE-100 began in 2008. After the full-power test, the design was reviewed carefully. The LLRF system uses multiple Direct Digital Synthesizers (DDSs) to control the phase relationships of independent cavities and implements a unique phase searching algorithm. With the help of {Digital Signal Processor (DSP) firmware, the LLRF system for CYCIAE-100 achieved fully automatic RF system startup, including automatic resonance searching, amplitude control, and auto phase alignment~\cite{yinzhiguo2016}. This design passed testing successfully on the CYCIAE-100 cyclotron RF system. The hardware structure of the RF control system is shown in Fig.~\ref{fig:2.2.14}. The unloaded Q value measured was 9300, reaching 92$\%$ of the calculated ideal value. Moreover, the loaded Q value measured using a network analyzer was 4404. The calculated unload Q value was closed to the one we get from two sampling ports directly~\cite{jibin2010}.

\begin{figure}[!htbp]
    \centering
    \includegraphics[width=0.8\linewidth]{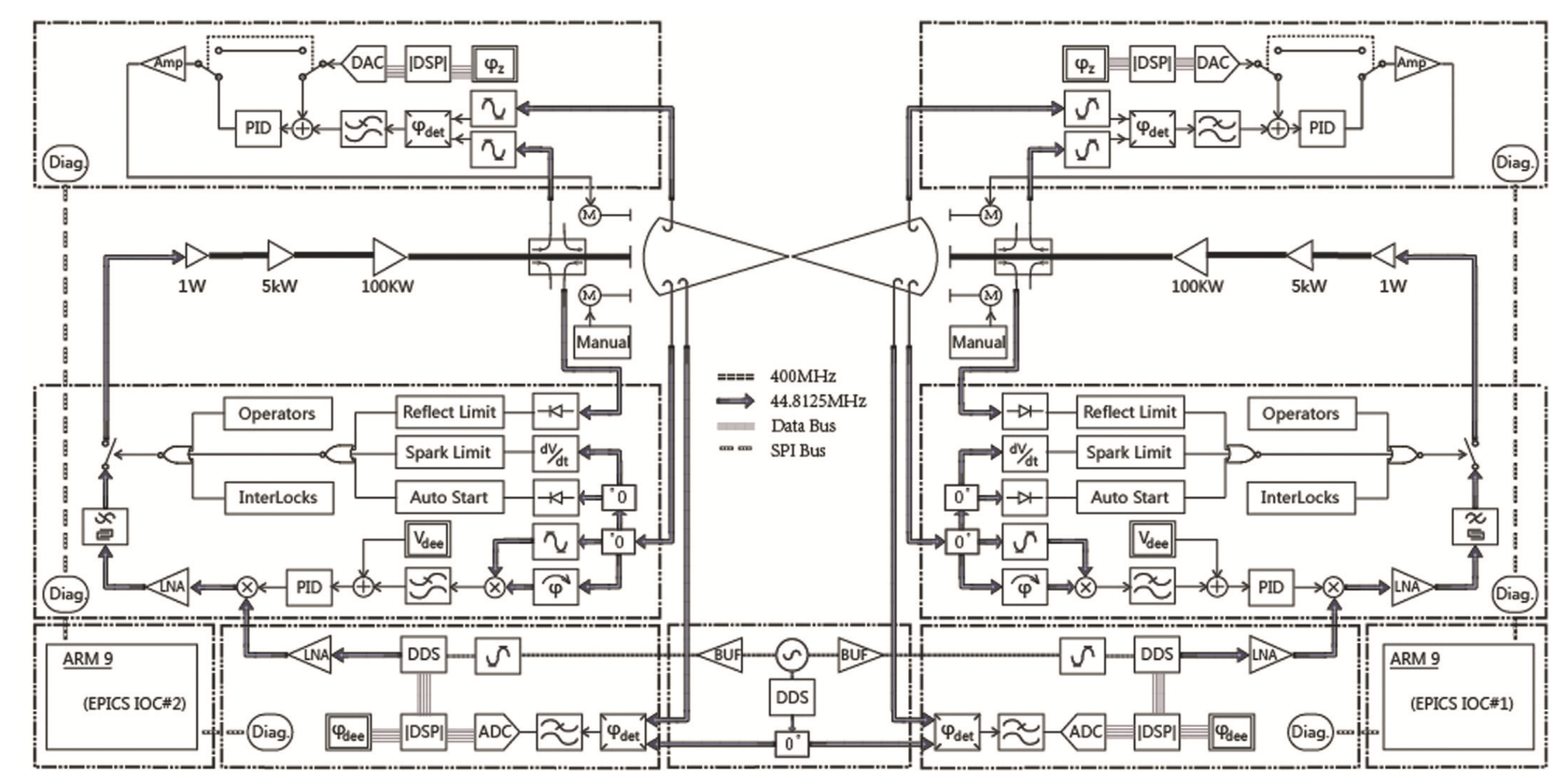}
    \caption{Block diagram for 100~MeV cyclotron RF control system hardware.}
    \label{fig:2.2.14}
\end{figure}

We also studied the impact of Dee plate deformation on beam dynamics. The deformation of the dees, with a maximum value of 0.35~mm at large radius, increased the amplitude of axial oscillation by 0.004~mm, resulting in a minimal impact on the beam. However, the offset of the dees, except for the central region, during manufacturing should be restricted to 0.2~mm to prevent significant disturbance to the beam~\cite{biyuanjie2007}.

\subsubsection{Injection and central region}\label{sub:2-2-6}

The design of the CYCIAE-100 aims for an extracted proton beam current of over 200~$\mu$A and plans to provide a pulsed beam. The cyclotron uses axial injection with two injection lines as shown in Fig.~\ref{fig:2.2.15}(a). Injection line $\#$1 neutralizes the negative hydrogen beam to enable the injection of high-intensity DC beams, using magnetic elements for transverse focusing to achieve a higher neutralization rate. Injection line $\#$2 is designed to provide a pulsed beam, utilizing electrostatic elements for transverse focusing, as the neutralization process was difficult to establish for pulsed negative hydrogen beams. The structure of the two lines allows for easy switching between operating modes. The beam envelopes including space charge for line $\#$1 and line $\#$2, calculated through TRANSOPTR~\cite{HEIGHWAY1981}, are shown in Fig.~\ref{fig:2.2.15}(b) and Fig.~\ref{fig:2.2.15}(c), respectively.

\begin{figure}[!htbp]
    \centering
    \includegraphics[width=0.8\linewidth]{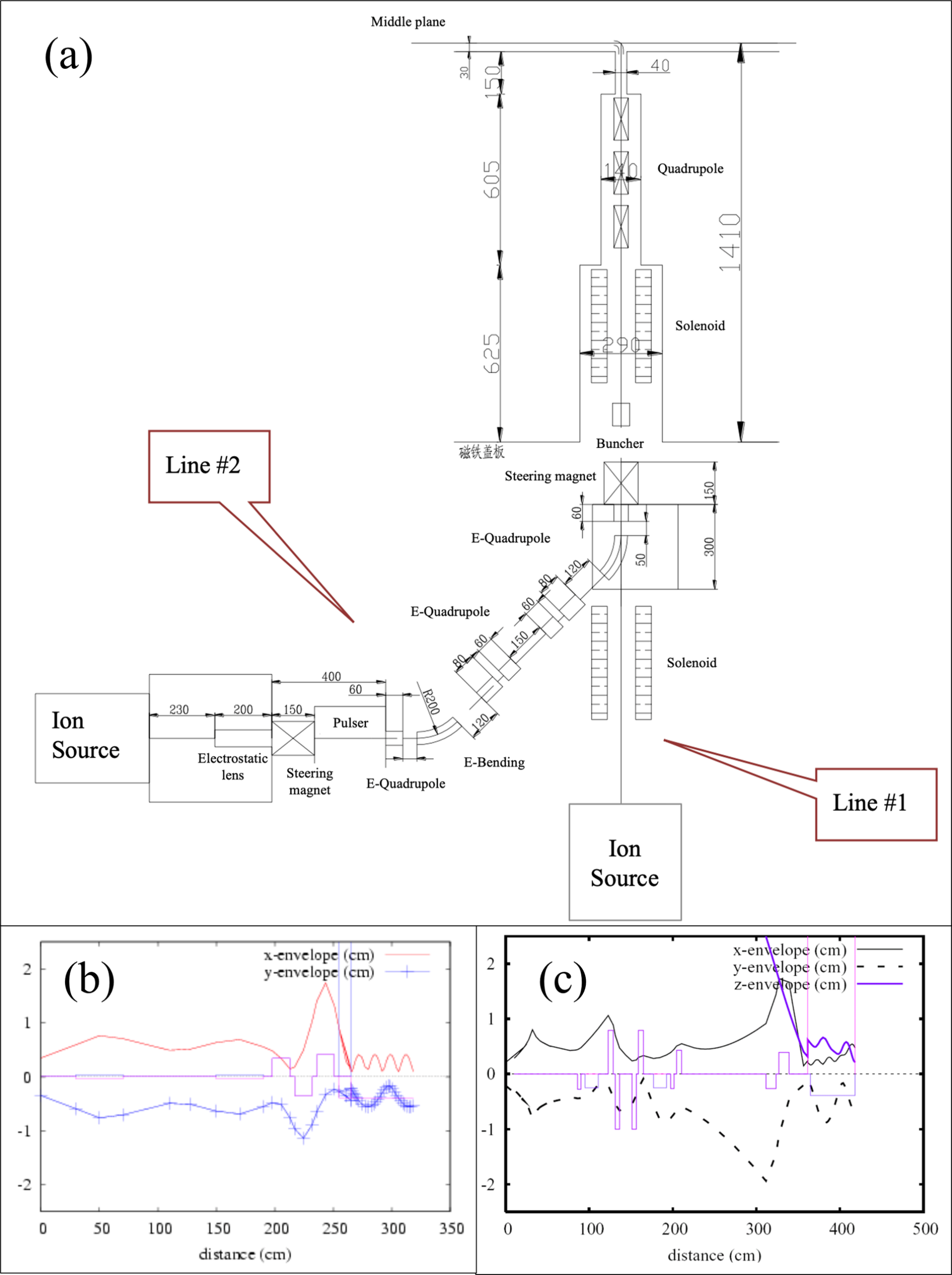}
    \caption{Layout of injection lines (a) and beam transmission transverse and longitudinal envelope diagram for injection line $\#$1 (b) and line $\#$2 (c).}
    \label{fig:2.2.15}
\end{figure}

In the central region, as low energy ions passed through the first two acceleration gaps, the curvature radius and the curvature center changed significantly. Therefore, the electric field shaped by the electrodes in the central region needed to be dedicatedly designed. We used Relax3D~\cite{houtman1994} for PC as the electric field calculation software and performed multi-particle orbit tracking based on the input electromagnetic fields, resulting in an optimized electrode structure that met the requirements for radial centering and axial focusing~\cite{yaohongjuan2007}. Using an optical calculation program that includes space charge effects, the beam optics were matched to different levels of neutralization, with RF phase acceptance of up to 40$^\circ$~\cite{ yaohongjuan2008}. The particle trajectories and phase history in the central region are shown in Fig.~\ref{fig:2.2.16}(a) and Fig.~\ref{fig:2.2.16}(b), respectively.

\begin{figure}[!htbp]
    \centering
    \includegraphics[width=0.6\linewidth]{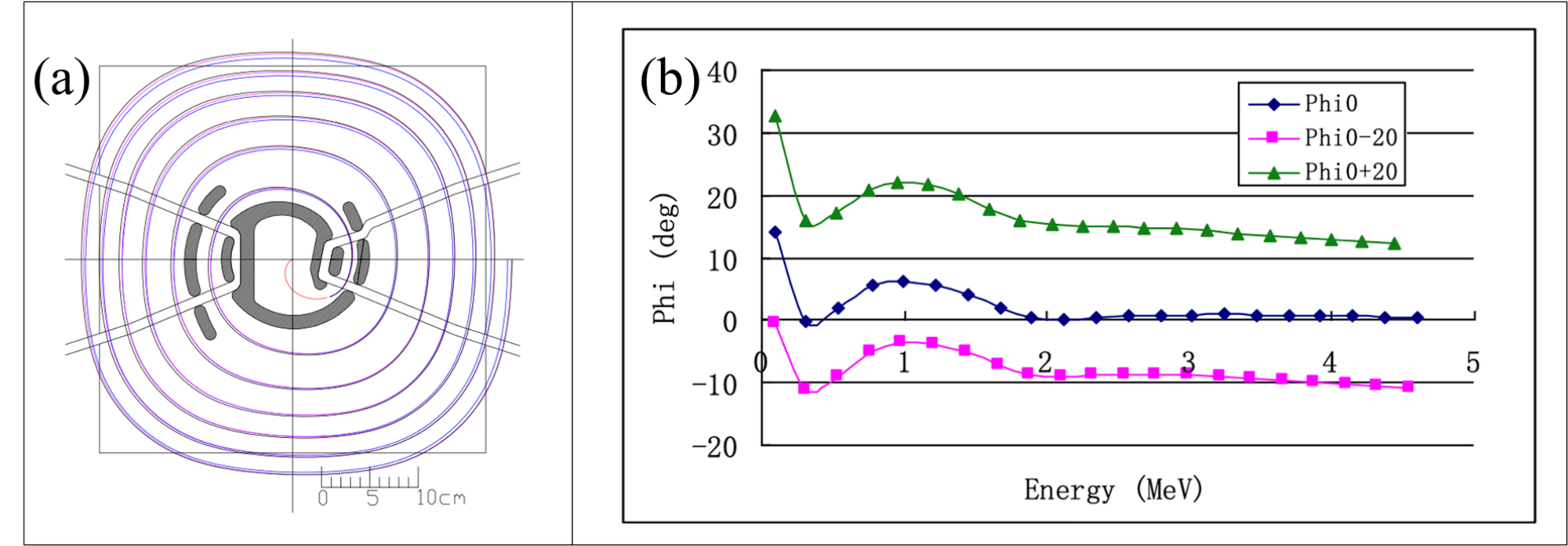}
    \caption{Particle trajectories (a) and phase history (b).}
    \label{fig:2.2.16}
\end{figure}

Continuous matching calculations from the ion source exit through the first 15 turns of acceleration showed that the designed injection system could effectively control the beam envelope and reduce beam losses. A relatively high RF acceptance in the central region enabled the 100~MeV proton cyclotron to accelerate high-intensity negative hydrogen beams.

\subsubsection{Extraction system}\label{sub:2-2-7}

The CYCIAE-100 extraction system utilizes two sets of stripping probes to extract the beam from the symmetry direction to various terminals. Two stripping probes with carbon foil were inserted radially in the opposite directions from the hill gap region and the two proton beams after stripping were transported into the crossing point in a combination magnet center separately under the fixed main magnetic field~\cite{weisumin2008}. For the CYCIAE-100, detailed simulations of the stripping extraction process were conducted for long bunches with a phase width of  $\Delta\Phi$~=~$\pm$20$^\circ$ and short bunches with $\Delta\Phi$~=~$\pm$2$^\circ$. The positions of the stripping points for different extracted beam energies were fixed after studying for the optical trajectories of the extracted beam, as shown in Fig.~\ref{fig:2.2.17}~\cite{anshizong2009}.

\begin{figure}[!htbp]
    \centering
    \includegraphics[width=0.6\linewidth]{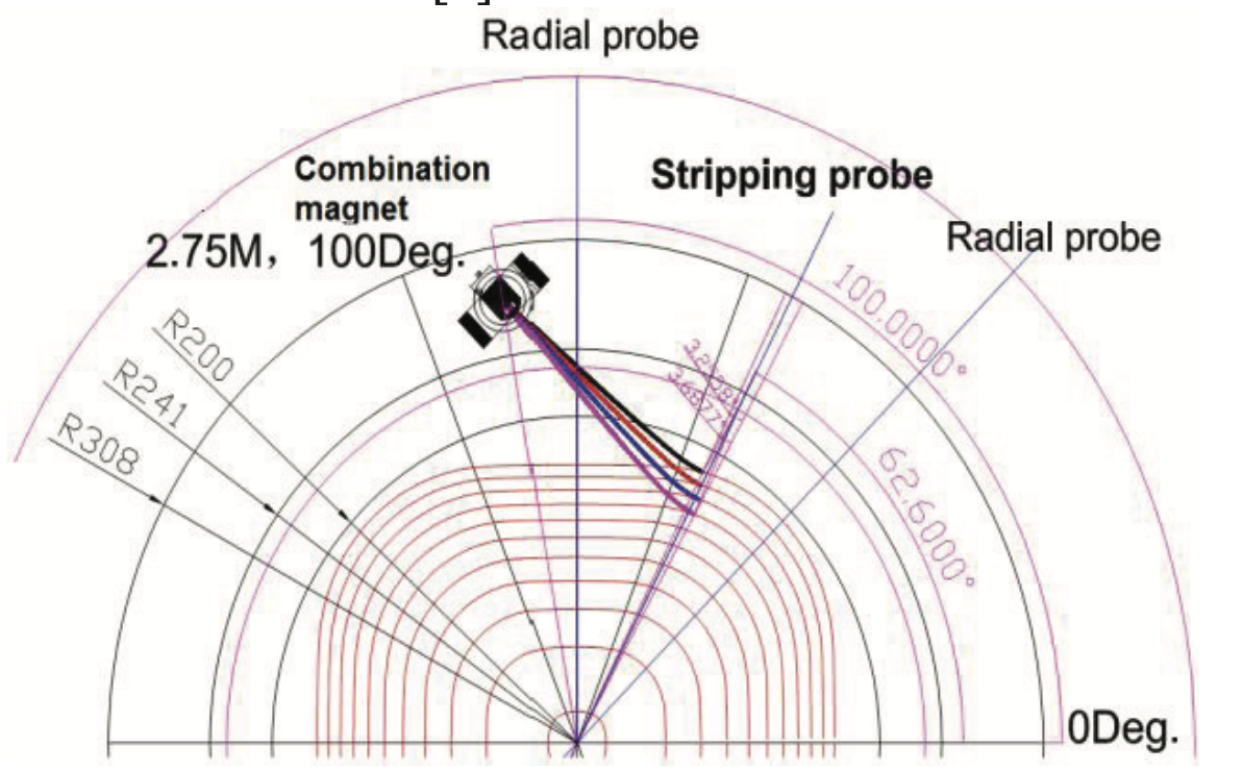}
    \caption{Positions of stripping points for extracted energy of 70~MeV, 80~MeV, 90~MeV and 100~MeV.}
    \label{fig:2.2.17}
\end{figure}

Simulation results showed that RF acceptance has almost no impact on the transverse beam distribution. A larger initial phase width resulted in more extracted turns, while a shorter phase width led to fewer turns. Achieving single-turn extraction was difficult, even with short bunch widths, when using sine waveform voltage for acceleration. The transverse space distributions and energy spread of the extracted beam were similar for both long and short phase widths, with larger phase widths corresponding to higher beam intensity.


         %
\subsection{ISOL system}\label{sub:2-3}
\subsubsection{General description}\label{sub:2-3-1}


The ISOL system produces medium- and short-lived radioactive ion beams, which can be injected into the tandem accelerator for post-acceleration or directly transmitted to the low-energy radioactive ion beam experimental terminal for physical research. The main design specifications of the facility are to produce a radioactive ion beam of 10$^{5}$$\sim$10$^{10}$~pps and a maximum mass resolution of 20,000.

\begin{figure}[!htbp]
    \centering
    \includegraphics[width=0.8\linewidth]{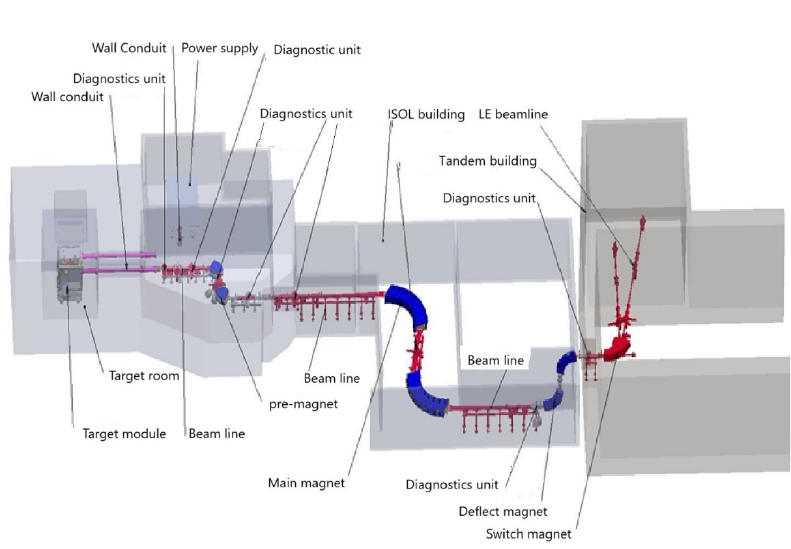}
    \caption{The layout of Isotope separator online.}
    \label{fig:2-3-1}
\end{figure}

The layout of the ISOL system is depicted in Fig.~\ref{fig:2-3-1}. It mainly consists of the following components: the target ion source system, the charge exchange cell, the pre-analysis system, the main analysis system, the low-energy experimental station, beam diagnostics, vacuum, control, and auxiliary systems such as water cooling and pneumatic systems. 

The proton beam from the cyclotron enters the 300~kV high-voltage platform through an acceleration column, where the proton beam first passes through the primary beam diagnostic module, in which information such as the beam intensity and beam profile of the primary beam can be obtained~\cite{mayingjun14}. The proton beam then bombards onto the target material within the target ion source module, where the proton beam reacts with the target material to produce radionuclides, and these radionuclides then enter the ion source to be ionized into singly-charged ions. The RIB with an energy of 30~keV from the ion source is adjusted by a set of steerers, and then enters the lens module to be focused by a set of three-unit quadrupole and an XY steerer to ensure that the ion beam can pass through the 3.5-meter shielding wall and the 2.5-meter high-voltage isolation zone, so that the beam can be transported to the pre-analyzing section with a high efficiency. The beam line is  at 300~kV potential, and when passing through the wall, the beamline passes through an SF6 conduit to isolate the beam line from the wall~\cite{huangqinghua08}. 

In pre-analyzing section, the RIBs are focused into a waist at the center of the following charge exchange cell by a quadrupole doublet, positive ion beam are converted into a negative ion beam in the charge exchange cell to meet the tandem injection requirement. The efficiency of the Charge Exchange Cell (CEC) has a vital effect on the intensity of RIB, normally only a few percent or less positive ions can be converted to a charge state of -1. To maximize the efficiency, the energy of extracted positive ions from ion source can be varied according to the elements to be converted, so the positive ions can pass through the CEC with adequate energy. When post-acceleration is not required, no cesium vapor is generated in the charge exchange cell; at this time, positive ions remain unchanged. 

Before the analyzing magnet, a quadrupole and a hexapole were placed to compress the envelope of the ion beam in the vertical direction, which allowed the analyzing magnet to have a smaller gap, to reduce the size of the analyzing magnet and power consumption. The hexapole was used to minimize aberration. After the second main analyzing magnet, a hexapole and a quadrupole were placed similarly to the first magnet. Isobar separation of the radioactive nuclear beam can be performed at the image point of this magnet.

The pre-analysis system used an asymmetric configuration consisting of two 90$^\circ$ analyzing magnets for isotope selection, also constraining the undesired RIBs produced by the ion source in the pre-analysis region. Immediately following the second analyzing magnet, a hexapole and a quadrupole were placed so that the ion beam forms the required image at its image point.

Behind the image point of the second analyzing magnet, a quadrupole doublet was placed for adjusting the beam shape, allowing the ion beam to pass through the acceleration column efficiently. After the quadrupole, a 300~kV acceleration column was used to accelerate the ion beam energy from 30~keV to a maximum of 300~keV.

After the acceleration column, two sets of quadrupole doublets were installed to focus the ion beam into the required beam shape at the object point of the analyzing magnet. Even at different accelerating voltages, the system can still effectively analyze and transmit the radioactive nuclear beam.

The main isotope separation system mainly consists of two analyzing magnets with a radius of 2.5~meters and a deflection angle of 100$^\circ$. This analyzing system is at ground potential. The arrangement of having two different sections of the analytical system at different potentials allows the system as a whole to partially eliminate the dispersion caused by energy spreads caused by the ion source and the charge exchange cell.

All the previous equipment was located in a 4-meter-deep underground room to reduce the radiation requirement. In order to inject the RIBs into the tandem accelerator, two deflection magnets with a radius of 1.2~m and a deflection angle of 90$^\circ$ were installed after the analysis slit to deflect the center of the beam from $\pm$2.24\,m below ground to 1.75~m above ground level.

Before the tandem, a switching magnet was installed to deflect the 300~keV RIBs to the two low-energy physical experiment stations. A pair of quadrupole lenses was placed after the switch magnet to ensure the effective transmission of the ion beam. When the switching magnet is not energized, the beam will directly enter the injection line of the tandem.

\subsubsection{The target station and ion source}\label{sub:2-3-2}

The target station, working in hot area, adopts modular design to make its maintenance possible. There are three modules, including entrance module, target module and exit module, shown in Fig.~\ref{fig:2-3-2}. The entrance module contains the diagnosis, including Faraday cup, beam position monitor and scanner, to measure the profile, position and current of the primary proton beam in front of the target. Target module consists of the target, ion source and a proton beam dump. The exit module contains a quadrupole triplet and a pair of steerers to transport and adjust the RIB beam from the ion source, a movable Faraday cup is followed to measure the current of the radioactive ion beam. An iron block with the thickness of 36~cm was used to reduce the radioactive level for each module, protecting the vacuum seals, service connections and turbo pump. A large "L" type aluminum vacuum tank contains all three modules. The vacuum, gas and electronics can be disconnected automatically when the module is pulled out from the vacuum tank remotely. There are three pins for each module, achieving the positioning accuracy better than 0.1~mm. 

Three specially designed turbo pumps were mounted on the iron block to evacuate the vacuum tank. The body of the pump was made from aluminum, plastic O-rings were replaced with indium, and semiconductor elements were eliminated. The space between target module and vacuum chamber, exit module and vacuum chamber was evacuated by the turbo pump on the entrance module.

There are 18 movable concrete blocks over the target station to shield against the neutron and $\gamma$-ray radiation produced from reactions between the primarily beam and the target. The weight for each concrete block is about 5 tons. The size of the hole for the concrete blocks is 3.4~m$\times$2.5~m$\times$2.6~m (long$\times$wide$\times$high). All of the concrete blocks will be pulled out by the crane remotely, monitored by CCD camera, before the maintaining of the target station. 

\begin{figure}[!htbp]
    \centering
    \includegraphics[width=0.6\linewidth]{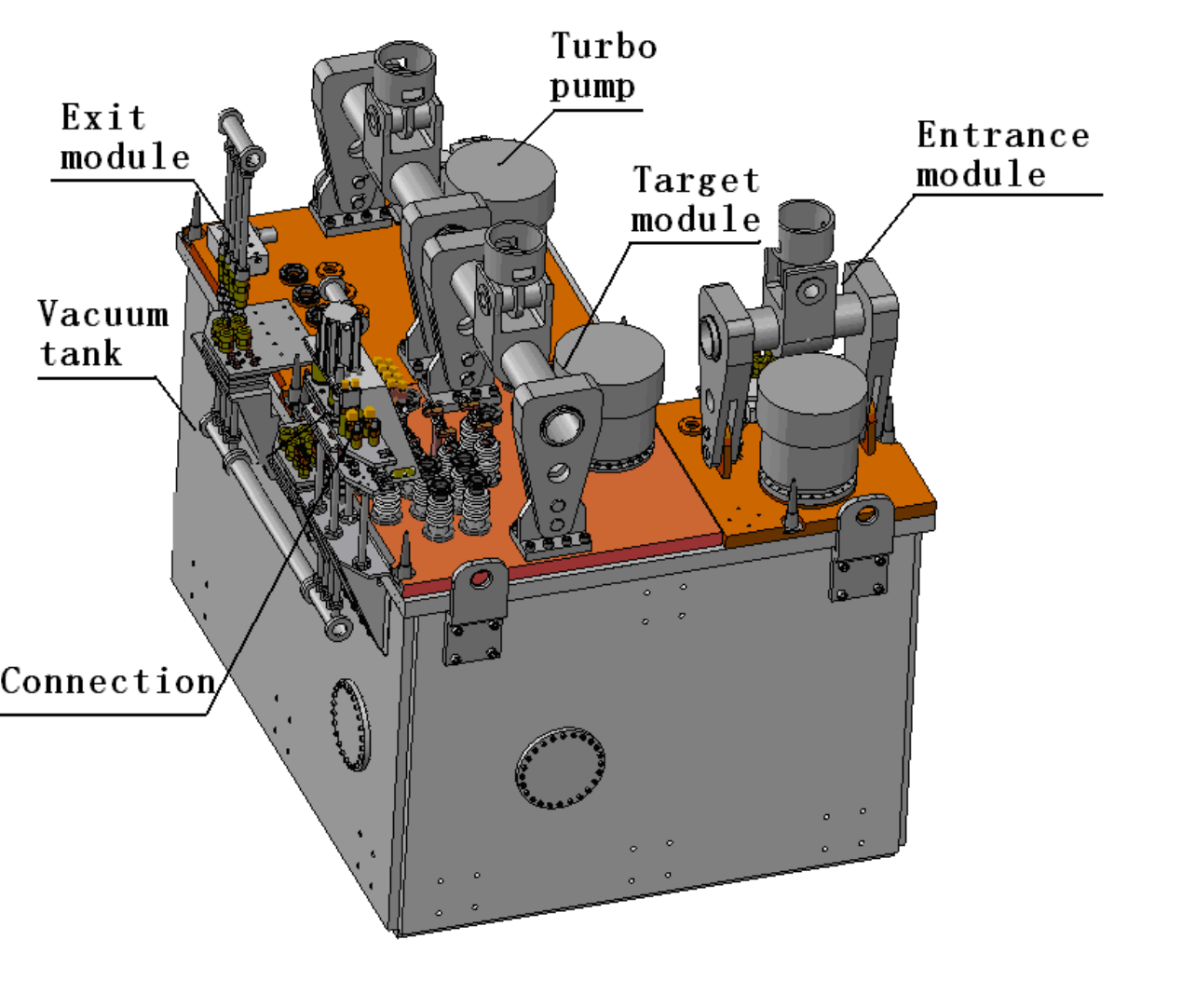}
    \caption{Schematic layout of the target station.}
    \label{fig:2-3-2}
\end{figure}

There are two types of ion sources at BRIF, one is surface ion source for low ionization potential elements, another is electron impact ion source for gaseous and high ionization potential elements.

The surface ion source developed for BRIF is shown in Fig.~\ref{fig:2-3-3}. The target and cathode are all made of tantalum material and can be resistively heated by passing high current direct current. The target tube is 18 mm in diameter and 100~mm long. The inner diameter of cathode is 3~mm. The target is surrounded by a water-cooled copper plate to shield the heat radiation of the target, and also a molybdenum disk with a single 3~mm extraction aperture is placed facing the cathode to shield the thermal radiation of the cathode. The ion source is supported by two insulator blocks and the maximum voltage of ion source is 50~kV. Following the extraction, a pair of electrostatic steerers was installed to guide the beam into the downstream beam line. Following the ion source an aluminium tube was installed to shield the electric field of high-voltage parts in the target module. The first radioactive beams ($^{37}$K$^{+}$, $^{38}$K$^{+}$, $^{42}$K$^{+}$, etc.) were produced by bombarding a CaO target with a 100~MeV proton beam from the cyclotron in 2014~\cite{tang2016}. The production yield of $^{38}$K$^{+}$ was 1$\times$10$^6$~pps when the proton beam intensity was 0.5~$\mu$A. 

\begin{figure}[!htbp]
    \centering
    \includegraphics[width=0.6\linewidth]{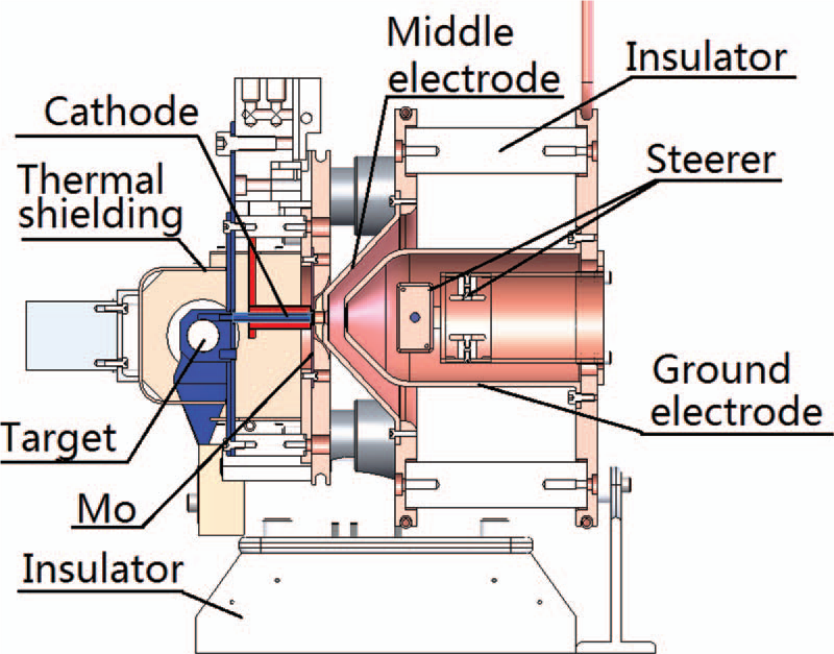}
    \caption{Schematic cutaway of the surface ion source.}
    \label{fig:2-3-3}
\end{figure}

For a RIB facility, there is no universal ion source that can ionize all elements with high efficiency and high selectivity, so several types of ion source must be developed, and changed one by one for different experiments. An innovative plasma structure based on conventional FEBIAD (Forced Electron Beam Induced Arc Discharge) ion source with the ability of dual ionization modes, including surface ionization mode and FEBIAD ionization mode, was developed for BRIF~\cite{tang2020}.

\begin{figure}[!htbp]
    \centering
    \includegraphics[width=0.6\linewidth]{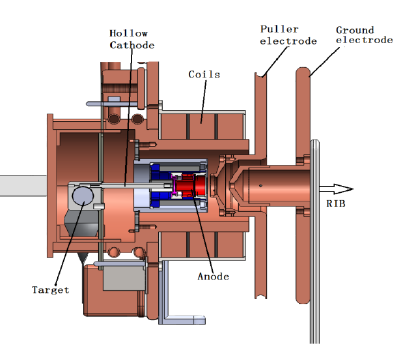}
    \caption{The FEBIAD ion source.}
    \label{fig:2-3-4}
\end{figure}

For conventional FEBIAD, when the neutral atoms pass through the cathode tube, most alkali atoms will be surface ionized. The voltage at the anode is positive, normally about +100$\sim$+300~V, then only neutral atoms, negative ions or electrons can be transferred to the plasma chamber. The positive ions from surface ionization of cathode will be suppressed by the positive potential of the anode, cannot be extracted. So the ionization efficiency of the FEBIAD ion source is much lower for alkaline elements, especially for the short-lived nucleus. But if the anode voltage is set to be zero or negative potential, the positive ions from the surface ionization of cathode will be extracted directly. 

An innovative plasma structure based on conventional FEBIAD ion source with the ability of dual ionization modes was developed. The structure of the anode and the puller electrode are modified to increase the extraction efficiency of the surface ionization mode, as shown in Fig.~\ref{fig:2-3-4}. A rhenium foil is rolled and then inserted inside at the tip of the cathode's inner tube to increase the work function of the cathode and increase the surface ionization efficiency. The tube is operated up to 2200~$^\text{o}$C. It means that all alkaline elements and other elements with ionization energy below about 6~eV can be ionized. A power supply with the ability of -400 V to 400 V output is fitted for the anode specially. When the anode is set to a positive potential, the ion source operates in FEBIAD mode similar to other conventional FEBIAD ion source. But if the anode is set to a negative potential, the ions from surface ionization of cathode and anode will be extracted efficiently. The voltage of the anode can also be varied from 0 to -400~V, optimizing the beam angle. Therefore, the new structure ion source has the ability of dual ionization modes, including surface ionization mode and FEBIAD ionization mode. 

The ionization efficiencies of the FEBIAD ion sources were measured with noble gases. The measured efficiencies of the conventional FEBIAD ion source were 14$\%$ for Xe and 3.98$\%$ for Kr respectively. And the efficiency of the new structure ion source was 6$\%$ for Xe, about half of the conventional one. But for $^{20}$Na, the beam current for surface mode was 20 times higher than FEBIAD mode. 

\subsubsection{The mass separator}\label{sub:2-3-3}

To reach the high mass resolving power of 20,000, a dual-energy system with big and small dipoles was used to reduce energy dispersion which is too large to get high mass resolving power. The target ion source was set on the high voltage platform No.1 (maximum 300~kV potential to ground), located inside a heavy shielded area. A beam guidance system including a quadrupole triplet and two pair of steerers lead the ion beam through a shielding wall with the thickness of 3.5~m to the high voltage platform No.2, which supports CEC and the pre-isotope separation system consisting of two 90$^\circ$ magnets which provides a mass resolving of 2000. This section was located in a light shielded area. The two platforms were connected by two 300~kV high voltage conduits, which were used to carry the beam pipe and electrical cables through the shielding wall respectively. Thus the accelerating tube was commercial product purchased from National Electrostatic Company. The beam line from target station to the main isotope separation system is 4.5~m below the ground for radiation protection. All the power supply for the elements of target station and the pre-isotope separation system were set in another room which is on 300~kV platform.

The overall mass separation was achieved in the main isotope separation system, which was placed on ground potential. The separator consists of two large dipole magnets with 100$^\circ$ deflection angle and 2.5~m radius, which finally allowed to achieve a mass resolving power of 20,000. The radioactive ion beam, with a maximum energy of 300~keV, will be sent to the beam guidance system consisting two guiding dipole magnets and a switching dipole magnet to the tandem or the other three RIB experimental terminals.

The beam envelope of the separator of ISOL is shown in Fig.~\ref{fig:2-3-5}. A surface coil of the type used by U.~Czok et al.~\cite{czok1977} was used in our situation to homogenize the magnetic field of the analysis magnet. The integral field uniformity of the analysis magnet was improved to 3$\times$10$^{-5}$ from 3.5$\times$10$^{-4}$. The entrance and exit field boundaries of magnet were already curved to correct second-order image aberrations. Additional, a pair of quadruple surface coil and hexapole surface coil similar as Ref.~\cite{camplan1981}, which can make a slightly inhomogeneous magnet homogeneous, were also equipped on the analysis magnet. The measured value of mass resolving power is 23800.

\begin{figure}[!htbp]
    \centering
    \includegraphics[width=0.6\linewidth]{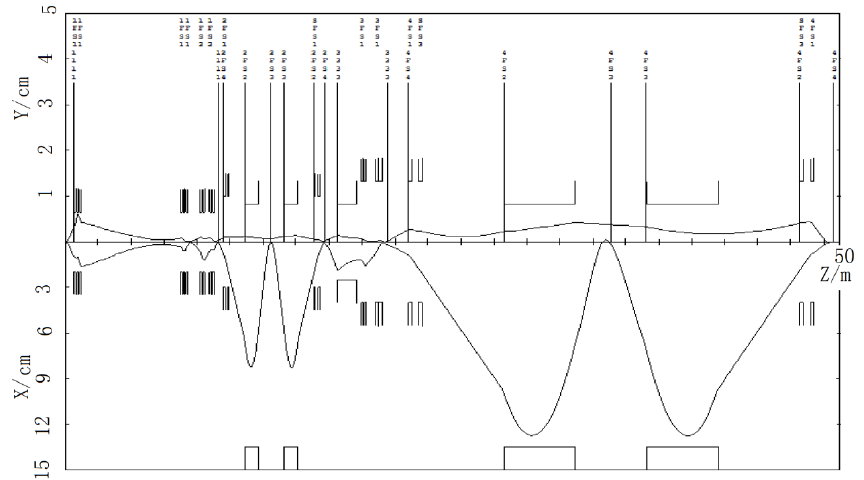}
    \caption{Calculated vertical and horizontal beam envelopes with TRANSPORT code.}
    \label{fig:2-3-5}
\end{figure}

	\subsection{Production and separation of radioactive ion beams}\label{sub:4-1}
Since the operation of the BRIF, several targets have been developed, tens of RIBs have been generated. The generated RIBs are introduced according to the targets.

The first RIB generated at BRIF was $^{38}$K which has a 7.71~minutes halflife. $^{38}$K can be generated by $^{40}$Ca(p,x)$^{38}$K reaction with reasonable 50~mb cross section.
Besides K isotope, many other isotopes such as P, S, Cl, Ar can be generated as shown in Fig.~\ref{fig:4-1-1}.

\begin{figure}[!htbp]
    \centering
    \includegraphics[width=0.6\linewidth]{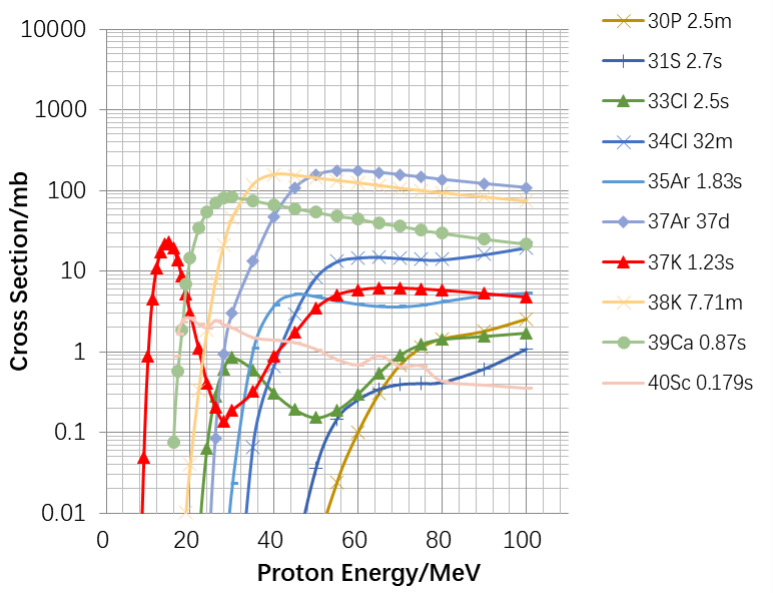}
    \caption{The cross section of different isotopes with CaO target.}
    \label{fig:4-1-1}
\end{figure}

CaO which has a 2572~$^\text{o}$C melting point was chosen as target material. For fast release of reaction product, CaO is made into a porous material by mixed Calcium hydroxide with carboxymethylcellulose sodium, compressed into pellets, heated to 1200~$^\text{o}$C. The final CaO target has a grain size of several micrometers and a porosity of 63\%, as shown in Fig.~\ref{fig:4-1-2}.

\begin{figure}[!htbp]
    \centering
    \includegraphics[width=0.6\linewidth]{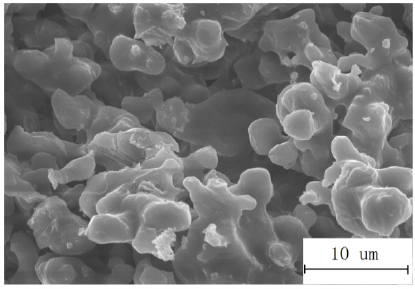}
    \caption{Micro-structure of the CaO target.}
    \label{fig:4-1-2}
\end{figure}

\begin{table}[!htbp]
     \caption{The radioactive isotope yield of CaO target measured in front of the 300 kV acceleration column.} 
     \centering
    \begin{tabular}{cccc}
    \hline
         RIB & Half life	&Proton/$\mu$A & Yield/pps \\
         \hline
      $^{36}$K$^{+}$ & 341~ms & 5 & 3$\times$10$^{4}$\\
  $^{37}$K$^{+}$ & 1.22~s & 15 & 4.62$\times$10$^{7}$\\
      $^{38}$K$^{+}$ & 7.65~min & 10 & 1.12$\times$10$^{10}$\\
      $^{42}$K$^{+}$ & 12.35~h & 9 & 9$\times$10$^{7}$\\      
$^{43}$K$^{+}$ & 22.3~h & 9 & 4.5$\times$10$^{7}$\\
$^{45}$K$^{+}$ & 17.81~min & 5 & 4.5$\times$10$^{5}$\\
$^{46}$K$^{+}$ & 105~s & 5 & 1.5$\times$10$^{5}$\\
$^{47}$K$^{+}$ & 17.5~s & 5 & 2.4$\times$10$^{4}$\\
       \hline
    \end{tabular}
    \label{table:4-1-1}
\end{table}

Although CaO has a relatively high melting point, when the target temperature is too high, there will be high-intensity Ca ions which can suppress the K production.

Porous MgO target was developed for Na isotopes generation. $^{20}$Na can be produced by $^{24}$Mg(p,x)$^{20}$Na. There is reasonable cross section for $^{20}$Na production with MgO target. The MgO with a melting point of 2850~$^\text{o}$C is a good target material candidate. The target pellets are a commercial product with a porosity of $30\%$. 
The measured yields of $^{20}$Na and other Na isotope generated using a MgO target are shown in Tab.~\ref{tab:4-1-2}.  

\begin{table}[!htbp]
     \caption{The radioactive isotope yield of MgO target measured in front of the 300 kV acceleration column.} 
     \centering
    \begin{tabular}{cccc}
    \hline
         RIB & Half life	&Proton/$\mu$A & Yield/pps \\
         \hline
      $^{20}$Na$^{+}$ & 0.448~s & 8 & 2$\times$10$^{5}$\\
  $^{21}$Na$^{+}$ & 22.5~s & 8 & 4$\times$10$^{8}$\\
      $^{22}$Na$^{+}$ & 2.6~y & 2.6 & 6$\times$10$^{9}$\\
      $^{24}$Na$^{+}$ & 14.9~h & 8 & 5$\times$10$^{7}$\\      
$^{25}$Na$^{+}$ & 59.1~s & 8 & 3.7$\times$10$^{7}$\\
$^{26}$Na$^{+}$ & 1.07~s & 6.5 & 1.6$\times$10$^{3}$\\
       \hline
    \end{tabular}
    \label{tab:4-1-2}
\end{table}

Although CaO and MgO have been used to produce RIBs successfully, there were still some issues to be overcome. The target materials decomposed at high temperatures, Ca and Mg are electron donors, electrodes deposited with Ca and Mg tend to spark which affects the reliable operation. The Al nuclides are also expected from the reaction $^{28}$Si(p,x)$^{26\text{g}}$Al. SiC has very suitable thermal and mechanical properties. The SiC target has been developed and tested on-line. The yields of RIBs are shown in Tab.~\ref{tab:4-1-3}.

\begin{table}[!htbp]
     \caption{The radioactive isotope yields of SiC target measured in front of the 300 kV acceleration column.} 
     \centering
    \begin{tabular}{cccc}
    \hline
         RIB & Half life	&Proton/$\mu$A & Yield/pps \\
         \hline
      $^{20}$Na$^{+}$ & 0.448~s & 15 & 2.1$\times$10$^{5}$\\
  $^{21}$Na$^{+}$ & 22.5~s & 15 & 1.6$\times$10$^{7}$\\
      $^{22}$Na$^{+}$ & 2.6~y & 15.5 & 4.4$\times$10$^{9}$\\
      $^{24}$Na$^{+}$ & 14.9~h & 15 & 2.3$\times$10$^{8}$\\      
$^{25}$Na$^{+}$ & 59.1~s & 15.5 & 2.9$\times$10$^{6}$\\
$^{26}$Na$^{+}$ & 1.07~s & 10 & 1.7$\times$10$^{4}$\\
$^{23}$Al$^{+}$ & 0.47~s & 13.5 & 2.2$\times$10$^{2}$\\
$^{25}$Al$^{+}$ & 7.183~s & 15.5 & 7.2$\times$10$^{3}$\\
$^{26\rm{g}}$Al$^{+}$ & 7.16$\times$10$^{5}$~y & 13.5 & 8.7$\times$10$^{7}$\\
$^{26\rm{m}}$Al$^{+}$ & 6.35~s & 10 & 1.5$\times$10$^{4}$\\
$^{28}$Al$^{+}$ & 2.245~min & 15.5 & 2.7$\times$10$^{4}$\\
       \hline
    \end{tabular}
    \label{tab:4-1-3}
\end{table}

Although reasonable intensity of RIBs has been obtained at the ISOL system, significant beam loss occurs due to post-acceleration schemes. A preexisting tandem, which requires negative ions injection, was used for post-acceleration. Normally the efficiency of the CEC is about a few percent or lower, the efficiency of post-acceleration by tandem and beam transportation is also about a few percent or lower, the beam intensity after the tandem is 4-5 orders magnitude lower than that at ion source. For example, for $^{21}$Na, the efficiency of CEC is 1$\%$, post-acceleration efficiency is 0.2$\%$, overall efficiency is only 2$\times$10$^{-5}$. Clearly, upgrading the post-accelerator from tandem to linac is urgent.

	\subsection{13-MV tandem}\label{sub:2-4}
The HI-13 tandem accelerator of China Institute of Atomic Energy was introduced from HVEC in the early 1980s, installed in 1983, passed national acceptance in 1987, and put into operation~\cite{tian1986}. For over 35 years, we have provided over 50 types of particles, 140,000 hours for operation and 120,000 hours of beam time to nearly 100 users both domestically and internationally. This accelerator is the main device for low-energy nuclear physics research in China, mainly used for basic nuclear physics research, nuclear data measurement, nuclear technology applications, and interdisciplinary research. In order to ensure the stable operation of the accelerator, multiple domestic research and technological upgrades have been carried out. 

The layout of the HI-13 tandem accelerator laboratory equipment is shown in Fig.~\ref{fig:hi-13}, including the injector, HI-13 tandem accelerator body and 15 experimental terminals. The experimental terminals, including the Accelerator Mass Spectrometry terminal, the Q3D magnetic spectrometer, the $\gamma$-angle analyzer, the Scattering target chamber with electrostatic deflection terminal and several others, are distributed in three halls. The injector is equipped with three types of ion sources: off-axis duoplasmatron, lithium charge exchange source and negative ion sputter source, which can generate over 50 commonly used ions.

 \begin{figure}[!htbp]
    \centering
    \includegraphics[width=0.8\linewidth]{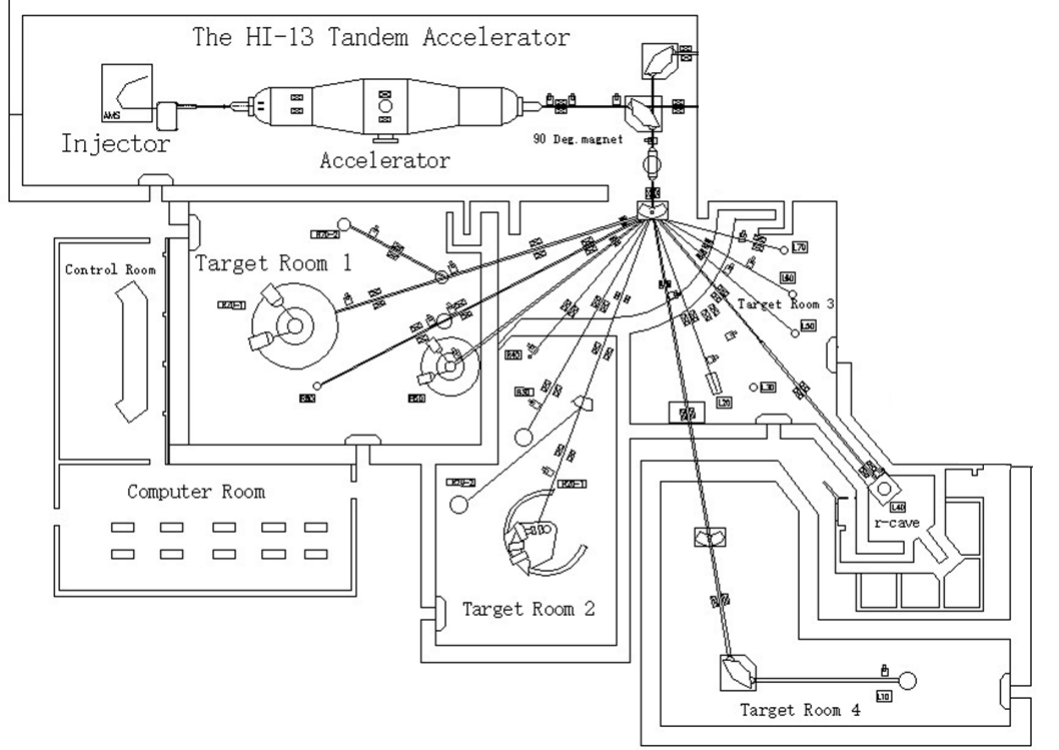}
    \caption{The layout of HI-13 accelerator and experimental terminal.}
    \label{fig:hi-13}
\end{figure}

The HI-13 tandem is an improvement on the standard MP tandem. It will be more flexible for heavy ion performance. The main specifications as follows Tab.~\ref{tab:13MV}:

\begin{table}[!htbp]
    \caption{the main specifications of HI-13 tandem accelerator.}
    \centering
    \begin{tabular}{p{40mm}p{30mm}p{30mm}}
\hline
Max terminal voltage& & 13~MV  \\
\hline
Terminal voltage ripple & & $\pm$1~KV \\
\hline
\multirow{4}{=} {Analyzed beam intensity} & H & 10~$\mu$A \\
& $^{35}$Cl & 0.25~p$\mu$A \\
& $^{79}$Br & 0.25~p$\mu$A \\
& $^{127}$I & 0.1~p$\mu$A \\
\hline
\multirow{3}{=} {Proton pulse beam} & fundamental & 4~MHz \\
&FWHM & 1~ns\\
&peak current&1~mA\\
\hline
    \end{tabular}

    \label{tab:13MV}
\end{table}

In order to ensure the smooth operation of the HI-13 tandem accelerator and improve its working condition, the localization and technological upgrading of key components around the accelerator have been ongoing. The main tasks completed successively include:

\emph{Laddertron}: New technology was adopted to develop nylon insulators and oil-free bearings, and a new assembly process for the laddertron was established, which improved the operation status of the laddertron and significantly reduced the debugging time.

\emph{Voltage divider resistor}: The frame structure was used instead of the original flat structure, and the domestically produced high-voltage resistor voltage divider system has stronger resistance to high-voltage spark and more stable resistance. 

\emph{Second Stripper}: A second stripper was installed between the accelerator tubes in the high-energy stages 5 and 6, significantly improving the performance of the HI-13 tandem accelerator. For the same ion current intensity, particle energy can be increased by 10-30~MeV.

\emph{Accelerator Tube Technical Reconstruction}: One 96 inch and seven 88 inch accelerator tubes were used to replace the original eight 72 inch tubes that had been in use for 16 years. At the same time, the accelerator dead zone structure, charging system, and resistance voltage divider system were renovated. After the renovation, the accelerator high voltage can reach 15~MV.

\emph{Injector Upgrade and improvement}: In order to meet the requirements of the HI-13 tandem accelerator upgrade project for superconducting linear post-acceleration and AMS nuclear data measurement for quality resolution, the original injector was upgraded and improved. New 300~KV injection line and 150~KV high-resolution injection line were built, and the two injection lines can be easily switched for use.

\emph{Control System}: In addition to the internal area of the steel cylinder, more than 80\% of the systems and equipment of the tandem accelerator were realized automatically controlled by computers, effectively improving the operation status of the accelerator and the efficiency of personnel beam adjustment.

With the efforts of accelerator operation and maintenance technicians, the performance of the HI-13 tandem accelerator has been maintained at a high level through the localization and technological transformation of key equipment, creating favorable conditions for scientific experiments and achieving fruitful results. As an intermediate acceleration link between the Beijing Radioactive Ion-beam Facility and the superconducting linear accelerator, the HI-13 tandem accelerator will continue to play an important role in the foreseeable future.

	\subsection{Superconducting booster}\label{sub:2-5}
A superconducting booster (SC Booster) for tandem is final part of the BRIF facility, and its development has gone through two stages. 

\subsubsection{Stage I: superconducting prototype}\label{sub:2-5-1}

In 2015, a prototype was installed downstream of the HI-13 tandem accelerator. This prototype includes a superconducting acceleration module and one beam pulse synchronization system. 

\begin{figure}[!htbp]
    \centering
    \includegraphics[width=0.8\linewidth]{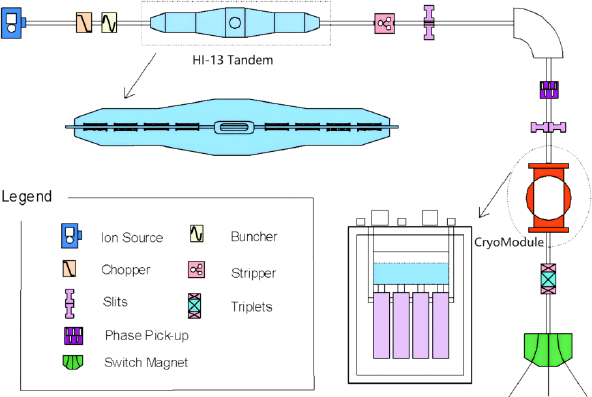}
    \caption{The layout of superconducting booster prototype.}
    \label{fig:2.5.1}
\end{figure}

The superconducting acceleration module contains four quarter-wave superconducting cavities, utilizing copper-based niobium sputtering technology, operating at a frequency of 150.4 MHz, with a designed $\beta$ value of 0.11, capable of boosting the beam energy by 2 MeV/q. 

The beam pulse synchronization system contains one traveling wave chopper, one double-drift buncher and one phase pick-up, operates at a fundamental frequency of 6 MHz, allowing for bunching across the full particle range by changing injection energy of beams and electrodes switching of buncher. 

\begin{figure}[!htbp]
    \centering
    \includegraphics[width=0.6\linewidth]{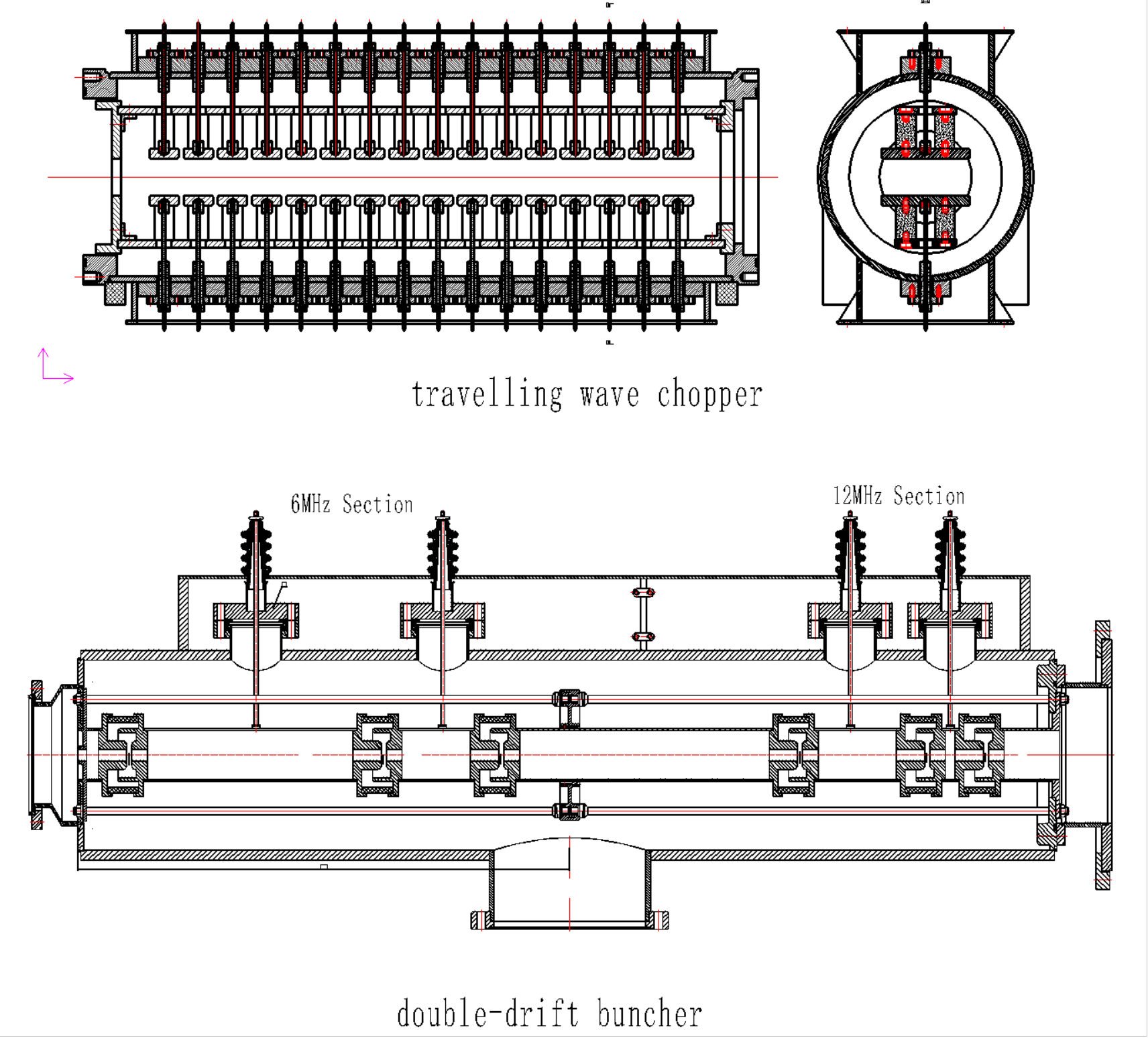}
    \caption{The traveling wave chopper and the double-drift buncher in beam pulse synchronization system.}
    \label{fig:2.5.2}
\end{figure}

\subsubsection{Stage II: superconducting booster}\label{sub:2-5-2}

In 2016, the China Institute of Atomic Energy imported a superconducting linear accelerator from Stony Brook University, State University of New York, USA, as a replacement for the superconducting acceleration module, as shown in the layout. 

After being accelerated by the HI-13 tandem accelerator, the beam passes through the original analyzing magnet, is transported to a new analyzing magnet, enters the SC booster. After acceleration, it passes through two newly added 90$^\circ$ bending magnets and returns to the original beam pipe of the tandem, finally being distributed to various experimental terminals via the original switching magnet. The beam pulse synchronization system adds a new phase detector after the new analyzing magnet, while the chopper and buncher continue to use the original equipment.  

The SC booster consists of 40 copper-based lead-film superconducting cavities, including 16 Quarter-Wavelength-Resonator (QWR) cavities with $\beta$  = 0.055 and 24 SpLit-Ring-resonator (SLR) cavities with $\beta$  = 0.1, operating at a frequency of 150.4~MHz and a temperature of 4.2~K, respectively installed in four low-$\beta$  cryostats and eight high-$\beta$  cryostats. 

\begin{figure}[!htbp]
    \centering
    \includegraphics[width=0.8\linewidth]{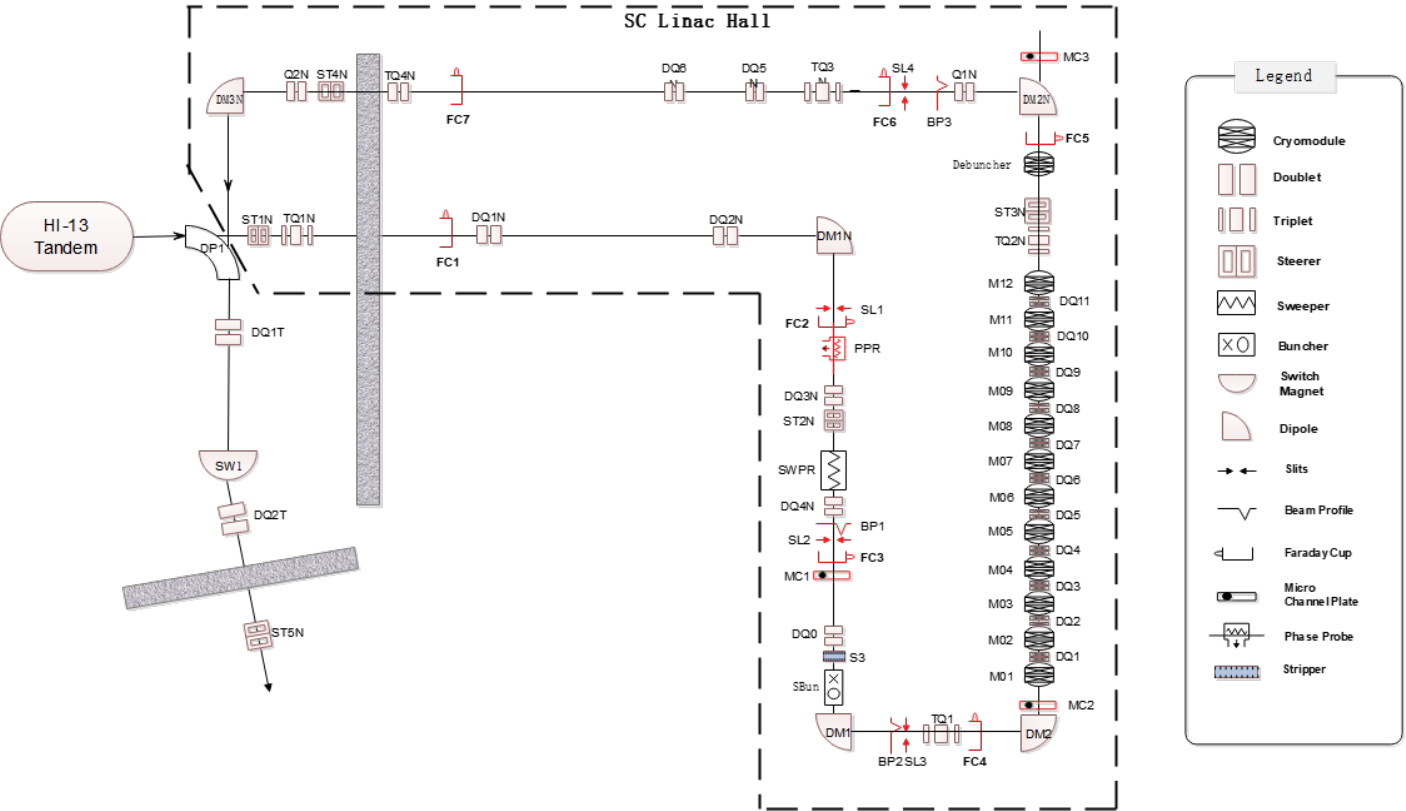}
    \caption{Composition Diagram of new superconducting booster.}
    \label{fig:2.5.4}
\end{figure}

The SC booster can theoretically provide a maximum accelerating voltage of about 17~MV. However, due to its high-frequency electric field, the variation of energy gain for ions with a maximum value of around 13~MeV/q depends on their speed and charge-to-mass ratio. Based on the output energy and charge state of different ions from the HI-13 tandem accelerator, reference values are given as Fig.~\ref{fig:2.5.5} for the final energy of different ions after being boosted by the SC Booster. 

\begin{figure}[!htbp]
    \centering
    \includegraphics[width=0.8\linewidth]{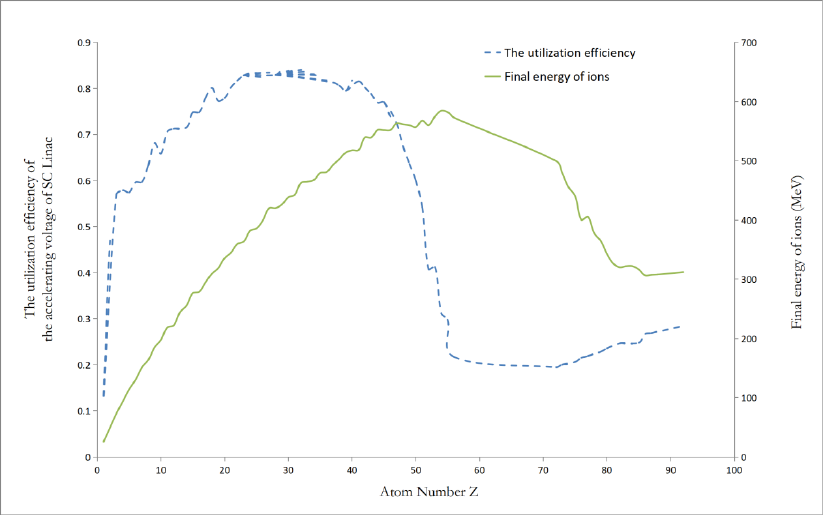}
    \caption{Final energy of different ions boosted by superconducting booster. The utilization efficiency $=$ real energy gain/(the charge state $\times$ the maximam accelerating voltage of Linac)}
    \label{fig:2.5.5}
\end{figure}

The reconstruction work of the SC Booster includes re-plating of superconducting cavities, repair of cryostats, replacement of the helium refrigeration system, replacement of the control system, replacement of the high-frequency  and LLRF control system, replacement of the vacuum system, and completion of the beam transport system. 

At the end of 2023, the reconstruction work was completed. In November 2023, online energy-boosting tests were conducted for the superconducting linear accelerator, with stable operation of beam transportation, helium refrigeration, vacuum, RF, and control systems. The response speed of low-level control needed improvement, but the performance of the superconducting cavities met requirements, with a maximum single-cavity accelerating voltage reaching 500 kV. The unloaded quality factor $Q_0$ of some superconducting cavities at low power is shown as Fig.~\ref{fig:2.5.6}. 

\begin{figure}[!htbp]
    \centering
    \includegraphics[width=0.6\linewidth]{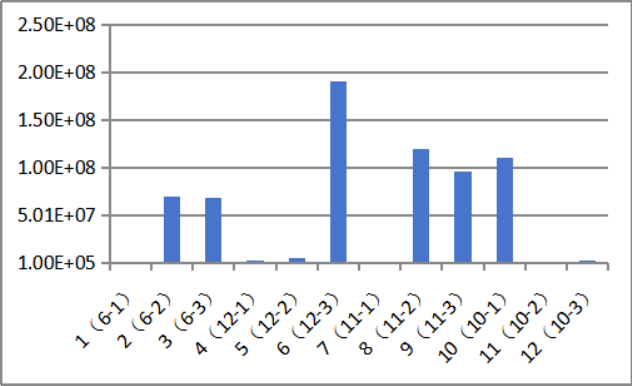}
    \caption{Unloaded quality factor $Q_0$ of some superconducting cavities at low power.}
    \label{fig:2.5.6}
\end{figure}

The domestically developed helium refrigeration system (both the helium compressor and cold box were made in China) has a heat load capacity of 420~W at 4.2~K, with the longest continuous operation time currently standing at 95 days. 

The SC booster is still under commissioning, with plans to conduct the first physics experiment by the end of 2024 and trial operation of the entire facility by the end of 2025.

	\newpage
	\section{Experimental instrumentation}\label{third}
At present, there are multiple experimental terminals at BRIF, such as collinear laser spectroscopy system~\cite{PKU-CLS-2022} and Q3D magnetic spectrograph~\cite{li1993}, total absorption gamma spectrometer LAMBDA~\cite{sheng2024large}, heavy-ion time-of-flight (HiToF) spectrometer~\cite{wanghaorui24} and gamma detector array. By utilizing these instrumentation, relevant studies of exotic nuclei, nuclear decay, nuclear structure and astrophysics can be conducted.
	\subsection{Collinear laser spectroscopy system}\label{sub:3-1}

Over the years, the laser spectroscopy technique has proven to be an indispensable tool for probing the basic properties of unstable nuclei, including nuclear spins, magnetic dipole moments, quadrupole moments and charge radii. These properties are important for studying exotic nuclear structures and nucleon-nucleon (NN) interactions~\cite{Neyens03,Cheal10,Neugart17,GarciaRuiz16,k-radii2021,Groote20,PPNP2023} and nuclear theories. To date, laser spectroscopy has been used to investigate approximately 1,000 isotopes, representing only one-third of the artificially produced unstable nuclei at RIB facilities. This leaves a vast region of isotopes on the current nuclear chart yet to be explored using this method, offering the potential for uncovering rich structural information about unstable nuclei~\cite{ye2024}.

\begin{figure*}[h!]
\centering
\includegraphics[width=0.99\linewidth]{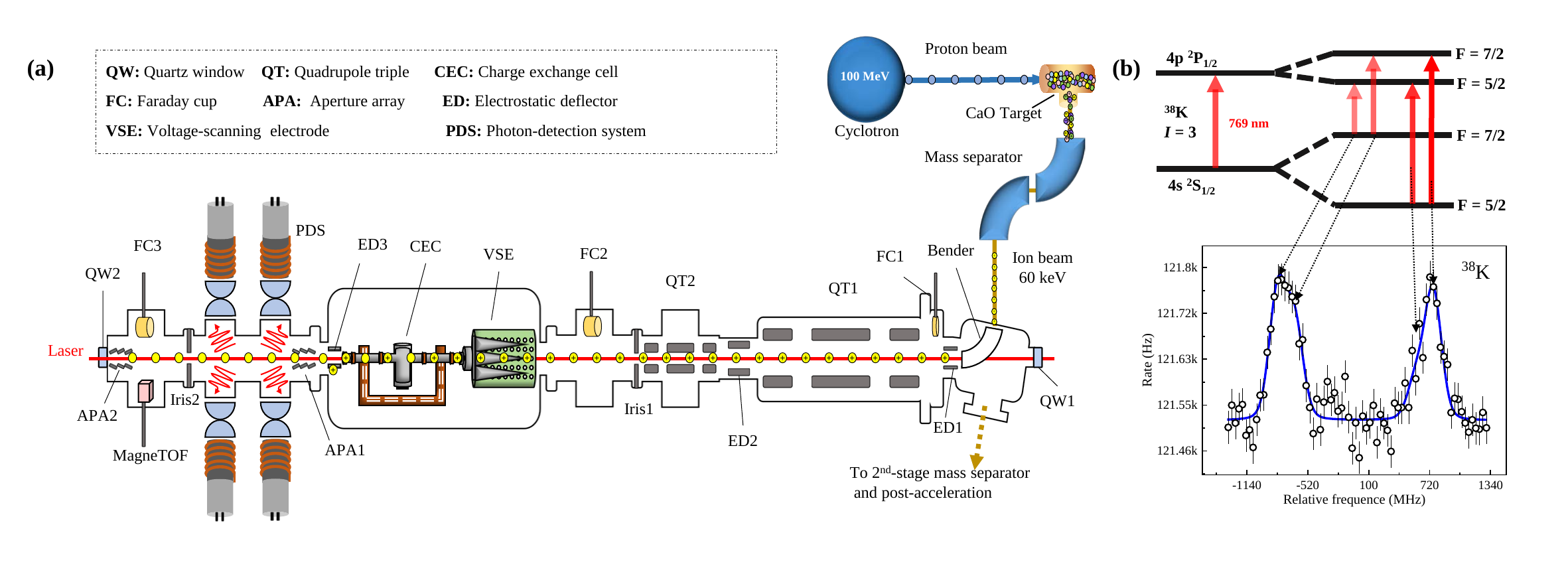}
\caption{(a) Layout of collinear laser spectroscopy setup at BRIF~\cite{PKU-CLS-2022,BRIF-CLS}. (b) Hyperfine structure spectrum of radioactive $^{38}$K isotope measured with laser spectroscopy at BRIF~\cite{CLS-DAQ-PKU}. }\label{figcls}
\end{figure*}

Laser spectroscopy primarily probes the subtle hyperfine splitting of electron energy levels resulting from the interaction between the atomic nucleus and surrounding electrons. Such hyperfine splitting is extremely small\textemdash typically on the order of one part in a million compared to the total optical transition frequency. Therefore, in order to extract the nuclear properties, a high-resolution laser spectroscopy technique is needed to measure the hyperfine structure (HFS) spectra of the atomic nuclei.

Among the laser spectroscopy techniques currently in use, collinear laser spectroscopy has the ability to achieve high-resolution HFS spectra measurements. This is achieved by overlapping a fast ion beam (e.g., 30$\sim$60~keV) with a laser in either collinear or anti-collinear geometry. In this configuration, Doppler broadening of the HFS spectra, which arises from the energy spread of the ion beam, is significantly suppressed to the same level of the natural linewidth (few to tens of MHz) of the optical transition being probed. Consequently, the final observed linewidth, $\delta \nu$, solely results from the residual Doppler broadening, given by:
\begin{equation}
\delta \nu = \nu_{0}\frac{\delta E}{\sqrt{2eUmc^2}}
\label{eq39}
\end{equation}
where $E$ is the total kinetic energy of the fast beam, and $\nu_{0}$ is the rest-frame transition frequency.
Due to its advantage, collinear laser spectroscopy has been widely used at various RIB facilities world-wide to investigate the exotic structure of radioactive isotopes across a broad mass region of the nuclear chart~\cite{PPNP2023}. 

In order to study the basic properties and exotic structure of unstable nuclei, a new collinear laser spectroscopy setup was developed for the first time in China~\cite{PKU-CLS-2022} and implemented at the BRIF facility~\cite{BRIF-CLS, CLS-DAQ-PKU}, as presented in Fig.~\ref{figcls}(a).

The collinear laser spectroscopy setup was successfully commissioned by measuring the HFS spectrum of the radioactive $^{38}$K isotope at BRIF. Potassium isotopes were produced by impinging 100-MeV protons on a CaO target, and the potassium atoms released from the target were ionized using a surface ion source. After being extracted from the ion source and accelerated to 60~keV, the $^{38}$K ion beam was mass-selected using separator magnets before being delivered to the experimental setup. As illustrated in Fig.~\ref{figcls}(a), the $^{38}$K ion beam was deflected into the collinear laser spectroscopy beamline via a pair of electrostatic deflecting plates. Its trajectory was controlled by the quadrupole triplet lens and additional electrostatic deflectors. The $^{38}$K ions in the collinear laser spectroscopy beamline were neutralized upon passing through a charge exchange cell (CEC) filled with dense Na vapors, populating the ground state of $3p^6 4p, ^2S_{1/2}$ in neutral potassium atoms. A frequency-fixed continuous-wave laser beam was overlapped with the potassium beam in an anti-collinear geometry to probe the $3p^6 4p, ^2S_{1/2} - 3p^6 4p, ^2P_{1/2}$ transition at 769.89~nm. The $^{38}$K atoms were resonantly excited to the $3p^6 4p, ^2P_{1/2}$ state through Doppler tuning, achieved by varying the potential applied to the voltage-scanning electrode upstream of the CEC to adjust the $^{38}$K beam velocity. The emitted fluorescent photons from the resonantly-excited atoms were detected as a function of the tuning potential using photomultiplier tubes (PMTs) in the photon-detection system. The resulting HFS spectrum of $^{38}$K is shown in Fig.~\ref{figcls}(b), which has been Doppler-corrected to the rest frame.

This experiment on the radioactive $^{38}$K ion beam at BRIF marks a significant milestone, paving the way for studying the fundamental properties of unstable nuclei using laser spectroscopy techniques in China. Nevertheless, we encountered two main issues that hindered our ability to measure the HFS spectra of exotic nuclei with high resolution and efficiency. Firstly, the ion beam provided by the BRIF facility operates in continuous mode. Under this condition, the experiment was affected by significant background noise caused by scattered laser light within the beamline. Secondly, the ion beam from the BRIF facility has a large energy spread of approximately 20~eV, which directly broadens the linewidth of the observed HFS spectrum, as shown in Fig.~\ref{figcls}(b). Fortunately, both issues can be overcome by using a Radio Frequency Quadrupole (RFQ) ion trap~\cite{RFQ-IGISOL,PKU-RFQ-2024}, which is expected to be installed soon.

        %
    \subsection{Total absorption gamma spectroscopy LAMBDA}\label{sub:3-2}
    \ \ \ \ The properties of $\beta$ decay have significant implications in the studies of nuclear astrophysics, reactor neutrino anomalies, and reactor decay heat. A common approach to obtaining these properties involves the measurement of $\beta$-delayed-$\gamma$-decay with high-purity germanium (HPGe) detectors, which have a high energy resolution. However, challenges arise as the nuclei under investigation move further from the $\beta$-stability valley. The increased $Q_{\beta}$ values facilitate the population of high-energy levels in the daughter nuclei, which typically deexcite to the ground state via complicated pathways, emitting multiple $\gamma$ rays with small branching ratios. Due to the limited detection efficiency of HPGe detectors, such low-intensity $\gamma$ rays are often omitted during measurements, leading to a systematic bias in the derived $\beta$-decay intensity distributions. This phenomenon, known as the "Pandemonium effect", was first reported by Hardy et al. in 1977~\cite{hardy1977essential} and is now recognized as one of the major obstacles in the accurate determination of reactor decay heat and neutrino energy spectrum.

	To address this issue, Greenwood proposed the use of total absorption gamma spectroscopy (TAGS) to perform measurements in the 1990s~\cite{greenwood1992total}. The first TAGS detector was developed by Duke et al. in 1970~\cite{duke1970strength}. The basic concept of TAGS is to use large-volume scintillators, such as NaI(Tl) or BaF$_2$, to enclose the source within a nearly 4$\pi$ solid angle, thereby achieving high detection efficiency and significantly improving the accuracy of $\beta$-decay measurements. Since the 2010s, several TAGS facilities have been developed~\cite{tain2010beta,tain2015decay,karny2016modular}, with a trend toward modular designs that enhance detection efficiency, energy resolution, and granularity. Today, the TAGS technique is not only employed for $\beta$ decay measurements of fission products~\cite{algora2010reactor, zakari2015total, guadilla2019large, rasco2016decays}, but has also been extended to other fields, such as direct measurements of the cross-sections of radioactive capture reactions~\cite{su2022first, zhang2021direct, zhang2022measurement, wang2023measurement, palmisano2022constraining, dombos2022measurement} and the indirect investigation of the (n, $\gamma$) reaction cross-sections~\cite{larsen2019beta, yang2022response}.

	A new TAGS facility, named LArge-scale Modular BGO Detection Array (LAMBDA), is currently under development. LAMBDA was designed specifically for measurements of $\beta$-delayed-$\gamma$-decay of neutron-rich medium-mass nuclei~\cite{sheng2024large}. LAMBDA featured a modular design with a capacity for up to 102 modules. Now there are 60 modules available, 48 of which were assembled into the LAMBDA-I detector shown in Fig.~\ref{fig3-2-lambda-modules}. The upper right corner illustrates the structure of a single detector module. The core of the module is a \qty{60}{mm} $\times$ \qty{60}{mm} $\times$ \qty{120}{mm} Bismuth Germanate (BGO) crystal, which was encased in a \qty{65}{um} thick enhanced specular reflector (ESR) foil, excluding the light output area. A 2-inch Photomultiplier Tube (PMT) was utilized to detect the scintillation light emitted by the BGO crystal. The crystal was installed within a \qty{1}{mm} thick carbon fiber housing, while the PMT was fixed by an aluminum base, which was equipped with threaded holes to facilitate the assembly of the array.

	\begin{figure}[h!]
		\centering
		\includegraphics[width=0.8\linewidth]{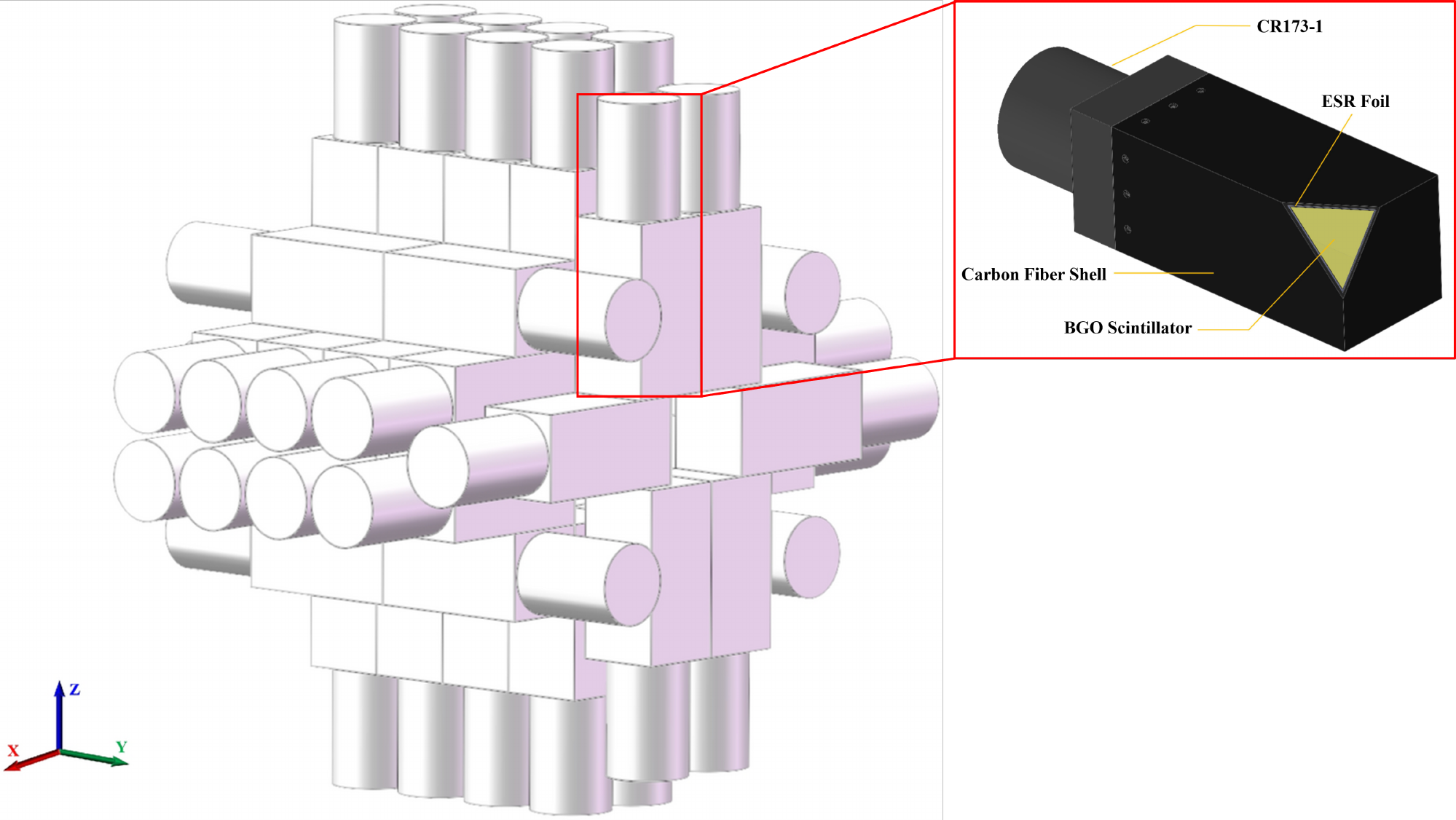}
		\caption{Schematic diagram of LAMBDA-I containing 48 modules~\cite{sheng2024large}. The inset illustrates the structure of a single detector module.}
		\label{fig3-2-lambda-modules}
	\end{figure}

LAMBDA-I used Pixie-16 modules manufactured by XIA Inc.~\cite{pixiexia} to record the experimental data. $^{137}$Cs and $^{60}$Co sources were used to evaluate the performance of LAMBDA-I. The energy resolutions were measured to be \qty{10.1}{\%}, \qty{7.3}{\%}, \qty{6.9}{\%} and \qty{5.2}{\%} for \qty{662}{keV}, \qty{1173}{keV}, \qty{1332}{keV} and \qty{2502}{keV} $\gamma$ rays, respectively. The summing efficiencies were \qty{82.5\pm1.5}{\%} and \qty{50.6\pm1.2}{\%} for \qty{662}{keV} and \qty{2502}{keV} sum peaks, respectively, which are consistent with \qty{82.8}{\%} and \qty{50.9}{\%} obtained by Geant4 simulation. 

	Furthermore, a $^{152}$Eu source was employed to validate the data analysis methodology. The $^{152}$Eu isotope exhibits a complex decay pattern with two distinct decay modes: electron capture to $^{152}$Sm, which populates 21 excited states ($Q_{\rm{EC}} = \qty{1874}{keV}$), and $\beta$-decay to $^{152}$Gd, which populates 12 excited states ($Q_{\beta} = \qty{1818}{keV}$). The measured sum and single spectra of the $^{152}$Eu source are shown in Fig.~\ref{fig3-2-eu-res}. Data were analyzed using the Bayesian Analysis Toolkit (BAT)~\cite{caldwell2009bat}, accounting for contributions from background and pile-up effects, which are particularly significant in the high energy region. The results of the analysis were in good agreement with the ENSDF data~\cite{sheng2024large}.

	\begin{figure}[h!]
		\centering
		\includegraphics[width=0.6\linewidth]{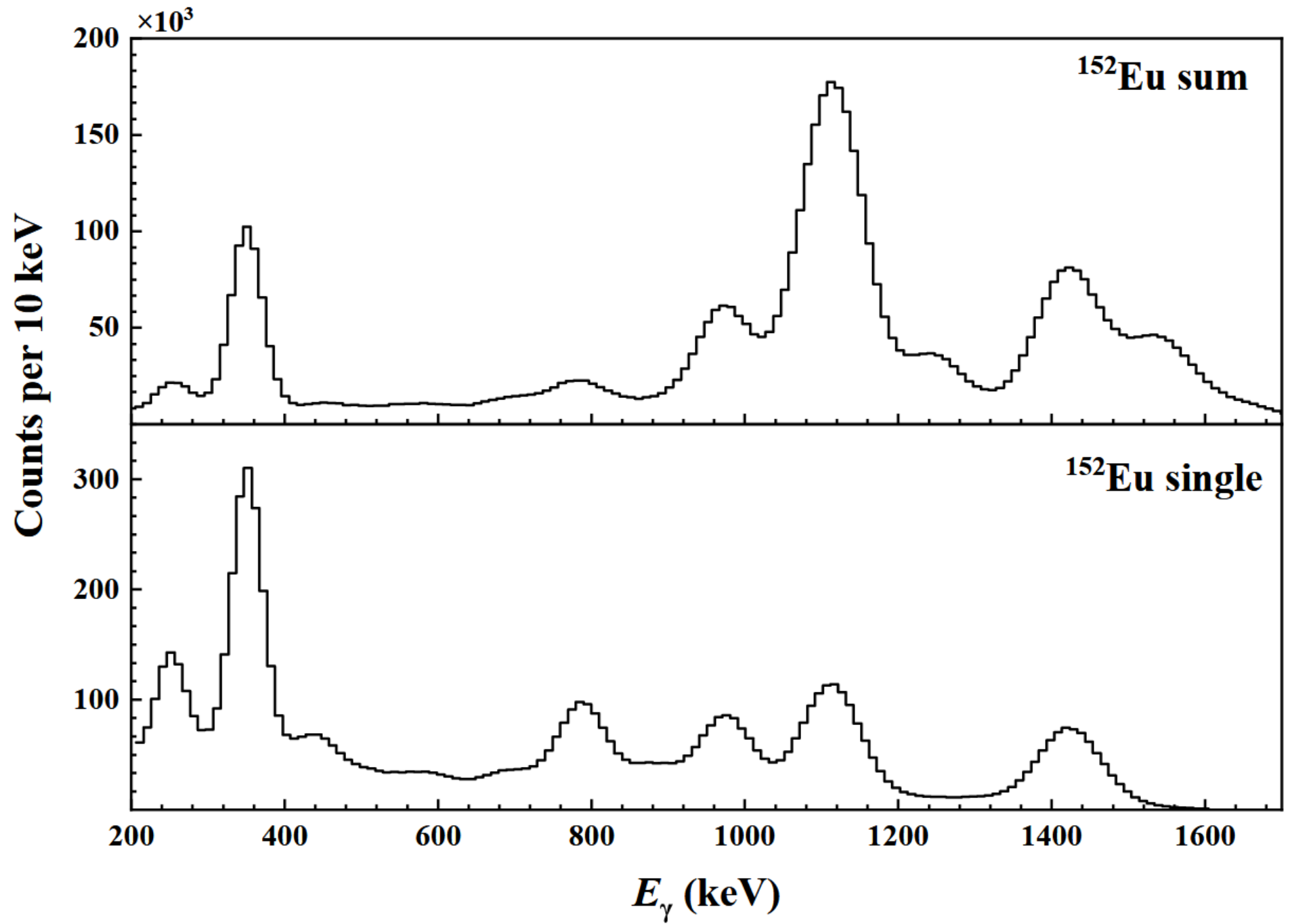}
		\caption{Measured spectra of $^{152}$Eu source with LAMBDA-I.}\label{fig3-2-eu-res}
	\end{figure}

	\subsection{Q3D magnetic spectrograph}\label{sub:3-3}
Q3D(Quadrupole-Dipole-Dipole-Dipole) magnetic spectrometer was equipped at the R20 beamline of the Beijing HI-13 tandem accelerator~\cite{li1993}, designed by H.~A.~Enge and manufactured by Scanditronix. The layout of the Q3D magnetic spectrometer is shown in Fig.~\ref{fig1}.
\begin{figure}[h!]
	\centering
	\includegraphics[width=1\linewidth]{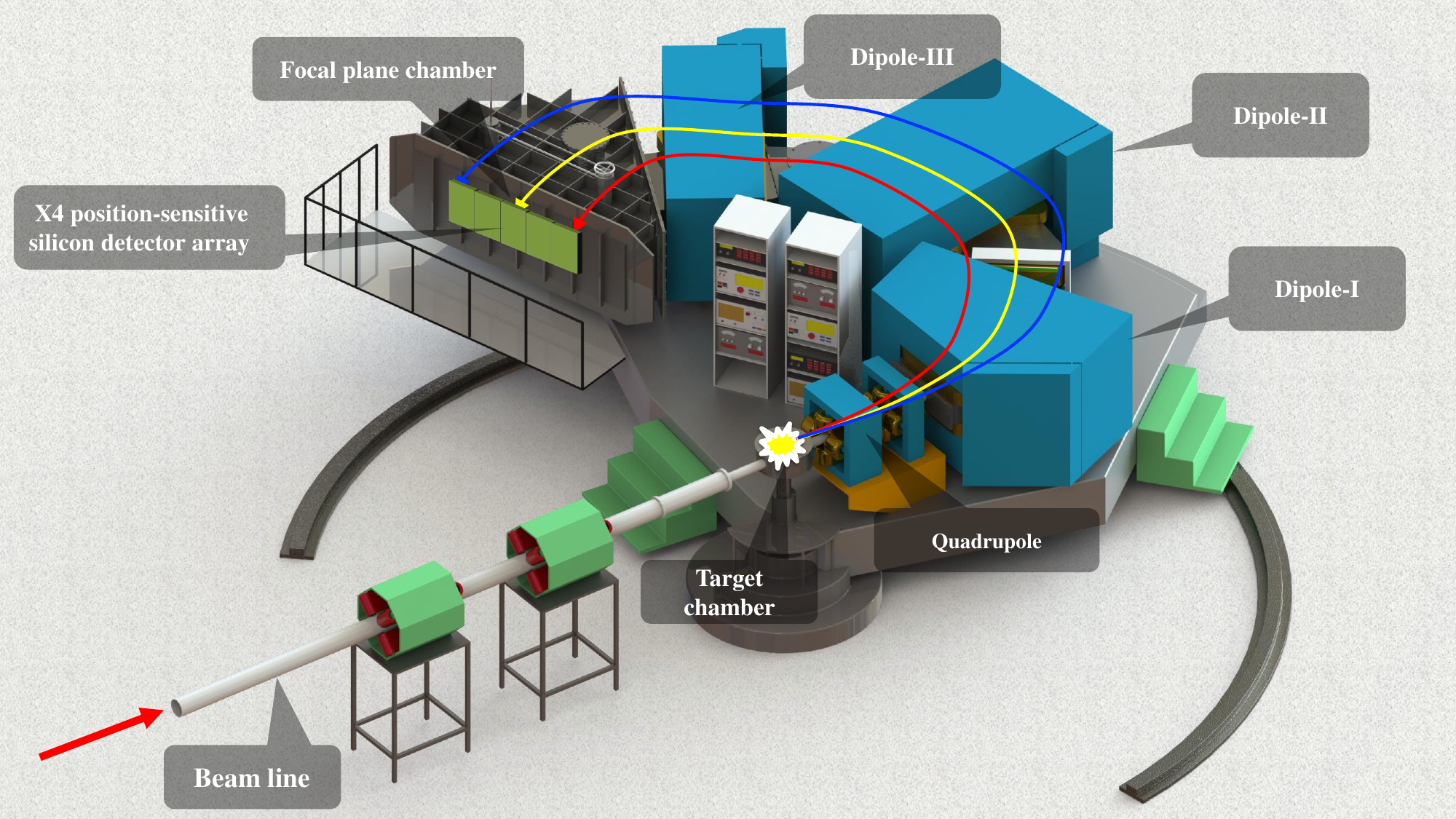}
	\caption{The layout of the Beijing Q3D magnetic spectrometer. }\label{fig1}
\end{figure}
The quadrupole magnet focuses the reaction products, and three dipole magnets deflect the reaction products to the focal plane. In addition to these main components, sextupole, octupole, and decupole magnets are used to correct aberrations. The designed maximum value of magnetic field for three dipoles is 1.7~T. The maximum spectrometer solid angle is 10~mSr, and the best momentum resolution at the maximum solid angle is $\Delta$p/p=10$^{-4}$, which decreases with increasing of the kinematic correction factor k=(1/p)$|dp/d\theta_{lab}|$. Under ideal conditions, when k=0 the momentum resolution can reach 1.0$\times$10$^{-4}$, and when k=0.3, the maximum momentum resolution is 2.0$\times$10$^{-4}$~\cite{drentje1974}. The rotation angle range of the spectrometer is from -10$^o$ to 150$^o$ with an angular precision of 0.05$^o$. The angular rotation and magnet settings have been automated and can be controlled remotely. The power supply of the dipole magnets has been upgraded and the magnetic field stability has reached down to 0.05~Gs.

In essence, the role of the Q3D spectrometer is to generate magnetic fields to control the path of ions. The ejectiles from different reactions or states have unique trajectories due to their varying magnetic rigidity B$\rho$=mv/q, where m, v, q, $\rho$ are the mass, velocity, charge, and radius of rotation of the ions, respectively. The position-sensitive silicon detectors are commonly used as the focal plane detector.

The Q3D spectrometer has been utilized for various experiments, including elastic scattering, inelastic scattering, transfer reactions, and AMS (Accelerator Mass Spectrometry) analysis. The transfer reactions such as ($^7$Li,$^6$He/$^6$Li), ($^{11}$B, $^{7}$Li), ($^6$Li, d), ($^9$Be, $^8$Li), ($^{11}$B, $^{12}$C) were used to obtain neutron, proton, and alpha spectroscopic factors or asymptotic normalization coefficients (ANCs) for multiple nuclei, and then the astrophysical S-factors or reaction rates of reactions such as $^{8}$Li(p, $\gamma$)$^9$Be, $^{12}$C($\alpha$, $\gamma$)$^{16}$O, $^{13}$C($\alpha$, n)$^{16}$O, $^{25}$Mg(p, $\gamma$)$^{26}$Al, $^{13}$C(p, $\gamma$)$^{14}$N, $^{11}$B(p, $\gamma$)$^{12}$C, $^{15}$N(n, $\gamma$)$^{16}$N were indirectly measured~\cite{li2013, shen2020, shen2023, nan2024, guo2012, li2020, li2012, li2014, guo2014}.

For the measurements of heavier ions, the X4 position-sensitive silicon detectors from MACRON were used as focal plane detectors. X4 silicon detector has a 75~mm$\times$40~mm effective detection area composed of eight strips with a width of 5 mm, with energy resolutions of less than 100~keV and a horizontal position resolution of 0.4~mm (FWHM).

When multiple states are measured at the same time, up to six X4 detectors can be used. The detectors can be arranged based on the position distribution of each excited state on the focal plane, or they can be set up as a detector array as shown in Fig.~\ref{fig2}. 
\begin{figure}[h!]
	\centering
	\includegraphics[height=7cm, width=12cm]{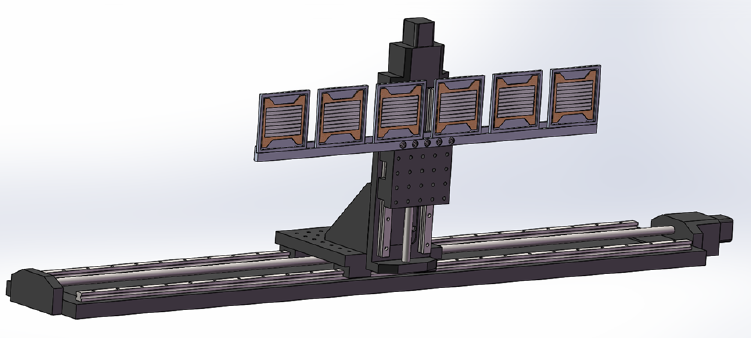}
	\caption{The schematic diagram of the X4 detector array.}\label{fig2}
\end{figure}
Six X4 detectors were arranged with gaps of 55~mm on a high-precision motorized linear stage. In this layout, the obtained spectra of two measurements with an interval of 65~mm for the detector array position can be joined to a continuous one with a length of about 800 mm, and the position distortion on the edge of the detectors can be eliminated effectively by the overlaps between two adjacent detection areas. The spectrum of the multiple states measurement for the $^{25}$Mg($^7$Li, $^6$He)$^{26}$Al reaction is shown in Fig.~\ref{fig3}, in which the events for different states can be identified distinctly by the detector array. 
\begin{figure}[h!]
	\centering
	\includegraphics[height=10cm, width=12cm]{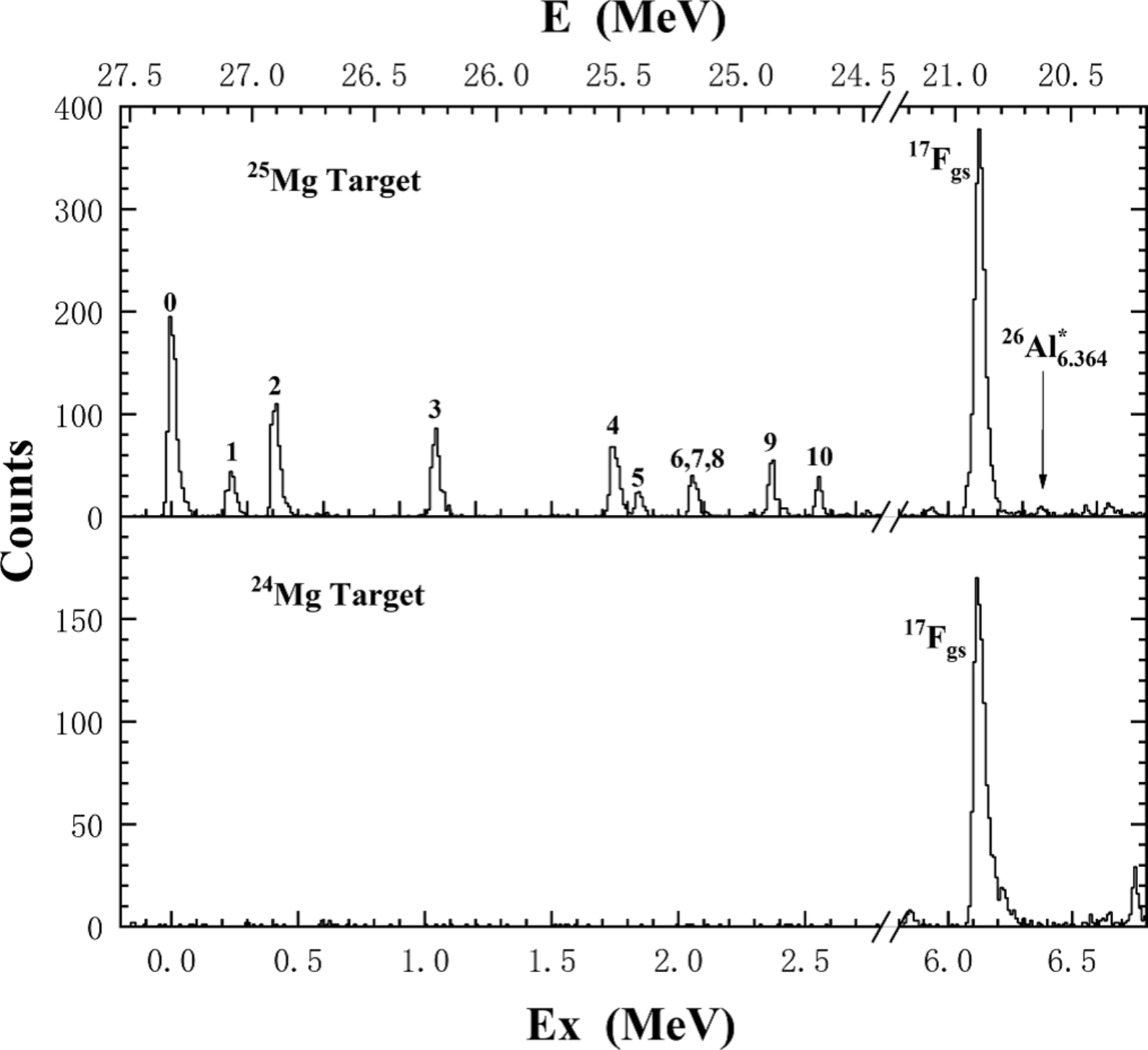}
	\caption{The spectra of the ($^7$Li, $^6$He) reaction for $^{25}$Mg and $^{24}$Mg targets at $\theta_{lab}$=12$^o$, 0$\sim$10 represent the ground and first ten excited states in $^{26}$Al~\cite{li2020}. }\label{fig3}
\end{figure}
To improve the sensitivity of AMS experiments, a Wien filter was installed in front of the Q3D magnetic spectrometer to purify the beam. The maximum voltage of the Wien filter is $\pm$60 kV, and the maximum magnetic field is 0.3 T. A test experiment with a $^{58}$Fe beam was performed to evaluate the performance of the Wien filter~\cite{zhang2024}, where an ionization chamber was used on the focal plane and the spectra are shown in Fig.~\ref{fig4}. 
\begin{figure}[h!]
	\centering
	\includegraphics[height=14cm, width=12cm]{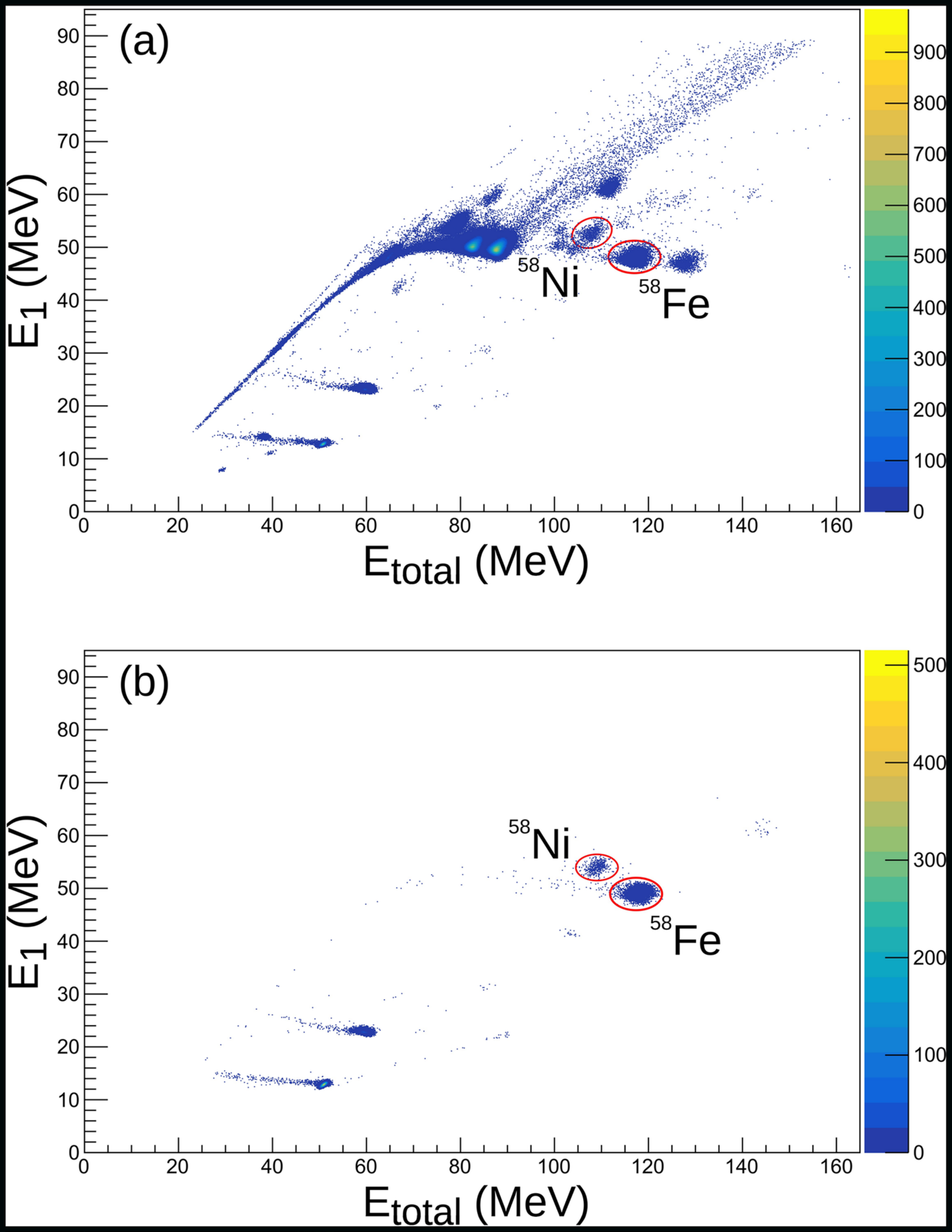}
	\caption{Two dimensional spectra of E$_1$ versus E$_{total}$. E$_1$ is the energy loss of the first anode for the ionization chambers, and E$_{total}$ is the total energy of the ions in the detector. \textbf{a} Spectrum measured without the Wien filter. \textbf{b} Spectrum measured with the Wien filter of $\pm$50 kV voltage and corresponding magnetic field. The figure comes from~\cite{zhang2024}. }\label{fig4}
\end{figure}
The significant presence of interfering beams was excluded after the Wien filter was activated. In the following experiment of $^{60}$Fe analysis, nearly all the contaminants in the beam of $^{60}$Fe were effectively separated. The above experimental method and results make future lunar sample measurements feasible. 
	\subsection{The HiToF spectrometer}\label{sub:3-4}

Multi-nucleon transfer (MNT) is regarded as a promising method to produce neutron-rich heavy or superheavy nuclei~\cite{heinz22,devaraja22} and therefore becomes an important topic nowadays in the field of low-energy nuclear reaction. Many experimental and theoretical investigations have been carried out~\cite{corradi09,dai24,zhu24}, however its mechanism is still unclear due to the complicated transfer and/or transport phenomena of many nucleons presented at energies close to the Coulomb barrier. Experimentally, explicit measurement of MNT products is rather difficult, which requires the equipment should have: i) a good mass ($\emph{A}$) and charge ($\emph{Z}$) identifications for validating enormous variety of reaction channels; ii) a large acceptance for detecting the rare products far from the projectile and target, given the steep decrease of cross sections with increasing number of transferred nucleons; iii) a good energy resolution for distinguishing a huge number of energy levels populated in a particular reaction channel. In order to study MNT reactions, a heavy-ion time-of-flight (HiToF) spectrometer was built in the R60 experimental terminal of the HI-13 tandem accelerator~\cite{wanghaorui24}. It consists of a quadrupole triplet lens and a detecting system, coupled to a rotating chamber with diameter of 40~cm, as shown in Fig.~\ref{fig:3.4.1}.

\begin{figure}[!htbp]
    \centering
    \includegraphics[width=0.5\linewidth]{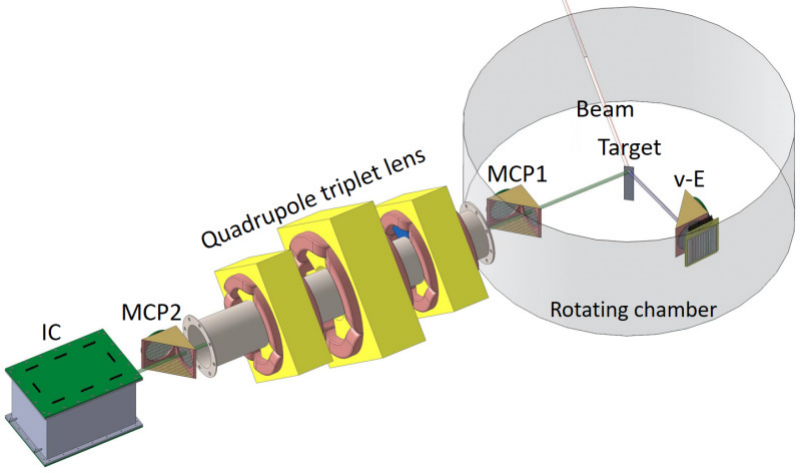}
    \caption{Schematic view of the HiToF spectrometer.}
    \label{fig:3.4.1}
\end{figure}

The HiToF spectrometer was designed for measuring the heavy ions produced in nuclear reactions at energies near the Coulomb barrier. The quadrupole triplet lens (Q$_{1}$-Q$_{2}$-Q$_{3}$) transports the entrance ions on the focal plane inside the ionization chamber (IC), about 2.70~m distance to the target. The Q$_{1}$ and Q$_{3}$ quadrupole have the same structure with aperture of $\Phi_{\text{ape}}$ = 100~mm and maximum magnetic flux density $\emph{B}_{\text{max}}$ = 0.373~T, while $\Phi_{\text{ape}}$ = 130~mm and $\emph{B}_{\text{max}}$ = 0.636~T for the Q$_{2}$ quadrupole. Ions with a magnetic rigidity up to $\emph{B}\rho$ = 0.95~Tm can be analyzed in the maximum angular acceptance $\Delta\theta$ = 3.3$^\circ$ and $\Delta\phi$ = 7.3$^\circ$. The HiToF spectrometer can be operated in three modes: a) both X and Y focusing, b) X focusing and Y parallel, and C) X parallel and Y focusing. The corresponding ion trajectories calculated by GICOSY code~\cite{berz87} are shown in Fig.~\ref{fig:3.4.2}. 

\begin{figure}[!htbp]
    \centering
    \includegraphics[width=0.5\linewidth]{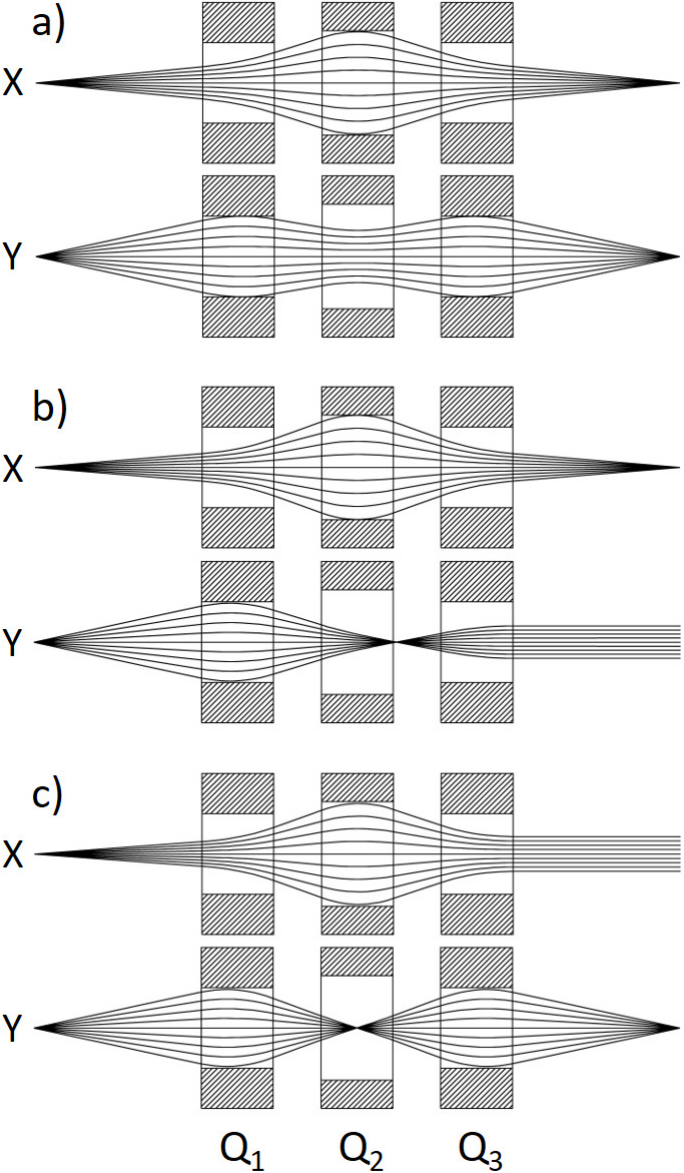}
    \caption{Ion trajectories of the HiToF spectrometer in three operating modes: a) both X and Y focusing, b) X focusing - Y parallel, and c) X parallel - Y focusing.}
    \label{fig:3.4.2}
\end{figure}

The detecting system includes: i) two micro-channel-plates (MCP1 and MCP2) for a ToF measurement with nominally flight length of 1.9~m and typical timing resolution of 120~ps; ii) a multi-sampling position-sensitive IC for a $\Delta\emph{E-E}$ measurement; and iii) a $\nu-\emph{E}$ detector including a MCP and a double-sided silicon-strip detector (DSSD) at the complementary angle for the velocity and energy measurement of target-like particles. The IC has 3-dimensional position resolution capability, as shown in Fig.~\ref{fig:3.4.3}. Its anode is divided into 7 sections, which can provide not only the energy loss $\Delta\emph{E}$ and residual energy $\emph{E}_{R}$, but also the trajectory ($\emph{z}$-direction) of the entrance ion. Each section is divided into two wedge-shaped parts to determine the x-position by the charge division method. The y-position is provided by the drift-time difference between anode and cathode. Typical $\emph{x}$, $\emph{y}$, and $\emph{z}$ position resolutions are about 1.0, 0.5 and 2.0~mm, respectively, mainly depending on the properties of entrance ion and also working gas. 

\begin{figure}[!htb]
    \centering
    \includegraphics[width=0.25\linewidth]{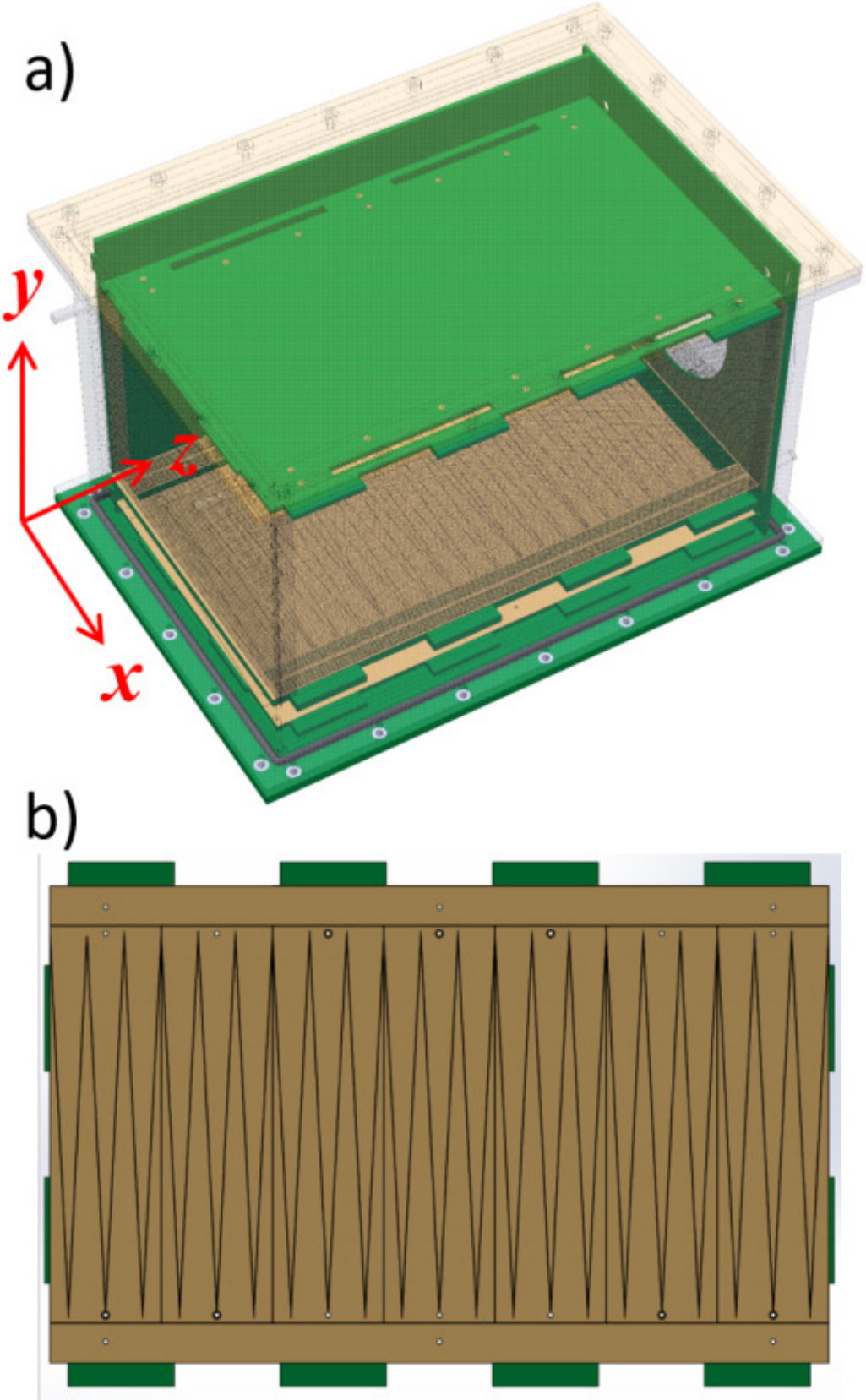}
    \caption{Schematic view of a) IC and b) its anode.}
    \label{fig:3.4.3}
\end{figure}

An experiment of $^{32}$S+$^{90,94}$Zr at beam energy of 135~MeV was performed to test the performance of the HiToF spectrometer. First, the transmission efficiencies were determined by measuring the product ions on the focal plane with and without magnetic fields. In Fig.~\ref{fig:3.4.4} we show the enhancement factors of yield ratios with quadrupole fields on and off for charge states of 11$^{+}$, 12$^{+}$, and 13$^{+}$ in the double-focusing operating mode. In this case, the transmission efficiency reaches its maximum at $\emph{B}_\text{Q2}$/$\emph{B}_\text{Q1}$ = 1.678 with $\emph{B}_\text{Q1}$ = $\emph{B}_\text{Q3}$, which is in good agreement with the GICOSY calculation. The three operating modes were also tested. As an example, in Fig.~\ref{fig:3.4.5} we show the images of X-Y distributions on the focal plane for double-focusing mode and X parallel - Y focusing mode. Finally, the particle identification capability was checked, and results are shown in Fig.~\ref{fig:3.4.6}, Fig.~\ref{fig:3.4.7}. Good resolutions in both $\emph{Z}$ and $\emph{A}$ have been achieved, which meet the requirements of low-energy MNT reactions.

\begin{figure}[!htb]
    \centering
    \includegraphics[width=0.3\linewidth]{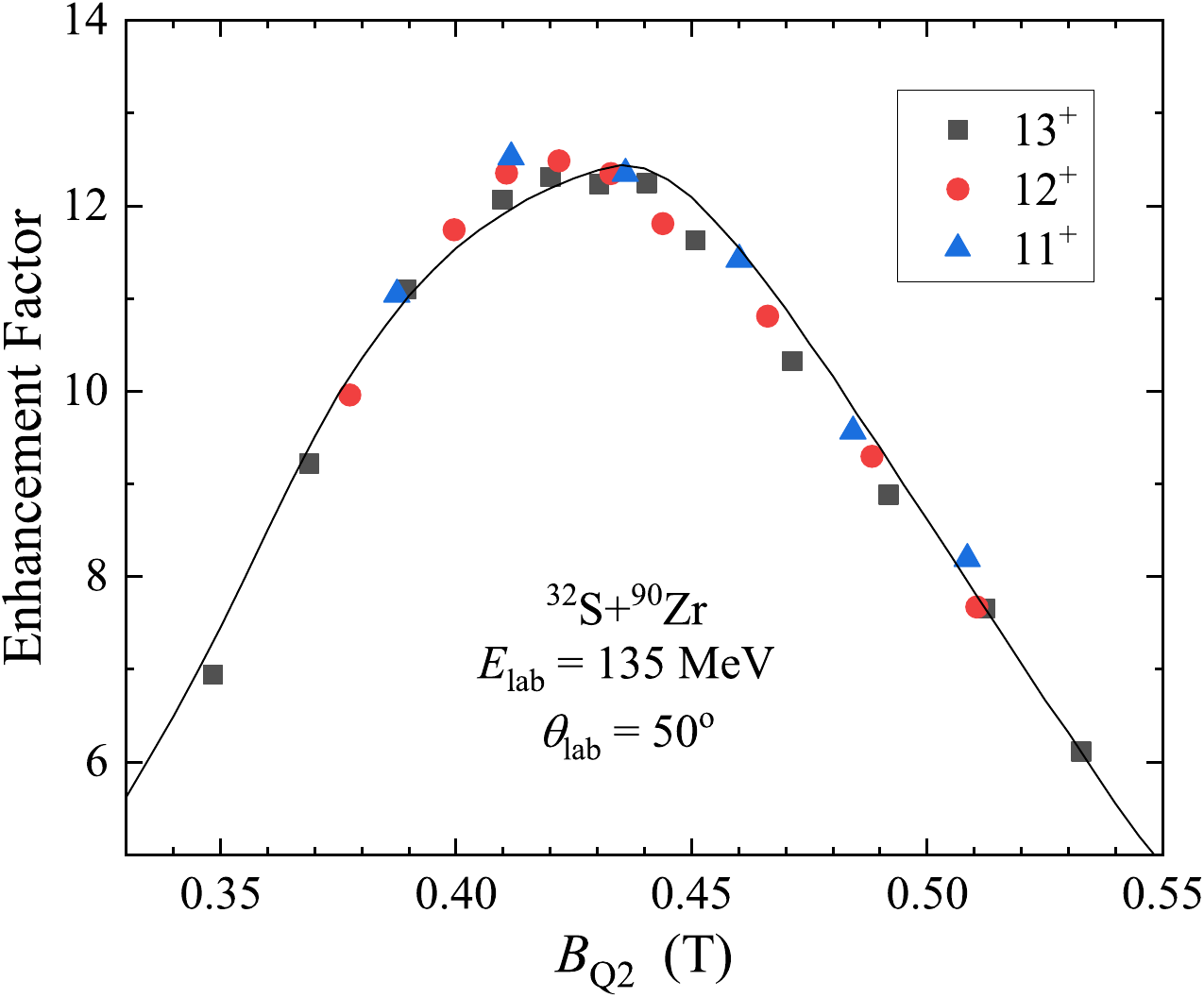}
    \caption{Yield ratios varying with the Q2 field for different charge states. The solid line represents the GICOSY calculation.}
    \label{fig:3.4.4}
\end{figure}

\begin{figure}[!htb]
    \centering
    \includegraphics[width=0.3\linewidth]{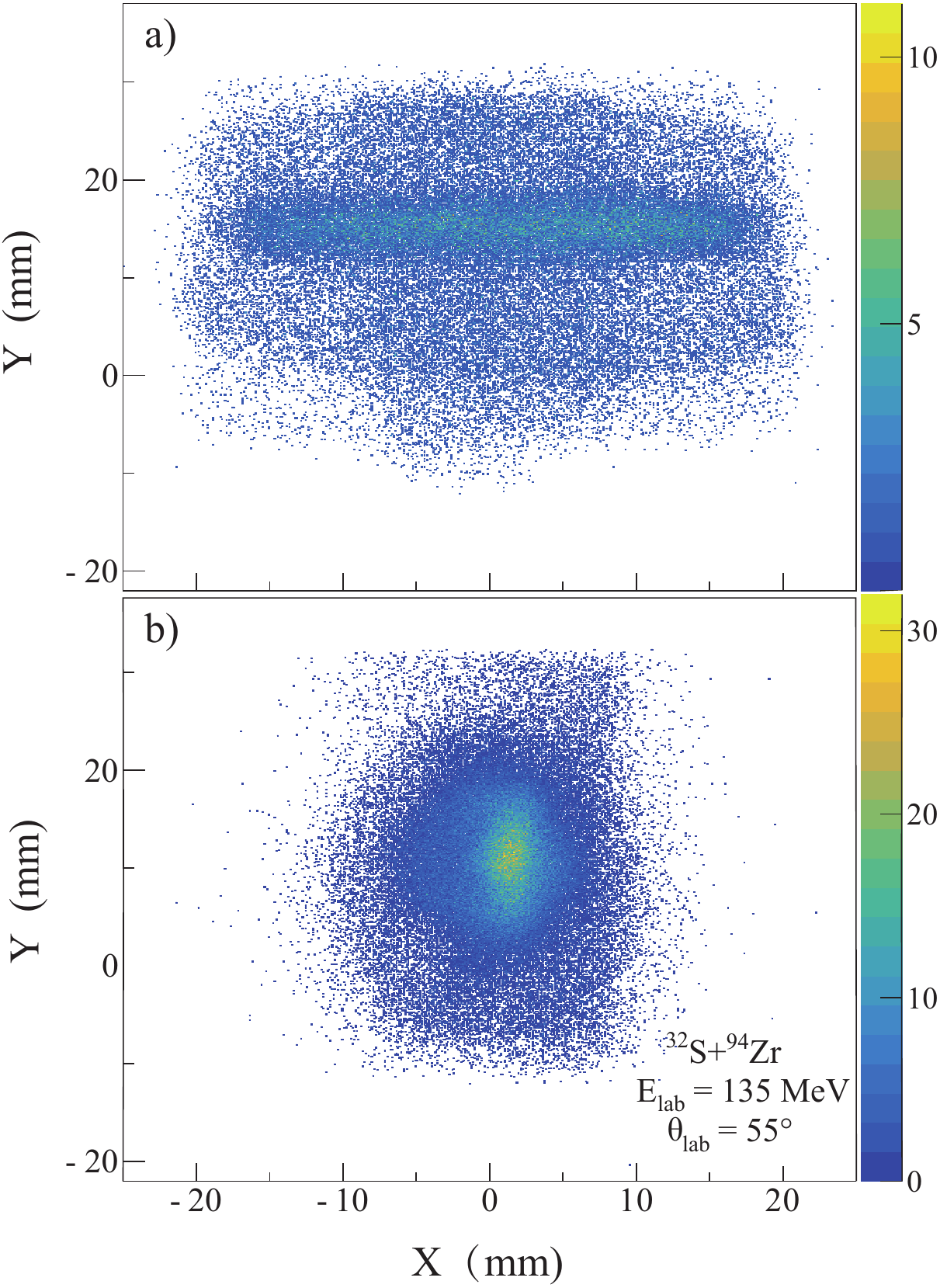}
    \caption{Images of X-Y distributions on the focal plane for (a) X parallel - Y focusing mode and (b) double-focusing mode.}
    \label{fig:3.4.5}
\end{figure}

\begin{figure}[!htb]
    \centering
    \includegraphics[width=0.5\linewidth]{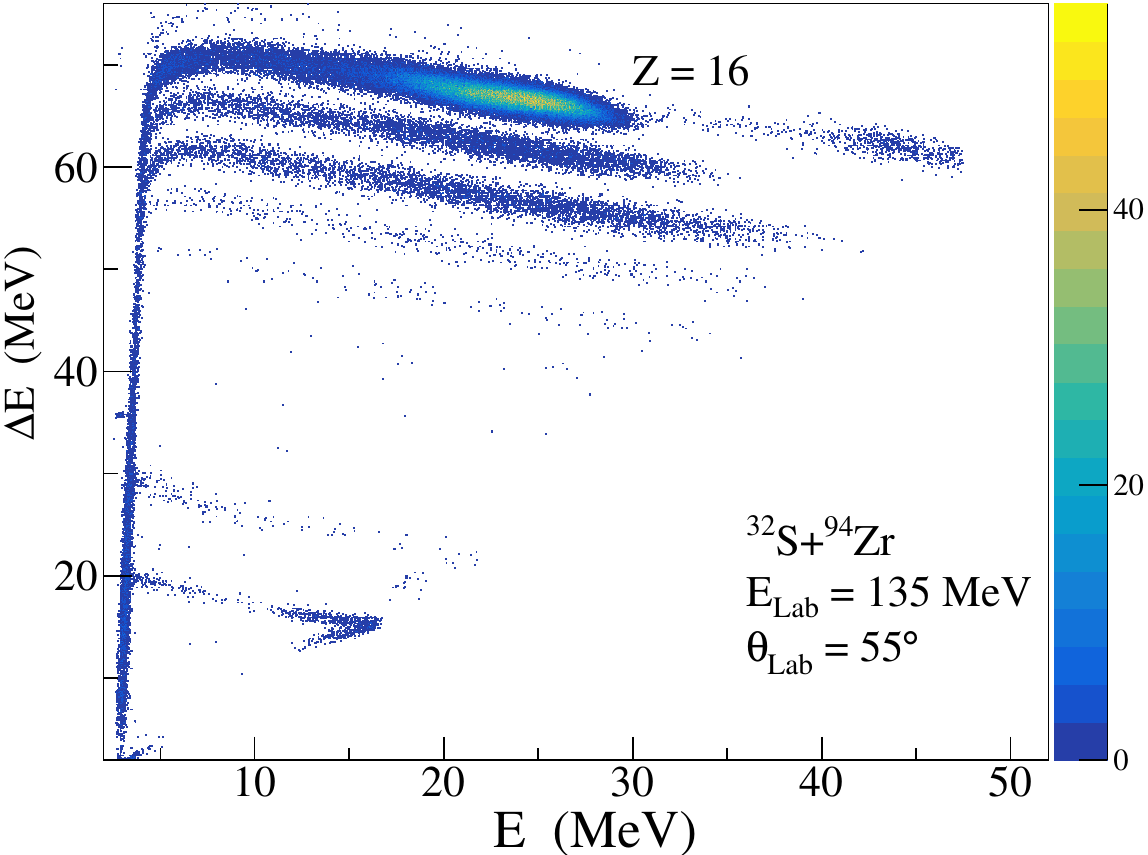}
    \caption{$\Delta\emph{E}$-$\emph{E}$ spectrum measured at $\theta_\text{lab}$ = 55$^\circ$.}
    \label{fig:3.4.6}
\end{figure}

\begin{figure}[!htb]
    \centering
    \includegraphics[width=0.5\linewidth]{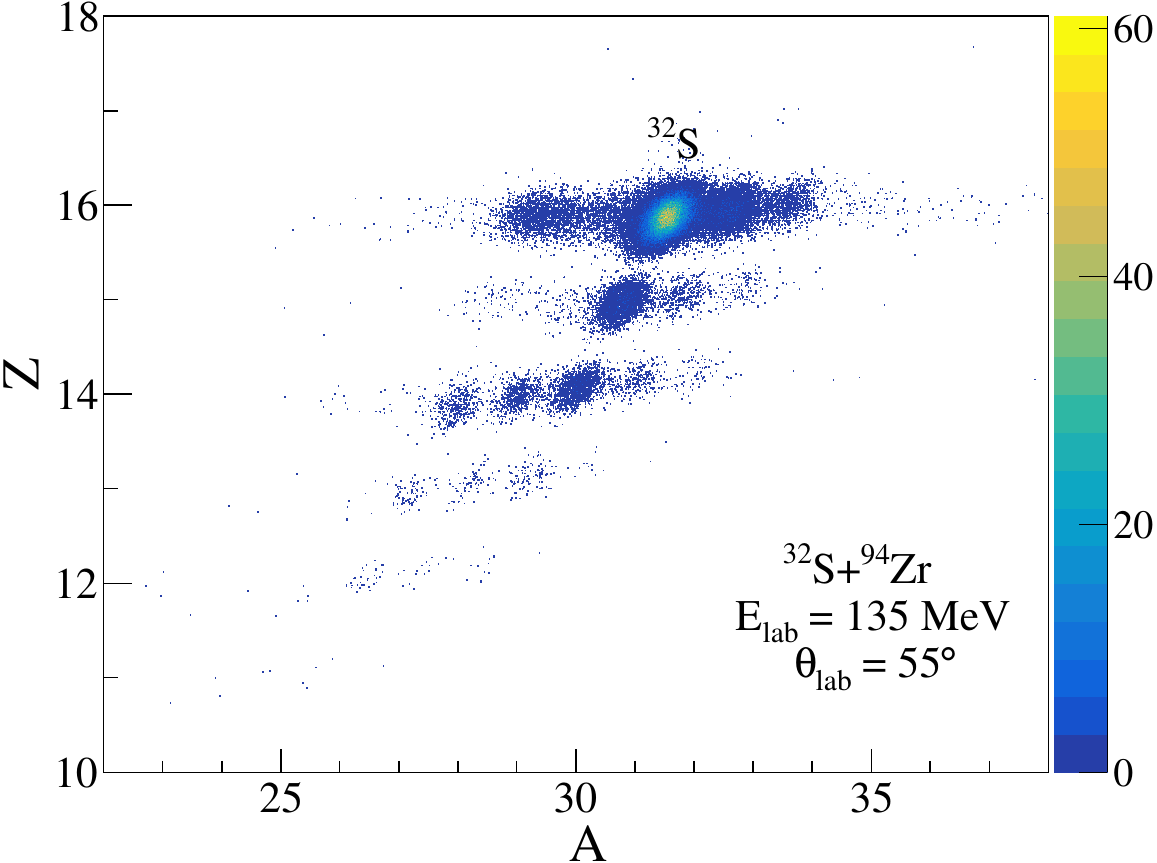}
    \caption{Mass-charge ($\emph{A-Z}$) spectrum of $^{32}$S+$^{94}Zr$ reaction products.}
    \label{fig:3.4.7}
\end{figure}

	\subsection{Gamma detector array}\label{sub:3-5}
	
 The study of atomic nucleus properties by means of $\gamma$-ray spectroscopy has achieved significant advancement with the evolution of $\gamma$-ray detector arrays. The initial attempts to establish detector arrays for $\gamma$-ray spectroscopy entailed the employment of NaI(Tl) crystals, which have high efficiency but exhibit poor energy resolution. In the mid-1960s, $\gamma$-ray spectroscopy research began to use germanium detectors, which significantly improved energy resolution. By the late 1970s, small arrays of germanium detectors, consisting of germanium crystals, were developed. The HPGe-BGO anti-Compton spectrometer, which integrates HPGe and BGO within the $\gamma$-ray detector array, represents a significant advancement in in-beam $\gamma$ spectroscopy. This anti-Compton spectrometer employs BGO crystals surrounding the primary HPGe detector to capture $\gamma$ rays that escape from the HPGe detector through Compton scattering. This mechanism effectively eliminates the Compton continuum within the HPGe, thereby improving the Peak to Total Ratio (P/T).

With advancements in detector technology, the energy resolution for $\gamma$ rays has significantly improved, making $\gamma$ rays a powerful tool for probing the nuclear structure. In particular, the integration of multiple $\gamma$ detectors into an array~\cite{LEE1990641,SIMPSON19971339} for $\gamma$-$\gamma$ coincidence measurements enhances the overall solid angle, improves granularity of the detector configuration, and substantially increases detection efficiency while maintaining a high level of selectivity. This effectively suppresses the background of the $\gamma$ spectrum, enabling the detection of $\gamma$ rays with extremely low intensity.

Starting in the 1980s, China embarked on a collaborative project at the Beijing Tandem Accelerator Nuclear Physics National Laboratory, leading to the establishment of a $\gamma$-ray detector array composed of 15 anti-Compton spectrometers. In 2018, the National Laboratory of Heavy Ion Research Facility in Lanzhou completed the construction of a detector array consisting of 15 HPGe detectors with 70\% relative efficiency and 8 clover detectors with 120\% relative efficiency, each equipped with BGO anti-Compton shielding. In response to the growing demands of $\gamma$-ray spectroscopy research, a consortium comprising multiple Chinese institutions and universities---including the China Institute of Atomic Energy, the Institute of Modern Physics of the Chinese Academy of Sciences, Peking University, Beihang University, Shandong University, Jilin University, Tsinghua University, Shenzhen University, and Shenzhen Technology University and so on---established the In-Beam Gamma Spectroscopy Collaboration Group in 2019. Collectively, they established the Conjoint Gamma Array in China (CGAC), as illustrated in Fig~\ref{fig:CGAC}.

\begin{figure}[!htb]
\begin{center}
\includegraphics[width=12.0cm]{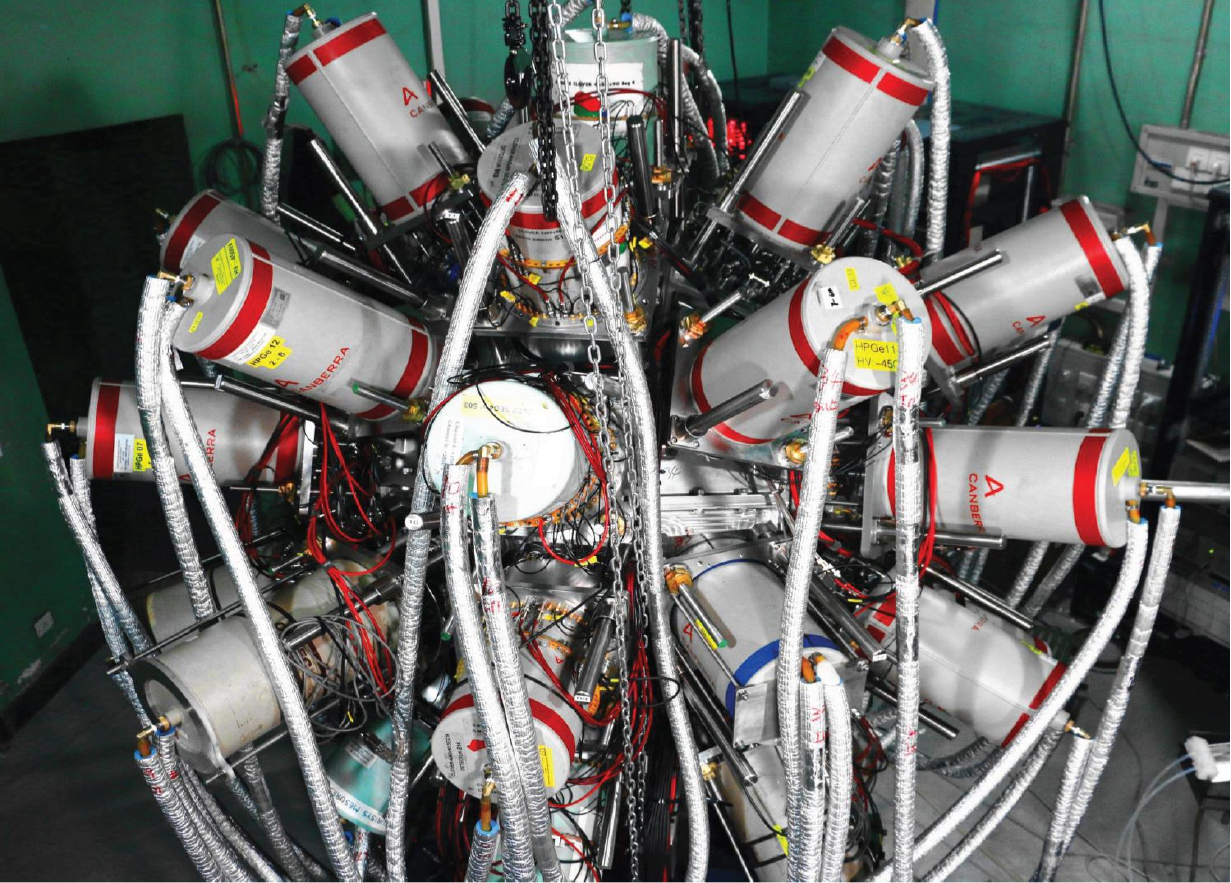}
\end{center}
\caption{The photo of the Conjoint Gamma Array in China (CGAC).}
\label{fig:CGAC}
\end{figure}

The CGAC comprises 12 clover HPGe detectors and 28 n-type coaxial HPGe detectors, totaling 40 detectors, all of which are equipped with BGO anti-Compton shielding. This array consists of 76 individual detection modules, each demonstrating an energy resolution between 2.0~keV and 2.8~keV at 1332~keV. The $\gamma$ detectors are arranged across five rings at angles of 31$^{\circ}$ (149$^{\circ}$), 61$^{\circ}$ (119$^{\circ}$), and 90$^{\circ}$ (as depicted in Fig~\ref{fig:CGAC-D}). This array features the highest number of detection units among similar Asian arrays, with a simulated detection efficiency of approximately 3\% for 1332~keV $\gamma$ rays.

\begin{figure}[!htbp]
\begin{center}
\includegraphics[width=8.0cm]{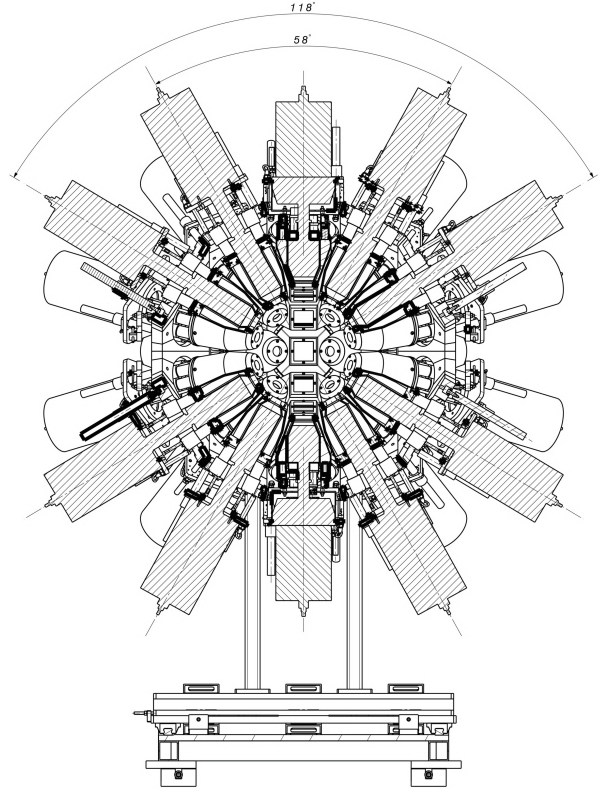}
\end{center}
\caption{The layout of the CGAC.}
\label{fig:CGAC-D}
\end{figure}

The data acquisition system of the CGAC employs a digital waveform sampling system, consisting of a 16-channel pulse processor (Pixie-16) and a programmable logic module (MZTIO) supplied by XIA LLC. The electronic signals generated by the HPGe and BGO detectors in the detector array, after being amplified by a preamplifier, are directly fed into the Pixie-16 processor for subsequent processing~\cite{WU2020164200}.

The digital data stream, after undergoing analog-to-digital conversion (ADC), is split into two components: one is rapidly filtered to generate trigger signals sent to the FPGA, while the other is directed into a delayed First-In-First-Out (FIFO) buffer to manage the delay between the rapid trigger signals and the anti-coincidence (VETO) signals. The data stream passing through the delayed FIFO is further segmented into three parts: energy filtering, pile-up detection, and constant fraction timing discrimination (CFD). The segment entering the CFD generates a trigger signal and records a timestamp to record the waveform. Once the signals from the BGO detectors reach the FPGA, they will undergo anti-coincidence processing with the signals from the HPGe detectors after waveform shaping to determine whether the event should be retained. Digital acquisition provides significant advantages, including enhanced energy resolution at high count rates compared to analog electronics, minimal dead time during acquisition, simplified electronic circuitry, and ease of parameter adjustment.

The auxiliary detectors of the CGAC include a Si(Li) detector for measuring internal conversion electrons, a Plunger device for lifetime measurements of excited states, an array of silicon detectors for charged particle measurements, and an array of CsI detectors, among others.

	\newpage
	\section{Current physical results}\label{fourth}
With BRIF, high-purity unstable ion beams can be produced. They can either be ISOL separated for direct decay experiments or be post-accelerated by the tandem accelerator and the superconducting linac for reaction dynamics experiments. Here we summarize current experimental results mainly on unstable Na and Rb isotopes, including exotic decay of $^{20}$Na from the ISOL system, elastic scattering of radioactive Na isotopes with post-accelerated ISOL beams, and decay properties of unstable Rb isotopes.

	\subsection{Exotic decay of $^{20}$Na from the ISOL system}\label{sub:4-2}
	Exotic nuclear structure and abnormal reaction dynamics of short-lived isotopes bring stringent challenge to nuclear theoretical models, and provide fresh input in the simulation of most violent cosmic events as well. As a result of weakening binding energy, unusual decay modes appear in nuclei at the drip line, or via highly excited states of daughter nuclei~\cite{pfutzner12,Hofmann82,Klepper82,Giovinazzo02,pfutzner02,cable83,fynbo05,hoff20,Blank2008,Qi2019}. Recently, exotic $\beta-\gamma-p$ decay modes were observed in $T_z=-2$ nuclei $^{32}$Ar~\cite{Bhattacharya2008} and $^{56}$Zn~\cite{Orrigo2014}, significant corrections to both $B(F)$ and $B(GT)$ were introduced that demonstrate the importance of the often omitted indirect feeding by $\gamma$ transitions to the particle intensity.

  $^{20}$Na decays with a half life of 447.9(23)~ms and a decay energy of 13892.5(11)~keV, to the daughter nucleus $^{20}$Ne. The spin and parity of the ground state of $^{20}$Na is $2^+$ and the isospin $T=1$, its decay mainly proceeds through the transitions to the $1^+$, $2^+$ and $3^+$ states in $^{20}$Ne with $T=0$ or 1. Many experimental investigations have been carried out for the decay of $^{20}$Na since 1960's, resulting in a detailed decay scheme of $^{20}$Na. Laursen \textit{et al.} made a high-statistics measurement of the $\beta$-delayed $\alpha$ spectrum of $^{20}$Na at the IGISOL facility, and observed two groups of $\alpha$'s below 1~MeV by applying the $\alpha$ and $^{16}$O coincidence~\cite{Laursen2013}. $\beta-\gamma-\alpha$ rare decay sequences were suggested based on these two new $\alpha$ particle lines. Such $\beta-\gamma-\alpha$ sequences are essential in determining the direct $\beta$ feedings to the low-lying $\alpha$ emitting states in $^{20}$Ne. As the first on-line experiment at BRIF, a direct measurement of exotic $\beta-\gamma-\alpha$ decay mode in $^{20}$Na was performed using a high-intensity pure $^{20}$Na beam and the $\beta-\gamma-\alpha$ coincidence spectroscopy.

 \begin{figure}[!htb]
    \centering
    \includegraphics
    [width=0.8\linewidth]
	{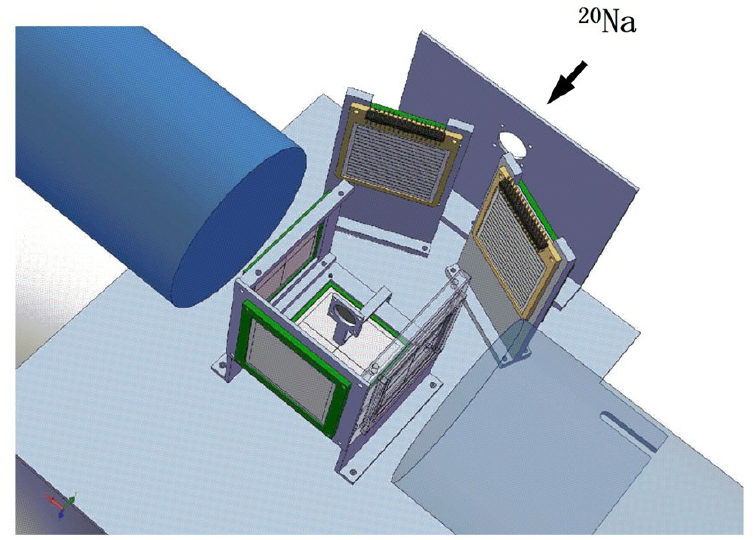}
	\caption{Experimental setup for the exotic decay of $^{20}$Na.}
	\label{20na setup}
\end{figure}

  The experimental setup for the exotic decay of $^{20}$Na is shown in Fig.~\ref{20na setup}. The $^{20}$Na ISOL beam of 100~keV was implanted into an aluminum collection foil surrounded by a silicon box formed by five pieces of Multiguard Silicon Quadrant (MSQ) of 1~mm thick to measure the $\beta$ particles. Opposite to the implantation point, two pieces of DSSDs were installed to measure the $\beta$-delayed $\alpha$ particles. Each DSSD was backed by an MSQ detector of the same thickness to reject the $\beta$ summing events in the $\alpha$ spectra. The DSSD has a thickness of about 70 $\mu$m  with a very thin dead-layer of only 100~nm to reduce the energy loss of low-energy $\alpha$ particles. Two HPGe detectors with relative efficiency of 175\% each were installed in the cylindrical flange to measure the $\gamma$ rays. The average implantation rate of $^{20}$Na $1^+$ ions is about $1.5\times10^4$ pps during the experiment.
\begin{figure}[!htb]
    \centering
    \includegraphics
    [width=0.6\linewidth]
	{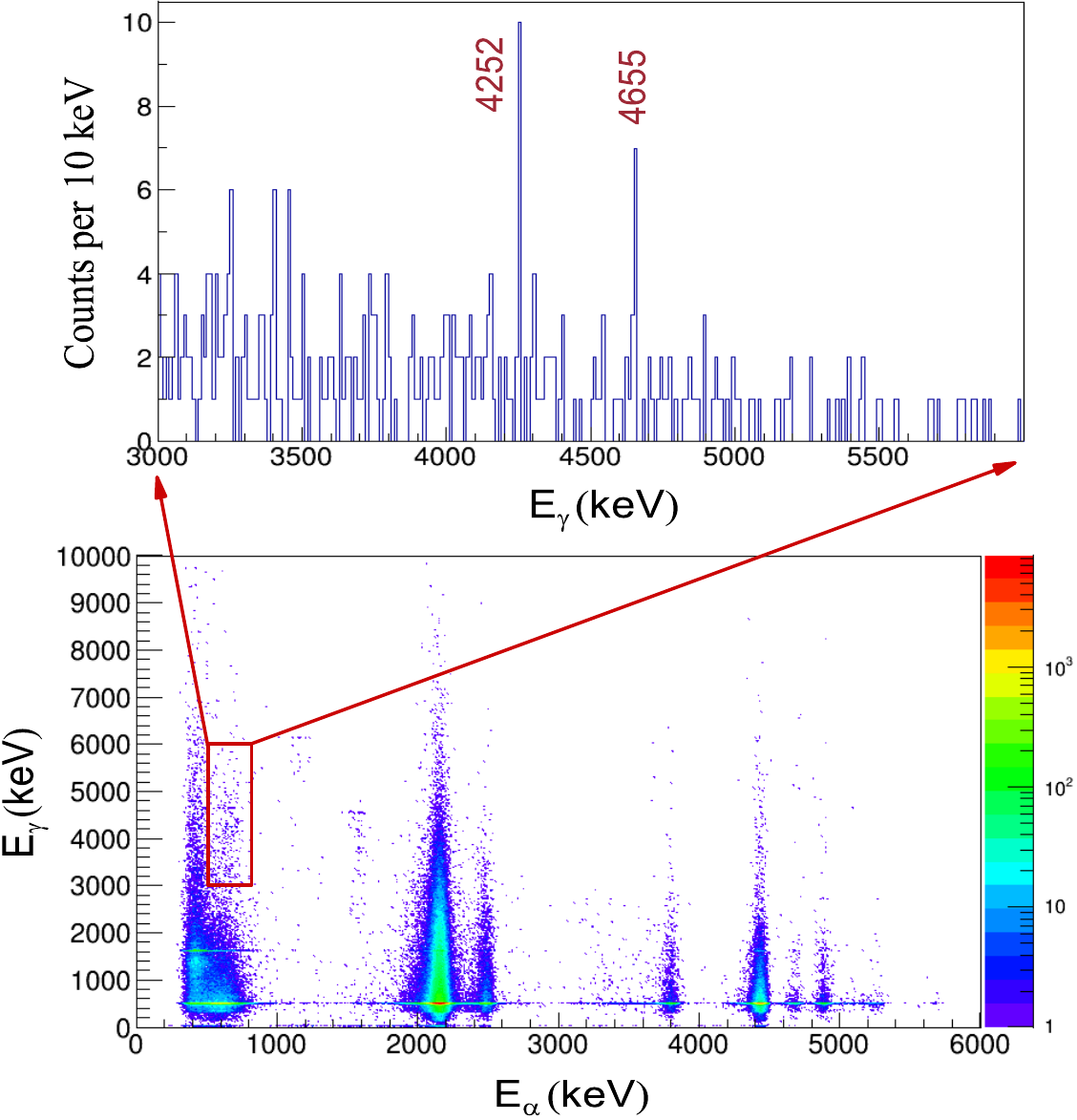}
	\caption{$\alpha$ gated $\gamma$ spectrum from the $\alpha-\gamma$ coincidence matrix of $^{20}$Na exotic decay.}
	\label{20na projection}
\end{figure}

  The energy calibration of $\beta$-delayed $\alpha$ singles spectrum of $^{20}$Na was realized by using the well-determined $\alpha$ particle lines, with the correction of energy-loss nonlinearity of $\alpha$ particle in aluminum and the dead layer of DSSD. Standard sources of $^{152}$Eu and $^{56}$Co were used to determine the absolute efficiencies of $\gamma$ rays up to about 3.5\,MeV. The relative detection efficiency curve up to high-energy region was measured by the well known resonance in the $^{27}$Al($p$,$\gamma$)$^{28}$Si reaction at $E_p=293$\,keV with the same HPGe detectors in several different distances. The extrapolation to high-energy $\gamma$ rays was then realized by GEANT4 simulation with the experimental geometry and the detector parameters based on the $^{27}$Al($p$,$\gamma$)$^{28}$Si reaction data. The intensity of $\gamma$ rays is determined relative to that of the 1634~keV $2_1^+\rightarrow0_{gs}^+$ transition in $^{20}$Ne, while the absolute intensity of 1634 keV $\gamma$ line is deduced to be 79.6(2.5)\% per 100 decays of $^{20}$Na from the net counts $N_\gamma(1634)$ by the following relation,
\begin{eqnarray}
I_\gamma(1634)=\frac{N_\gamma(1634)}{N_\gamma(1634)+N_\gamma(11259)+N_\alpha(total)}.
\end{eqnarray}
The total $\alpha$ particle emission probability $I_\alpha$(total) is deduced to be 20.3(6)\% from the total $\alpha$ counts $N_\alpha$(total) by the similar relation.

An $\alpha-\gamma$ matrix was constructed taking all $\alpha-\gamma$ coincidence events within 140\,ns to explore the exotic $\beta-\gamma-\alpha$ decay sequences of $^{20}$Na. The $\alpha$ gated $\gamma$ spectra were projected at different $\alpha$ particle energy intervals from the $\alpha-\gamma$ matrix, as shown in Fig.~\ref{20na projection} as an example. Two $\gamma$ lines are seen in Fig.~\ref{20na setup} at 4252 and 4655\,keV, which coincide with the $9875 \rightarrow 5621$\,keV and $10277 \rightarrow 5621$\,keV transitions in $^{20}$Ne, respectively. The $\alpha$ particle energy gate of $E_\alpha$=640-780~keV agrees with the recently reported 714~keV $\alpha$ particle line emitted from the 5621~keV $3^-$ state in $^{20}$Ne. Two $\gamma-\alpha$ exotic sequences are thus assigned that correspond to $^{20}$Ne(9875)$\rightarrow$$^{20}$Ne(5621)$\rightarrow$$^{16}$O$_{gs}$ and $^{20}$Ne(10277)$\rightarrow$$^{20}$Ne(5621)$\rightarrow$$^{16}$O$_{gs}$, respectively. A 4542~keV $\gamma$ line is also observed in coincidence with a known $\alpha$ particle line of $E_\alpha$=1592\,keV, which makes up a $^{20}$Ne(11262)$\rightarrow$$^{20}$Ne(6720)$\rightarrow$$^{16}$O$_{gs}$ $\gamma-\alpha$ exotic sequence. %

A decay scheme of $^{20}$Na is shown in Fig.~\ref{20na decay scheme} on the basis of the $\gamma$, $\alpha$ lines and the $\alpha-\gamma$ coincidence observed in the present work~\cite{Wang_2021}. Three $\beta-\gamma-\alpha$ exotic decay sequences in $^{20}$Na are shown in Fig.~\ref{20na decay scheme}. Two $\gamma$ transitions de-exciting from the 10277\,keV isobaric analog state (IAS) and 9875\,keV $3^+$ state to the 5621\,keV level in $^{20}$Ne, respectively, contribute to the $\alpha$ particle intensity observed at $E_\alpha=714$\,keV~\cite{Laursen2013}. After subtraction, the direct $\beta$ feeding has an upper limit ($1\sigma$) of 0.0029\%, which corresponds to a lower limit of log$ft>8.5$. For the 5788\,keV $1^-$ level, no $\gamma$ line is observed in coincidence with the $E_\alpha$=847\,keV region and the intensity of 847\,keV $\alpha$ particle line is thus from the direct $\beta$ feeding. The 5621~keV $3^-$ level and 5788~keV $1^-$ level have similar $\beta$ feedings and log$ft$ values, which are consistent with the first forbidden $\beta$ transition.

\begin{figure}[!htb]
    \centering
    \includegraphics
    [width=0.6\linewidth]
	{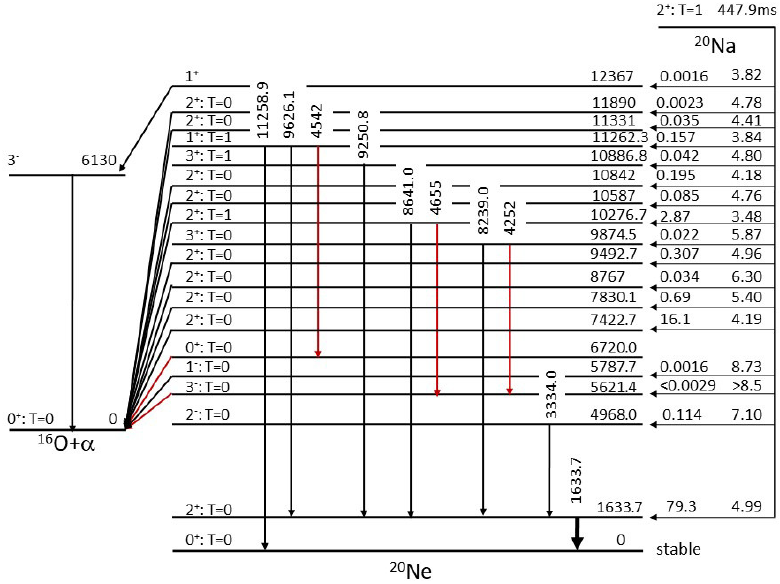}
	\caption{Decay scheme of $^{20}$Na. The newly discovered exotic decay sequences are shown with red arrows. }
	\label{20na decay scheme}
\end{figure}

The $\beta$ decay of $^{20}$Na was calculated with USDB interaction for the charge independent part plus the Coulomb, charge-dependent and charge-asymmetric nuclear Hamiltonian for the $sd$ shell. The calculated $B(GT)$ values scaled with a quenching factor of $q^2_{(GT)} = 0.60$~\cite{Richter2008} agree well with the experimental ones. For the 10277\,keV IAS in $^{20}$Ne, the calculated total $\beta$ transition strength has $B(GT)=0.093$ and $B(F)=1.994$. The reduction in $B(F)$ is due to isospin mixing with other $2^+$ $T=0$ states, which can account for the observed isospin-forbidden $\alpha$ decay of the IAS in $^{20}$Ne.
	\subsection{Elastic scattering of radioactive Na isotopes with post-accelerated ISOL beams}\label{sub:4-3}
	The reaction dynamics of light loosely bound nuclei near the drip lines is being studied extensively in nuclear physics. Because of the low nucleon binding energies, those nuclei could exhibit exotic properties. Elastic scattering is a useful probe for studying the size and the surface diffuseness of exotic nuclei by comparing the angular distributions and optical potential parameters of the reactions induced by different nuclei~\cite{yang13,acosta11,cubero12,lemasson09,marquinez16,cook18,pietro10,pesudo17,duan20,Tavora24,yan18,mazzocco19,blackmon05,yang21,zhang18,liang02,romoli04,yang22,ovejas23}. The Coulomb barrier energy region is a transition region between classical and quantum mechanisms. Studies of the mechanism of nuclear reactions at energies around the Coulomb barrier are helpful for understanding the nuclear structure and nuclear reaction kinetics~\cite{yang23}. To study the properties of Na series isotopes $^{21,22,23}$Na systematically, a series of measurements of $^{21,22,23}$Na+$^{40}$Ca elastic scattering angular distributions were performed above the Coulomb barrier energy, and also for verifying the abilities of BRIF to perform experiments with post-accelerated beams.

The proton cyclotron served as the driving accelerator. The proton beam then bombarded a MgO target and the Na isotopes were produced. The products diffusing out of the thick target were collected by an ion source and separated by a two-stage ISOL system for subsequent post-acceleration. Finally, the ISOL beams were accelerated by the HI-13 tandem accelerator and transported to the R60 scattering target chamber for the angular distribution measurement. The energies of the ISOL $^{21,22,23}$Na$^{7+}$ beams were 91~MeV ,87~MeV and 83.25~MeV, and the currents were 1.2~$\times$~10$^{4}$~pps ,2.5~$\times$~10$^{5}$~pps and 0.5~enA, respectively.

The setup in the target chamber is shown in Fig.~\ref{fig:chamber}. In order to identify the different beams, two micro-channel plates (MCPs) were placed in front of and behind the target to measure the time-of-flight (TOF) signals of the beams. After the identification of the beams, the MCPs were removed to avoid the interference to the measurement of the angular distribution. A quadrant silicon detector (QSD) with thickness of 1~mm was placed at the end of the chamber to measure the energy and intensity of the beams. The target was made by evaporation of natural CaF$_{2}$ on a Au substrate. In the $^{23}$Na experiment, the thicknesses of target and substrate were 118~$\mu$g/cm$^{2}$ and 200~$\mu$g/cm$^{2}$, respectively. Because of the low intensities of the $^{21,22}$Na beams, the thicknesses of target and substrate in the corresponding experiment were 1120~$\mu$g/cm$^{2}$ and 400~$\mu$g/cm$^{2}$, respectively.

\begin{figure}[!htb]
    \centering
    \includegraphics
    [width=0.8\linewidth]
	{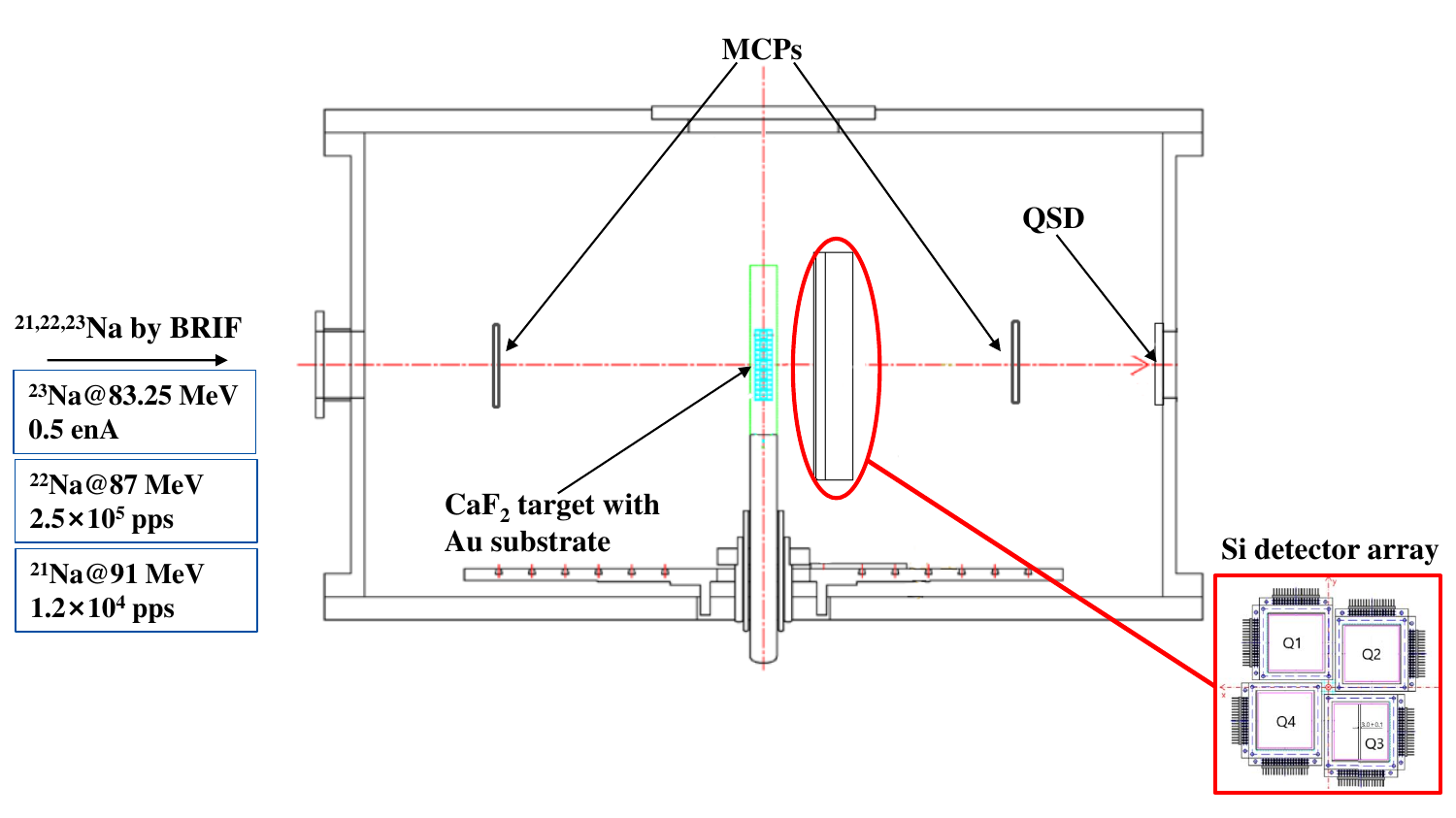}
	\caption{Experimental setup in the target chamber.}
	\label{fig:chamber}
\end{figure}

The silicon detector array consists of four detector telescopes with 3 layers and was placed 10 cm behind the target. The first and second layers of each telescope are two single-sided silicon strip detectors (SSSDs) with thickness of 20~$\mu$m and 300~$\mu$m, respectively. Each SSSD has a typical energy resolution of 0.3\% and has 16 strips with a width of 3~mm, which were spaced 0.1~mm apart. These two SSSDs were placed in horizontal and vertical directions, respectively. The third layer is the QSD with a thickness of 1~mm and an area of 50$\times$50~mm$^{2}$ for emitted light reaction products such as proton and $\alpha$. Such a configuration is able to identify reaction products including light and heavy ions. The angular coverage range of the detector array in the laboratory system is from 10.2$^{\circ}$ to 39.6$^{\circ}$.

The first part of the experiments was the measurement of $^{23}$Na+$^{40}$Ca elastic scattering angular distribution~\cite{nan21}. The angular distribution measured by a silicon strip detector array in a scattering chamber using the ISOL beam from BRIF is in good agreement with that measured by the high-precision Q3D magnetic spectrograph using the non-ISOL beam at nearly the same energy. The result is shown in Fig.~\ref{fig:spec_23na}. 

\begin{figure}[!htb]
    \centering
    \includegraphics
    [width=0.6\linewidth]
	{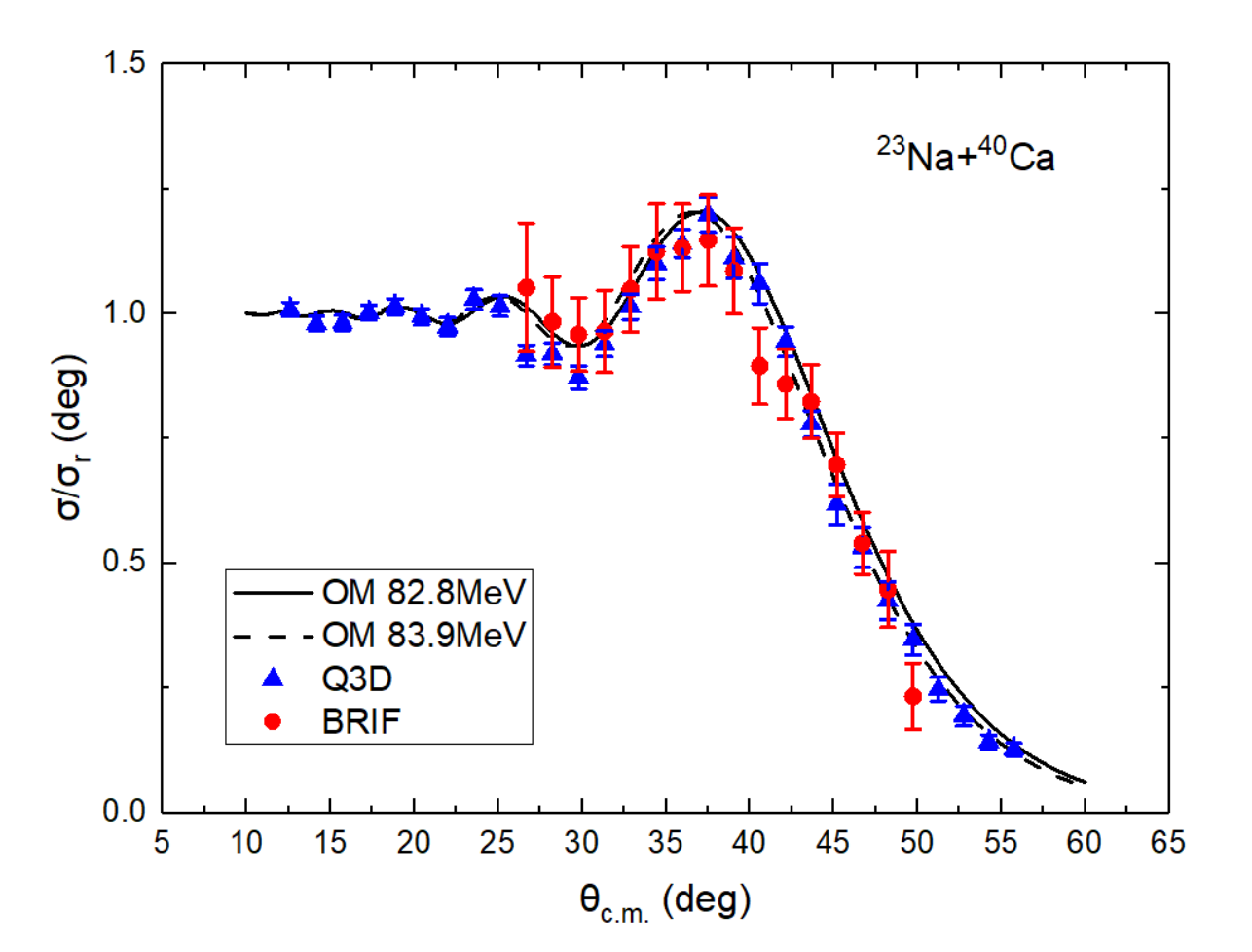}
	\caption{Angular distributions of $^{23}$Na+$^{40}$Ca elastic scattering. Circles and triangles denote the results based on BRIF and Q3D. Solid and dashed curves represent the optical model calculations at 82.8 and 83.9~MeV.}
	\label{fig:spec_23na}
\end{figure}

\begin{figure}[!htb]
    \centering
    \includegraphics
	[width=0.6\linewidth]
	{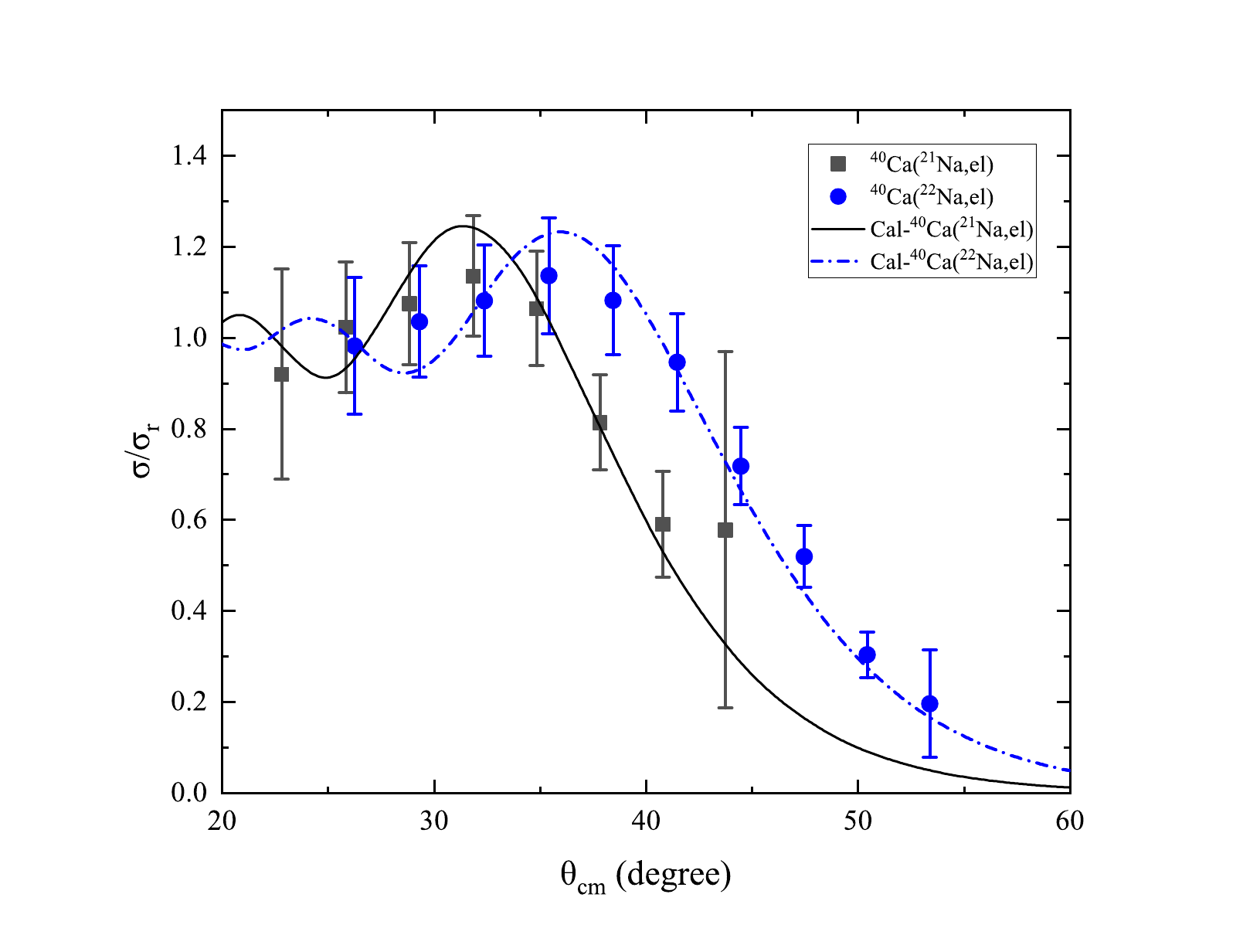}
	\caption{Angular distribution of $^{21,22}$Na+$^{40}$Ca elastic scattering. Black squares and blue circles are experimental data for $^{21}$Na and $^{22}$Na systems, respectively. Blue dot-dashed and black curves denote the optical calculations with systematic optical model potentials for the corresponding systems.}
	\label{fig:spec_2122na}
\end{figure}

The other part of the experiments was the measurement of the $^{21,22}$Na+$^{40}$Ca elastic scattering angular distribution~\cite{nan24}. The experimental results are in good agreement with the systematic optical potential theory, proving that there is no significant breakup effect in the $^{21,22}$Na+$^{40}$Ca reaction channels, which is in agreement with the conclusion in~\cite{yan18} that there is no significant breakup effect in the reaction channels of nuclei near the proton drip-line. The result is shown in Fig.~\ref{fig:spec_2122na}. These works have proved the ability of BRIF to carry out experiments with post-accelerated beams, especially radioactive beams. 

	\subsection{Measurement of $\beta$ intensity distribution of $^{88}$Rb}\label{sub:4-4}
	\ \ \ \ Reactor decay heat, which comes from the decay of radionuclides in fission products and neutron transmutation products, is the main contributor of heat in the reactor core~\cite{Tobias1980}. Due to the "Pandemonium Effect"~\cite{hardy1977essential}, beta decay intensity distribution derived from HPGe detectors is shifted toward the lower energy region, which leads to a systematic bias in $\overline{E}_\gamma$ and $\overline{E}_\beta$, where $\overline{E}_\gamma$ and $\overline{E}_\beta$ represent the average energies of the emitted gamma rays and beta particles, respectively.
	
	To obtain more accurate data to reduce the uncertainty of reactor decay heat, the Working Party on International Nuclear Data Evaluation Cooperation (WPEC) of the OECD-NEA~\cite{WPEC_25}, the International Nuclear Data Committee (INDC) of the IAEA~\cite{IAEA_2010}, and Nichols et al.~\cite{DH_2023} have identified several fission products that urgently require TAGS measurement. Their recommendations included priorities for isotopes such as $^{88}$Rb and $^\mathrm{90m}$Rb, which are easily produced by BRIF and have high purity. In addition to their contribution to reactor decay heat, these two isotopes are also critical for understanding the reactor neutrino anomaly~\cite{sonzogni2015nuclear}.

	The experimental setup for the decay measurements is shown in Fig.~\ref{rbcs_setup}. An aluminum foil, \qty{18}{um} in thickness, was placed in the center of LAMBDA-I to block the radioactive nuclear beam produced by the BRIF-ISOL facility. A MSQ detector was placed after the foil to detect the electrons emitted during $\beta$ decay, covering a solid angle  of approximately 40\%. The vacuum chamber is constructed from Polyether Ether Ketone (PEEK), which also serves to block electrons from entering the LAMBDA-I detector and interfering with the $\gamma$-ray measurement. At the front of the LAMBDA-I detector, two lead (Pb) shields, each \qty{10}{mm} thick, were installed, along with a collimator (\qty{10}{mm} $\times$ \qty{20}{mm}) positioned at the detector entrance. Two plastic scintillators were used in anti-coincidence with the LAMBDA-I detector to reduce muon-induced background. The HPGe detector was used to verify the isotope type and purity. The CsI target and quarter plate were used to monitor and control the isotope beam. Data were recorded using Pixie-16 modules manufactured by XIA Inc.~\cite{pixiexia}.

	\begin{figure}[!htb]
		\centering
		\includegraphics[width=0.6\linewidth]{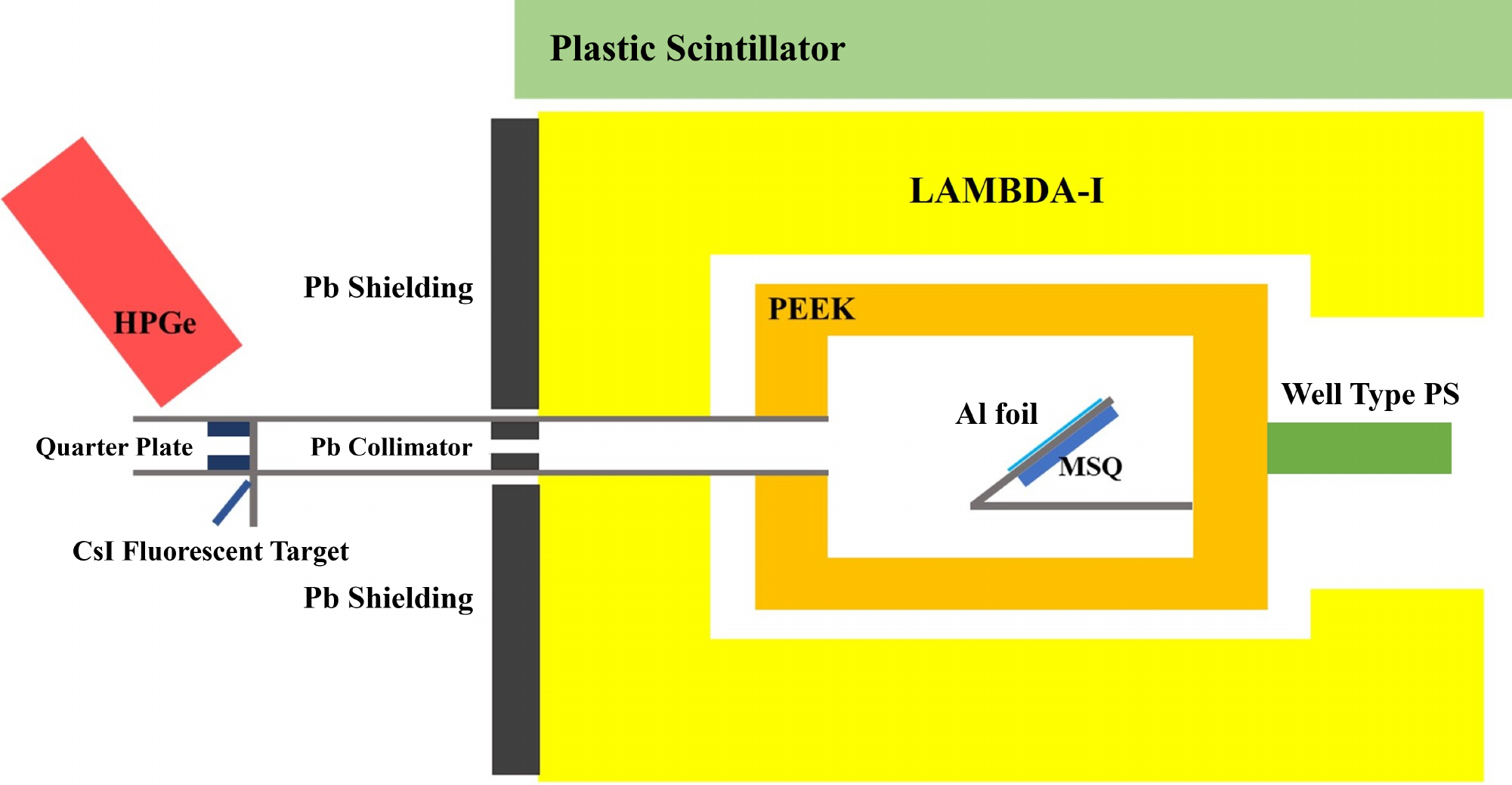}
		\caption{Experimental setup for decay measurements.}
		\label{rbcs_setup}
	\end{figure}

	The Rb isotopes were produced via proton-induced fission of a UC$_x$ target, which is the first time this is used at BRIF. In Fig.~\ref{rb88_purity_a} we show the $\gamma$ spectrum measured by the HPGe detector when the CsI fluorescent target was aligned with the beam line. After beam shutdown, the $\beta$ particles emitted by the remaining $^{88}$Rb nuclei on the aluminum foil were measured by the MSQ detector, as shown in Fig.~\ref{rb88_purity_b}. The half-life of $^{88}$Rb was determined to be \qty{17.93\pm0.15}{min}, which is consistent with the literature value. 

	\begin{figure}[!htb]
		\centering
		\begin{subfigure}[b]{0.41\linewidth}
			\centering
			\includegraphics[width=\linewidth]{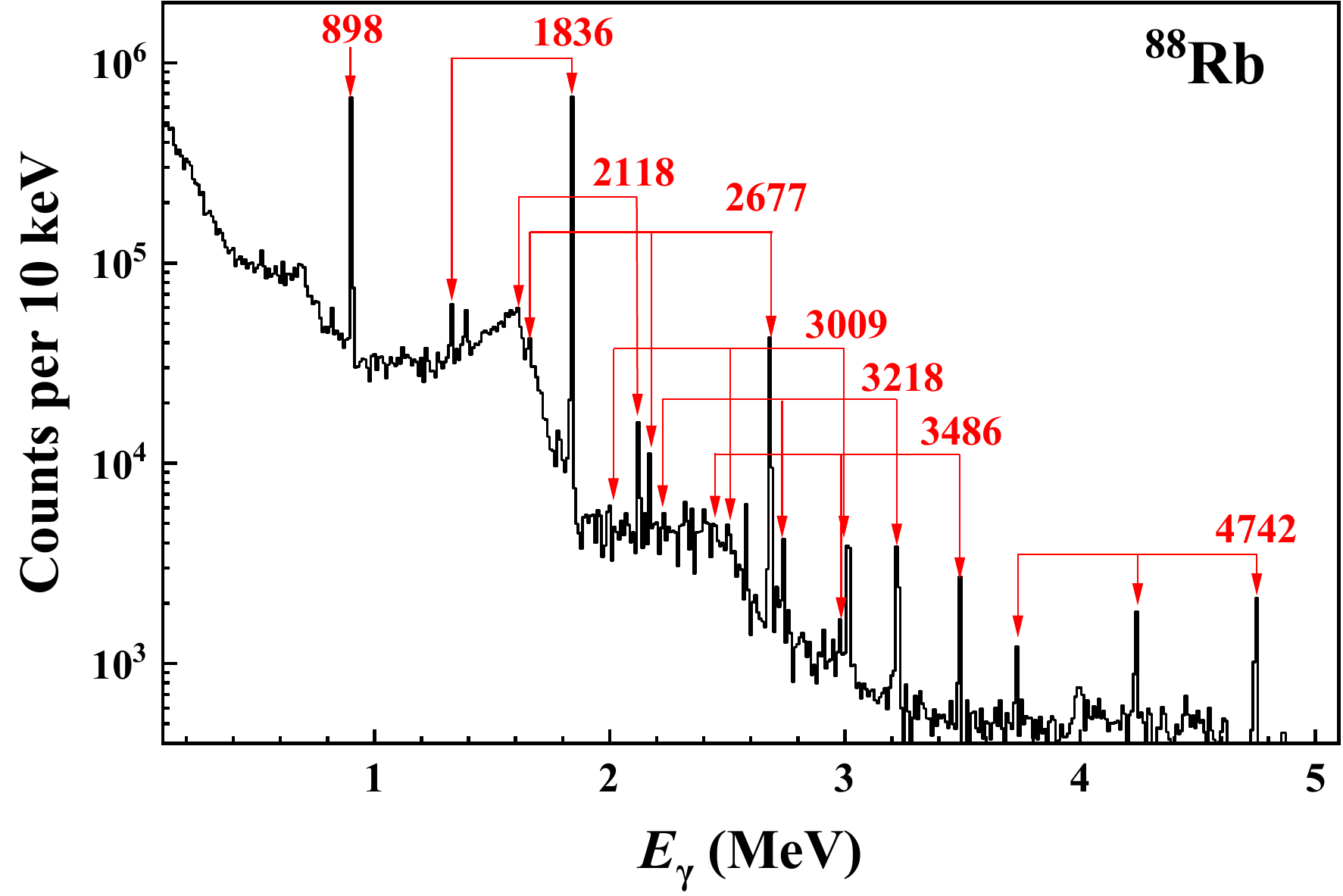}
			\caption{$\gamma$ spectrum measured by the HPGe detector.}
			\label{rb88_purity_a}
		\end{subfigure}
		\quad
		\begin{subfigure}[b]{0.45\linewidth}
			\centering
			\includegraphics[width=\linewidth]{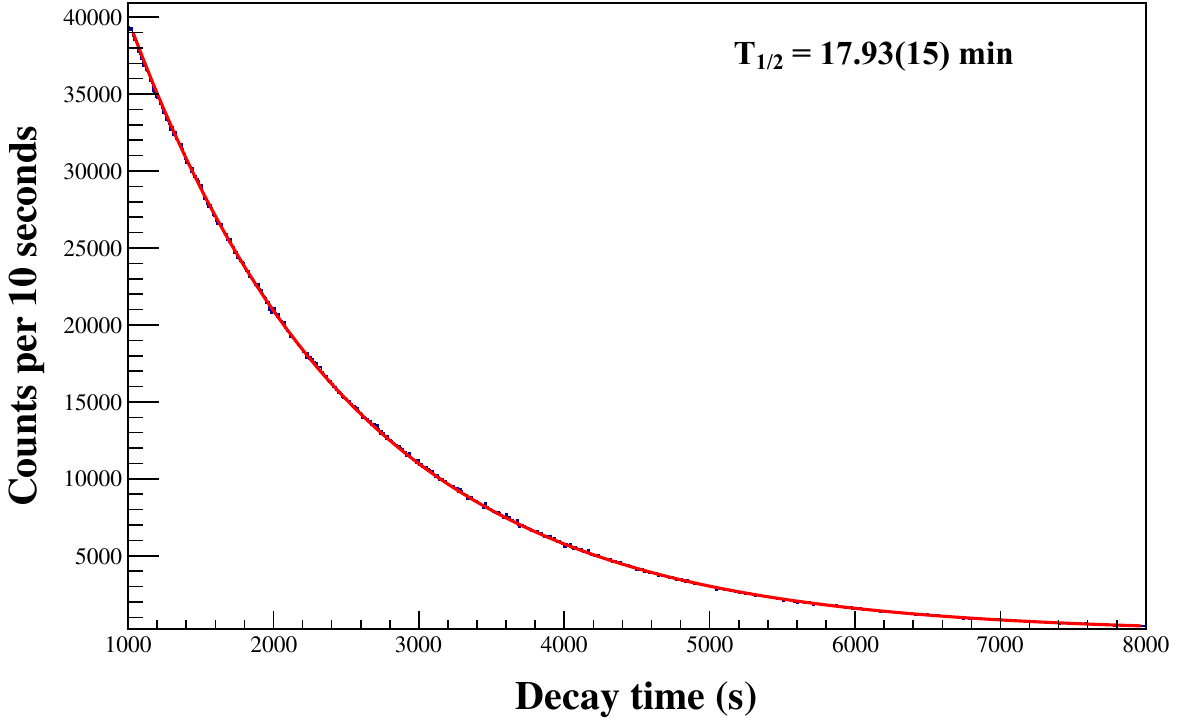}
			\caption{Count rate measured by the MSQ detector after beam shutdown.}
			\label{rb88_purity_b}
		\end{subfigure}
		\caption{Results of auxiliary detectors for $^{88}$Rb.}
		\label{rb88_purity}
	\end{figure}

	The data were analyzed using two channels fitted method introduced by Sheng et al.~\cite{sheng2024large}, simultaneously constraining the sum and single channel. The results shown in Fig.~\ref{rb88_result}, in which the black points represent the experimental data and each $\beta$ decay branch is represented by color shading (the red shading represents the ground state branch), can reproduce the experimental data well.
	
	It should be noted that the energy levels at \qty{4845}{keV} and \qty{4853}{keV} differ by a small amount, making them difficult to distinguish in the sum channel. However, the decay paths of these two levels are different, and this distinction can be observed in the single channel. In Fig.~\ref{rb88_response} we show the response function of \qty{4845}{keV} and \qty{4853}{keV} energy levels in both the sum and single channels.

	\begin{figure}[!htb]
		\centering
		\begin{subfigure}[b]{0.45\linewidth}
			\centering
			\includegraphics[width=\linewidth]{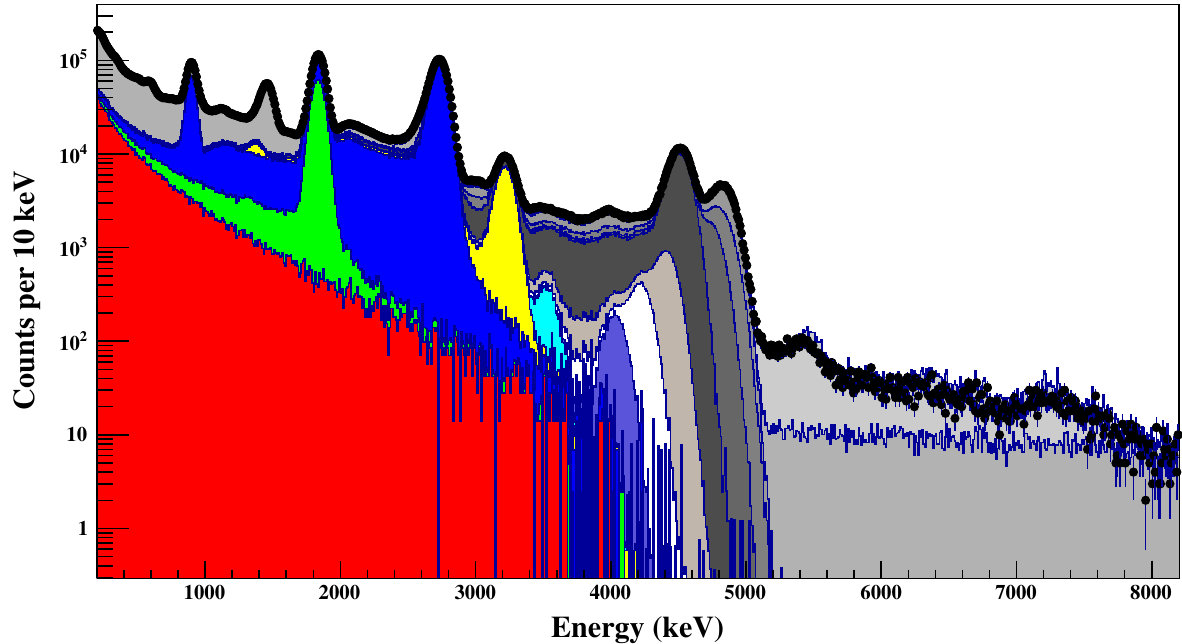}
			\caption{Fitted result of sum channel.}
			\label{rb88_result_a}
		\end{subfigure}
		\quad
		\begin{subfigure}[b]{0.45\linewidth}
			\centering
			\includegraphics[width=\linewidth]{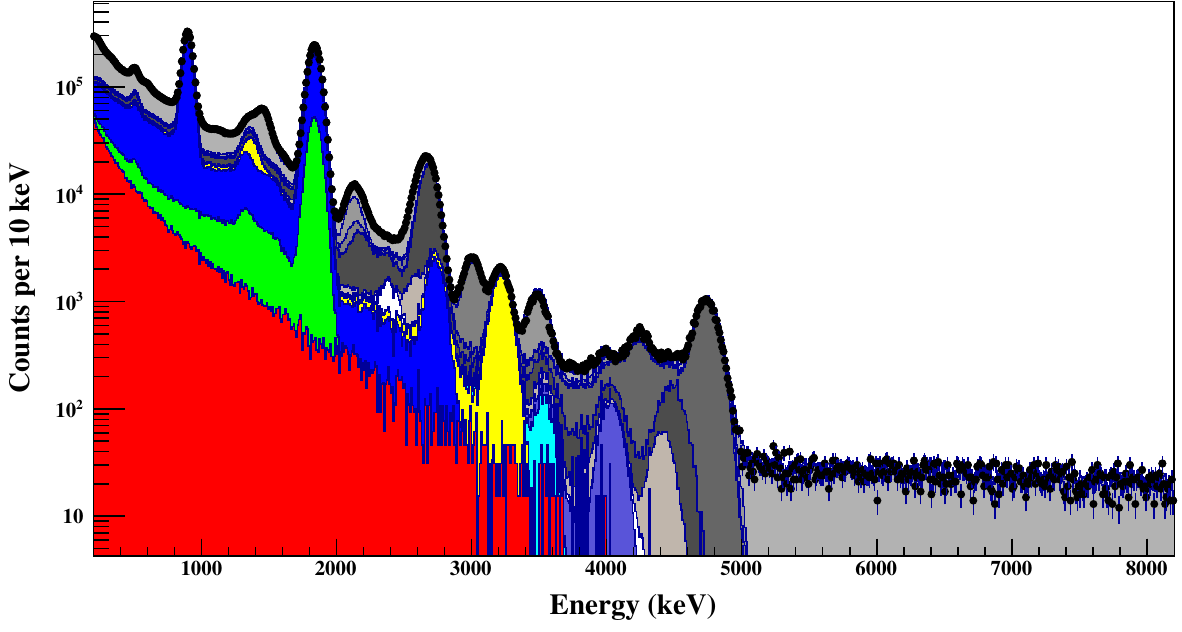}
			\caption{Fitted result of single channel.}
			\label{rb88_result_b}
		\end{subfigure}
		\caption{Fitted results of LAMBDA-I for $^{88}$Rb decay.}
		\label{rb88_result}
	\end{figure}

	\begin{figure}[!htb]
		\centering
		\begin{subfigure}[b]{0.45\linewidth}
			\centering
			\includegraphics[width=\linewidth]{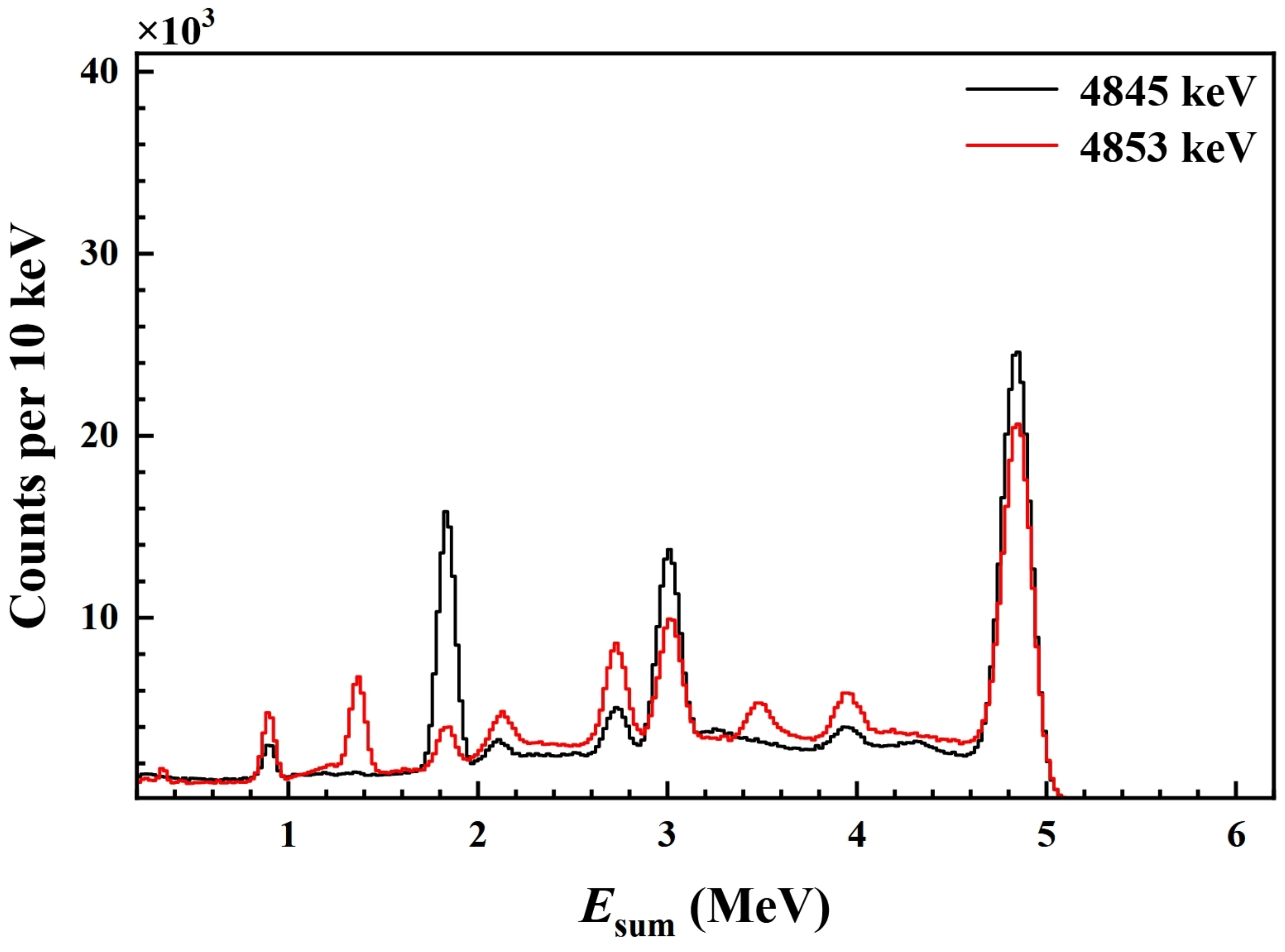}
			\caption{Simulated response function of sum channel.}
			\label{rb88_response_a}
		\end{subfigure}
		\quad
		\begin{subfigure}[b]{0.45\linewidth}
			\centering
			\includegraphics[width=\linewidth]{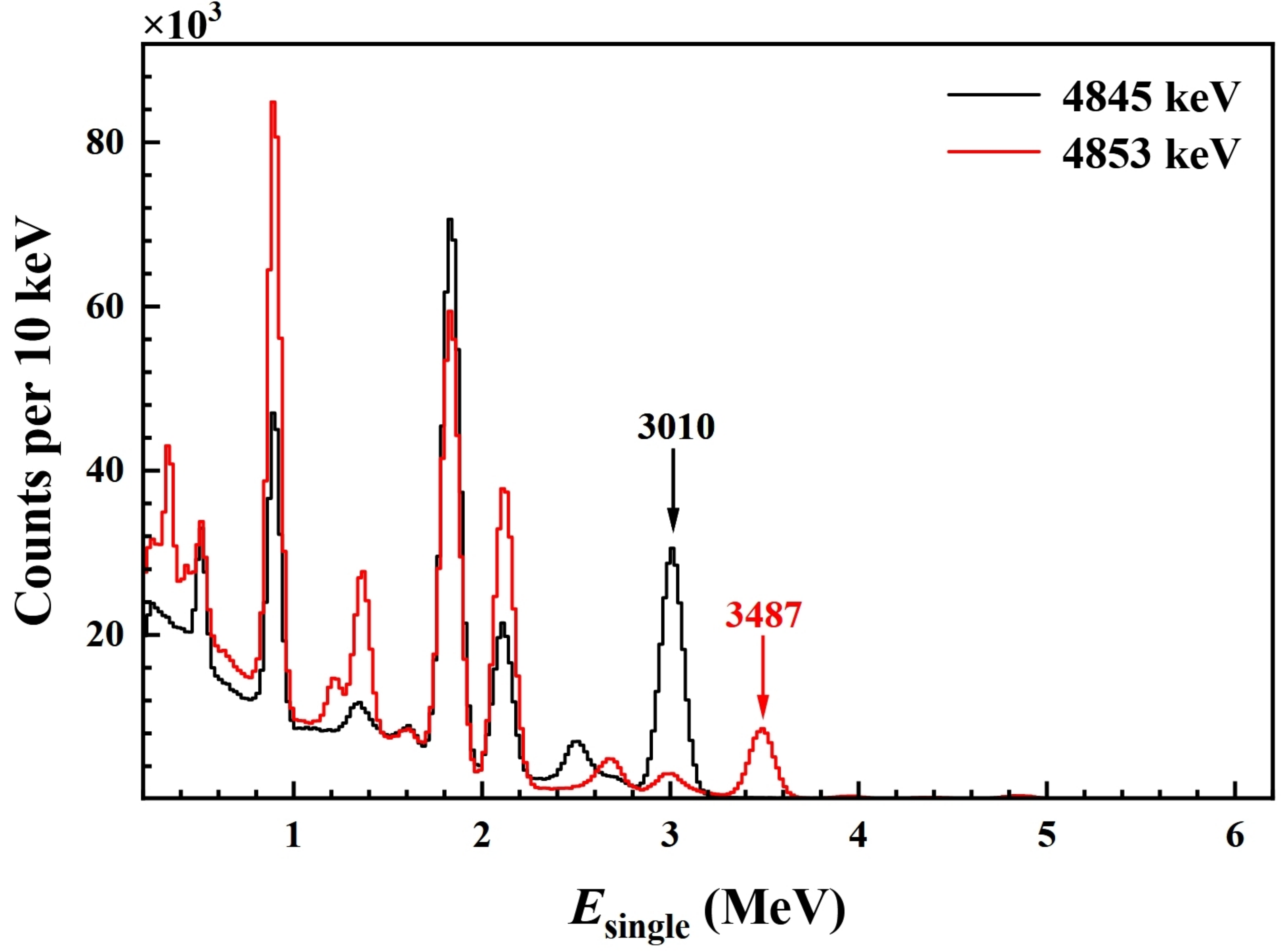}
			\caption{Simulated response function of single channel.}
			\label{rb88_response_b}
		\end{subfigure}
		\caption{Simulated response function of \qty{4845}{keV} and \qty{4853}{keV} energy levels in sum and single channel.}
		\label{rb88_response}
	\end{figure}

	In Fig.~\ref{fig:88RbResults} we compare the $\beta$ intensity distribution of $^{88}$Rb  obtained by LAMBDA-I with existing data~\cite{shuai2022determination}. Compared to ENSDF data~\cite{MCCUTCHAN2014135}, our results show larger uncertainties. This is because, although the TAS detector corrects systematic biases from the "Pandemonium effect", the precision of its branching ratios remains limited by poorer energy resolution and stronger non-linearity compared with HPGe, and uncertainties in Geant4-simluated response functions. Future improvements will focus on improving the resolution of BGO via cooling, enhancing the detection efficiency by increasing modularity, implementing energy calibration and validating response functions at high-energy region by $\gamma$-$\gamma$ coincidence. The $\overline{E}_\gamma$ value of $^{88}$Rb is \qty{658\pm65}{keV}, which is in agreement with the literature value \qty{677\pm15}{keV}~\cite{MCCUTCHAN2014135}. 

	\begin{figure}[!htb]
		\centering
		\includegraphics[width=0.7\linewidth]{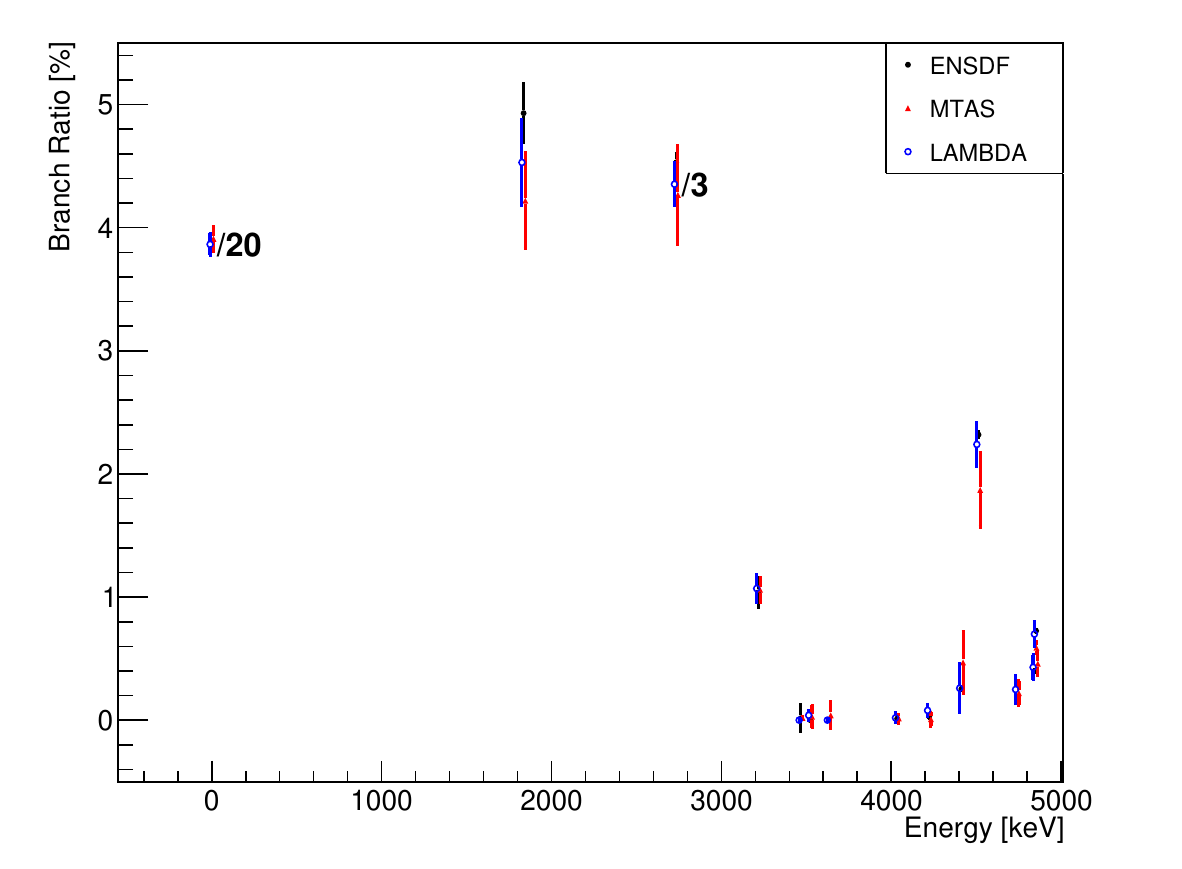}
		\caption{Comparison of $^{88}$Rb $\beta$ intensity distribution obtained by LAMBDA-I and existing data. The branching ratios of beta decay to the ground state and second excited state in $^{88}$Sr are divided by a factor of 20 and 3, respectively. The x-coordinates (energies) of MTAS and LAMBDA data points are shifted with $\pm$\qty{10}{keV}.}
		\label{fig:88RbResults}
	\end{figure}

    \newpage
	\section{Further development of BRIF }\label{fifth}
Although the present facility starts producing some unstable ion beams and contributing to the investigation
of nuclear structure physics, major improvements are desirable in the future.
	\subsection{mA-level extracted proton beam}\label{sub:5-1}

\subsubsection{Approach to increasing the beam intensity to the mA-level.}\label{sub:5-1-1}

To achieve a mA-level beam, several factors were thoroughly evaluated, including RF system beam loading, space charge effect limitations, and beam matching from the ion source to the central region. Improvements were also made to subsystems such as the ion source and buncher system.

The CYCIAE-100 employs a multi-cusp H- ion source capable of generating a maximum operational beam current of 10~mA~\cite{zhangtianjue2014rsi}. To enhance beam stability and quality, several modifications were made to the extraction system due to the frequent occurrence of sparks during testing. The extraction system of the ion source features an axially symmetric design with three electrodes: the plasma electrode, lens electrode, and ground electrode. The structure of the multi-cusp ion source and extraction electrode is shown in Fig.~\ref{fig:5.1.1}. A new x-y steering magnet was developed with a smaller height and additional turns to decrease the distance between the lens and the magnet. A protection cover was added to the steering magnet to prevent electron strikes, with 20 holes of 10~mm diameter drilled for gas flow. A smaller ground electrode was designed with a beam limit hole of 14~mm and a total size of 36~mm by 117~mm, with a 14~mm extraction hole. Six holes (12~mm by 5~mm) were drilled into the head of the ground electrode to reduce gas stripping and maintain optimal gas pressure. The distance between the lenses was also optimized. These improvements enabled an increase in extraction energy from 35~keV to 40~keV, achieving stable operation without sparks during a week-long stability test. 

The DC beam produced by the ion source is bunched using a first harmonic non-intercepting 2-gap buncher located about 1.1~m away from the inflector, between the first solenoid and the triplet. After increasing the injection energy to 40~keV, the distance between the two buncher gaps was adjusted to match $\beta\lambda$/2, and RF power calibration was also conducted to optimized efficiency.

\begin{figure}[!htbp]
    \centering
    \includegraphics[width=0.8\linewidth]{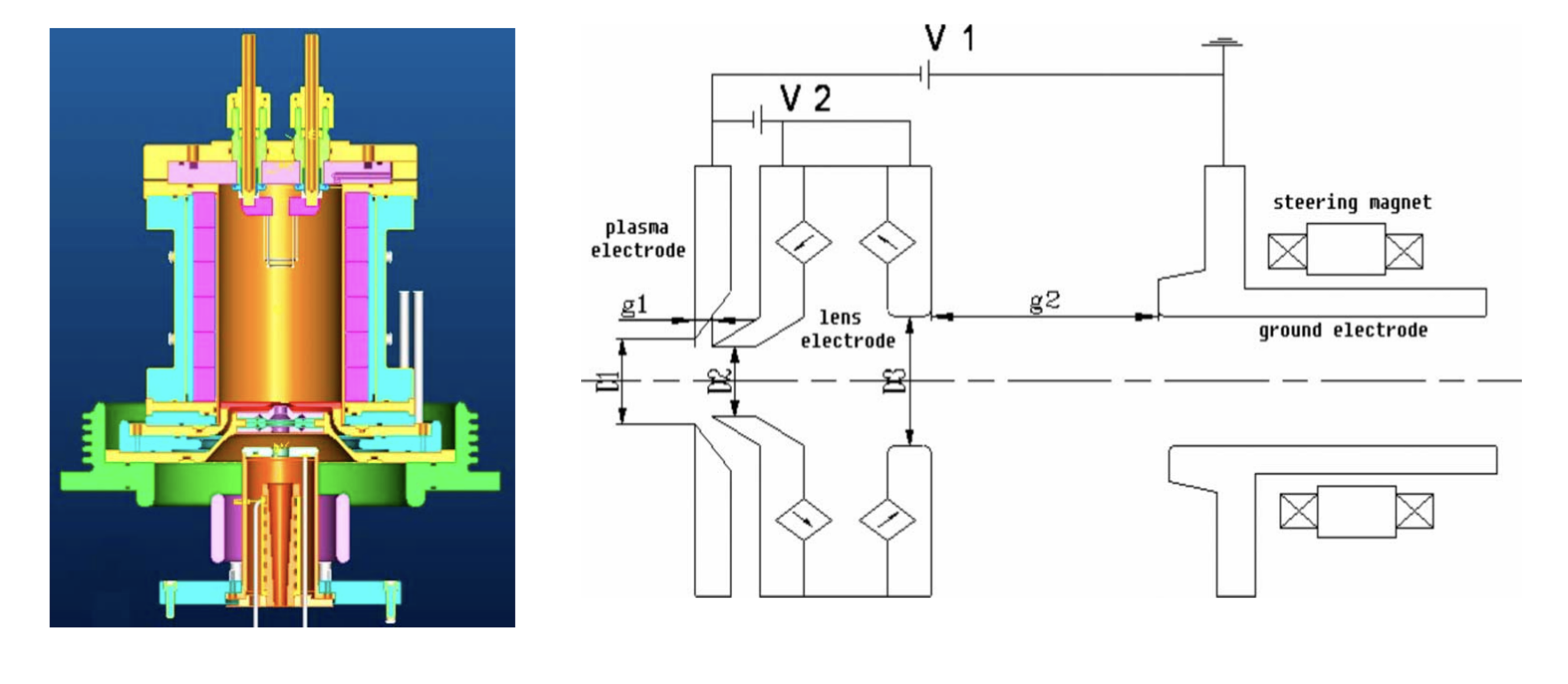}
    \caption{The structure of CYCOAE-100 ion source and extraction system.}
    \label{fig:5.1.1}
\end{figure}

A collimator before the buncher limits the beam radius, while the effective buncher radius is 10~mm, made from gold-plated tungsten wires. The RF system for the buncher includes an LC matching circuit, a 600~W solid-state RF amplifier, and an amplitude/phase control unit. The structure of the buncher and the block diagram of the system are shown in Fig.~\ref{fig:5.1.2}.

\begin{figure}[!htbp]
    \centering
    \includegraphics[width=0.8\linewidth]{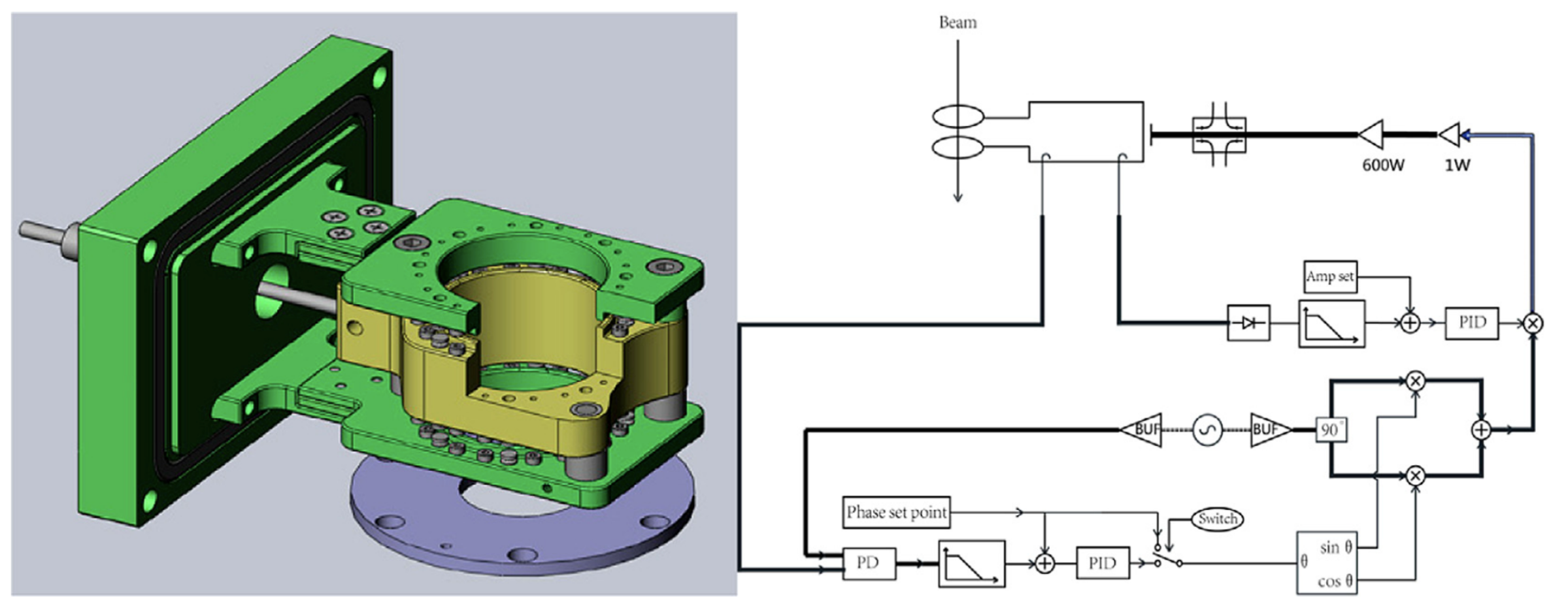}
    \caption{The buncher CST model and its RF system diagram.}
    \label{fig:5.1.2}
\end{figure}

Prior to beam commissioning, experiments were conducted to enhance the power supervising system, measure the buncher's shunt impedance, and fine-tune the LC matching circuit. An EPICS-based power monitor system was developed to oversee the RF power flow to the buncher, consisting of a digital control board based on an ARM Cortex-A7 CPU running Linux, and an RF analog board that demodulates forward and reflected power using RF power detectors. The shunt impedance of the buncher was measured using an Agilent E5070B vector network analyzer, which measures the S21 parameter with a 50~$\Omega$ loaded probe connected to the buncher's grid as port 2, while port 1 served as the conventional coupling port. The shunt impedance is calculated based on these measurements. The LC matching circuit features a coarse capacitor that experiences significant temperature drift, necessitating manual online adjustments. To protect the 600~W RF amplifier, a pulse signal with a duty factor of 1/10 is used to drive the buncher system. The capacitor is fixed when the amplitude of the pickup signal reaches its maximum, after which the system switches to automatic mode.

The RF system has been improved in two aspects before the mA level beam commissioning, including LLRFs and cavities. The mA-level beam current is a heavy load for the RF system and may cause the Dee voltage regulation to enter an open-loop condition. After the amplitude loop is closed, the amplitude of the RF drive signal is determined by the DSP and DDS of the LLRF system. In other words, to achieve precise amplitude control, the LLRF needs to adopt an adaptive strategy to ensure that the amplitude control loop remains closed unless the power requirement exceeds 120$\%$ of the nominal value. As for the RF cavities, the fine tuner was changed to a smaller disk before the beam commissioning in order to achieve better tuning performance. This effort was evident during commissioning [20]. For low currents, approximately $\pm$1~kW of power modulation stabilizes the Dee voltage, with minimal phase variation from amplifier crosstalk. However, the non-orthogonality of the amplification chain could cause a Dee phase shift of up to 30$^\circ$ by a 500~$\mu$A beam, with $\delta\phi$ potentially reaching 10$^\circ$, leading to instantaneous effects from intensity variations. Careful phase control was implemented for stable operation, with manual adjustments aiding high-intensity beam tuning. After continuous efforts, the residual tuning error of both cavities in the CYCIAE-100 cyclotron was reduced to below 3$^\circ$.

During the high-power beam commissioning, most RF optimizations aimed at mitigating the impact of 50~kW beam loading on the cyclotron's Dee circuit were completed. The final stage of the RF amplifier can deliver up to 2$\times$100~kW, requiring about 35~kW of RF power for low-intensity acceleration and increasing to 65~kW for beams over 0.5~mA. Higher anode voltages were applied to enhance amplification efficiency, with adjustments made to the output capacitance to handle load variations. These modifications were verified with dummy loads prior to commissioning. 

Another preparatory work before high-power beam commissioning included improving the cooling of the stripping foil, enhancing the beam dump cooling for 50~kW and 100~kW power, and adding water cooling for the radial probe. The main cavity's vacuum was also improved, achieving a static vacuum of 6.4$\times$10$^{-8}$~mbar before the mA-level beam commissioning.

\subsubsection{Commissioning of mA-level beam on the internal target}\label{sub:5-1-2}

After making improvements in the aforementioned aspects, we have carried out beam commissioning work for mA level proton beam. A removable internal target positioned at the center of the magnet valley is adopted for beam diagnosis. It is designed to withstand a minimum of 1~MeV/1~mA. Simulation results indicate that with a water flow of 10~L/min and a beam power of 1.5~kW, the maximum temperature in the center of the irradiation region reaches 360~K, while the cooling wall reaches 347~K. To mitigate the impact of secondary electron emission on internal target measurements, a wing with two thin copper pieces was added to the internal target. 

The proton beam was commissioned on the internal target for different currents to evaluate the bunching factor. According to the beam commissioning results, for a 3~mA beam current extracted from the ion source, the bunching factor can reach a factor of 2, while remaining at 1.66 for a nearly 9~mA beam current extracted from the ion source. It is evident that the bunching factor decreases as the extraction current from the ion source increases~\cite{zhangtianjue2017nimb}. 

Fig.~\ref{fig:5.1.3} demonstrates the measured mA-level beam current on the internal target at about 1~MeV. The beam current was maintained stably at 1.1~mA in July 2016. In Tab.~\ref{tab:5.1.1} we list the bunching factors and acceleration efficiencies of the H$^{-}$ ion source at different current levels. As shown in Tab.~\ref{tab:5.1.1}, the measured beam is increased about 6.3$\%$ on the internal target after the elimination of the secondary electron emission.

\begin{figure}[!htbp]
    \centering
    \includegraphics[width=0.8\linewidth]{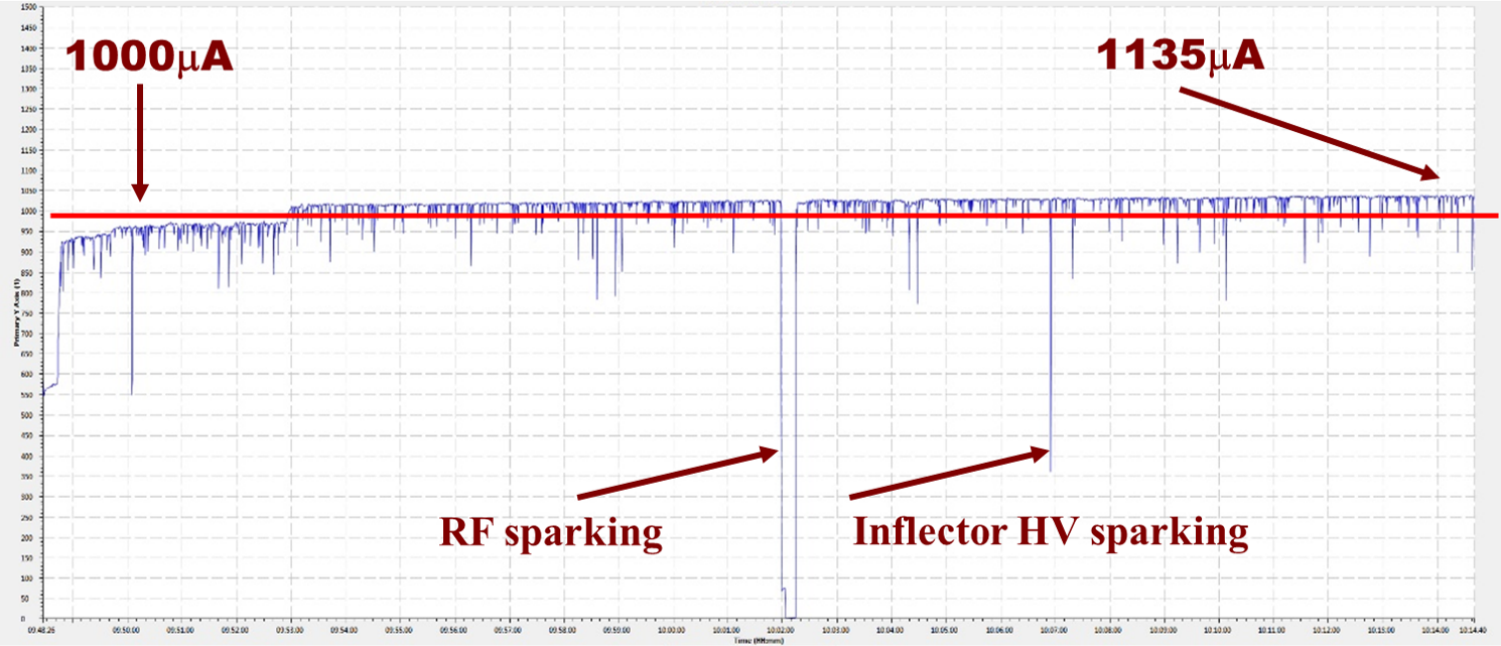}
    \caption{The mA proton beam history on the internal target at 1~MeV.}
    \label{fig:5.1.3}
\end{figure}

\begin{table}[!htbp]
     \caption{The bunching factor and acceleration efficiency with different beam current from the ion source.} 
     \centering
    \begin{tabular}{ccccc}
    \hline
         Ion source (mA) & Without buncher & With buncher & Bunching factor & Acceleration efficiency ($\%$) \\
     \hline
1.33& 100& 201&	2.01&	15.1\\
1.91&	145&	310&	2.14&	16.2\\
3.25&	201&	399& 1.99&	12.3\\
4.27&	258&	490&	1.90&	11.5\\
8.69&	610&	950&	1.56&	10.9\\
8.67&	608&	1010&	1.66&	11.6\textsuperscript{*}\\
\hline
    \end{tabular}
\begin{tablenotes}
\small
\item[*]\textsuperscript{*}The measured beam on the removable internal target which has a wing with two thin copper pieces to eliminate the secondary electron emission.
\end{tablenotes}
    \label{tab:5.1.1}
\end{table}

Collimators at the ion source exit and reflector entrance helped limit beam emittance, reduce bombardment on the reflector, and minimize sparking, contributing to stable operation. The overall efficiency, including acceleration from 1~MeV to 100~MeV, stripping extraction, and beam transmission, is approximately 99.07$\%$ from the project acceptance testing.

\subsubsection{Preparation of mA-level beam commissioning on the external target}\label{sub:5-1-3}

Before beam commissioning on the external target, a higher-power beam dump and a new central region for mA-level beam intensity were designed and fabricated. 

For high-power beam dump, it is necessary to optimize the angle between the target surface and the beam as much as possible to increase the beam absorption area and reduce energy deposition per unit area. The 100~kW high-power beam dump uses a V-shaped flat target structure instead of the commonly used conical structure~\cite{liujingyuan2022}. It consists of the target and fixed components, vacuum components, signal measurement, motion devices, and support structures. The V-shaped structure is made by welding two copper thick plates, with a cooling water channel machined inside. As shown in Fig.~\ref{fig:5.1.4}, the beam dump is movable, which is also a technical challenge for such a 100~kW high power. Due to this mobile design, the beam line where the beam dump is located can also be used downstream as a white light neutron source. The high-power beam dump has now completed fabrication and testing, with all performance meeting design requirements.

\begin{figure}
    \centering
    \includegraphics[width=0.8\linewidth]{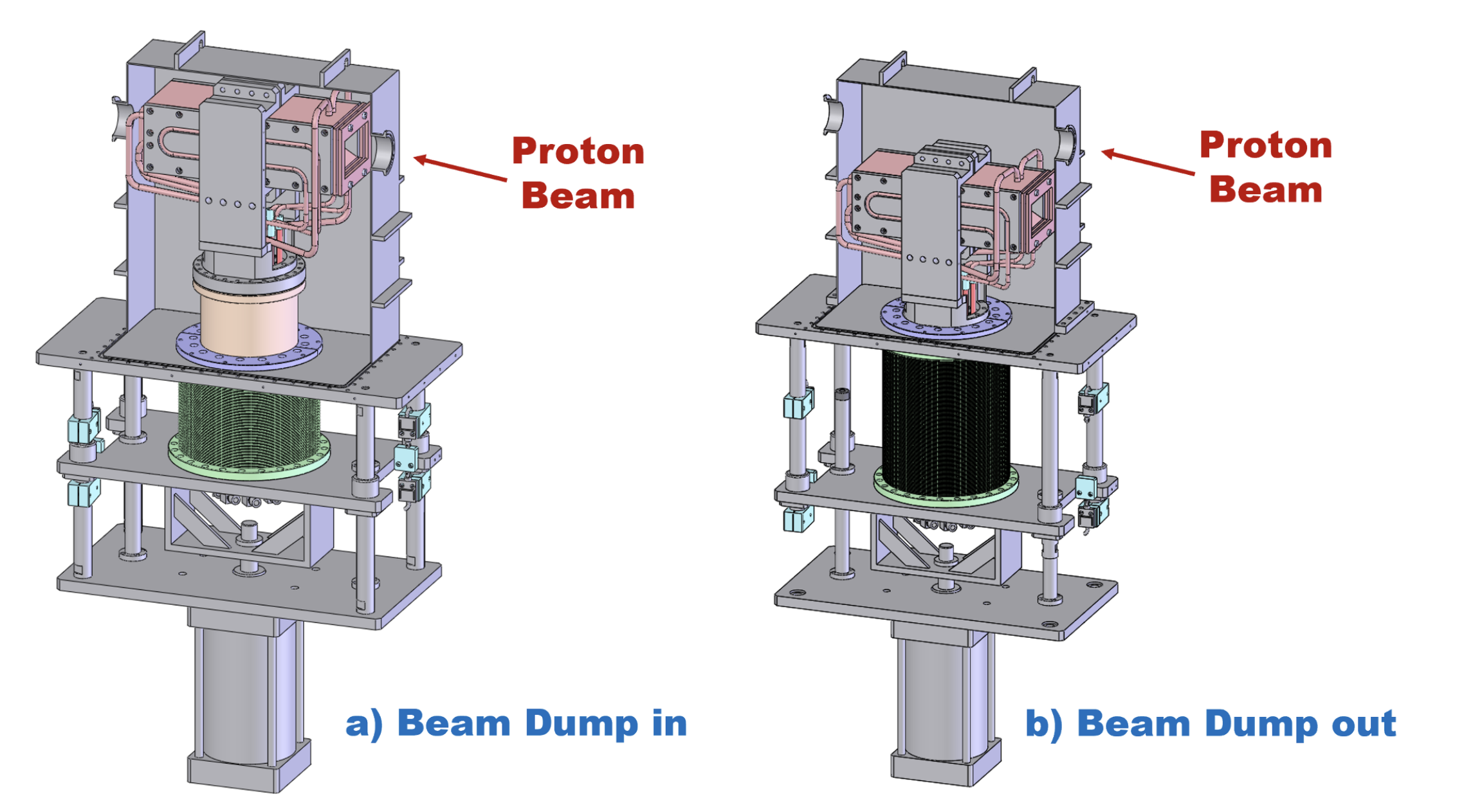}
    \caption{The 100-kW high power beam dump.}
    \label{fig:5.1.4}
\end{figure}

For the new central region, we have strengthened the thermal conduction structure between the inflector with high voltage and the grounding copper base. And a water-cooled structure was designed in the central region to avoid overheating and thermal deformation, as it needs to sustain higher beam intensity loosed in the inflector, central region and Dee tips. For example, the lower flange of the central region has been equipped with a water-cooling structure, allowing for indirect cooling of the spiral inflector. The cross section of the support column connecting the two inflector plates has also been increased, improving the heat conductivity efficiency. Another important improvement is that all the support columns of the inflector are covered with ceramic, reducing secondary electron emission. The design and machining of the new central region have been completed, the structure is as shown in Fig.~\ref{fig:5.1.5}. The installation in the cyclotron and commissioning were completed in early October 2024.

\begin{figure}[!htbp]
    \centering
    \includegraphics[width=0.8\linewidth]{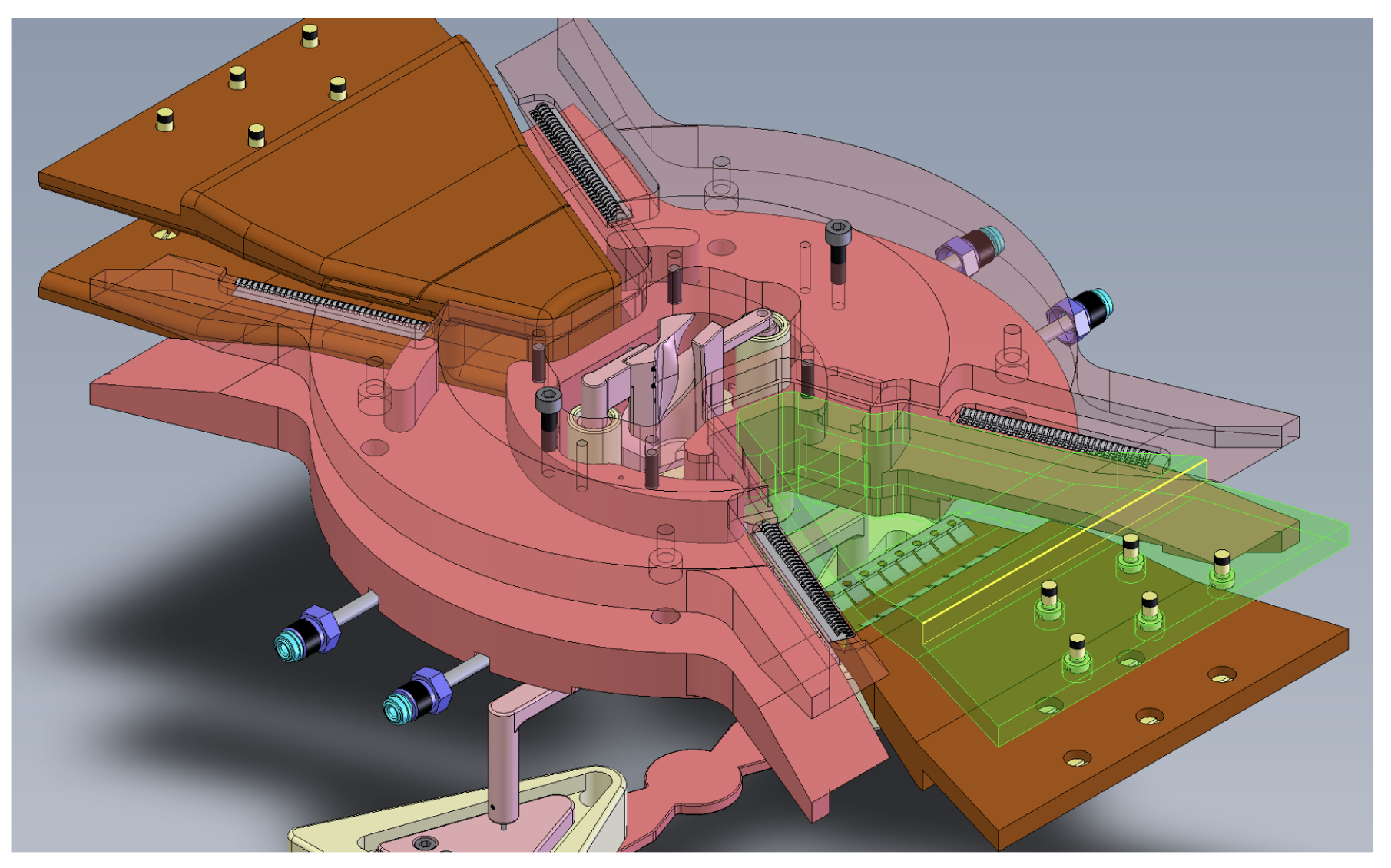}
    \caption{New structure of the central region.}
    \label{fig:5.1.5}
\end{figure}

As mentioned above, a new central region and high-power beam dump have recently been installed for the commissioning of extracted mA-level proton beams. Meanwhile, we have also completed the improvements to the cooling water system for the cyclotron, and the buncher structure and 2$\times$100~kW RF amplifier. We are now almost at the stage of mA beam commissioning on the external target. At present, this commissioning is waiting for the permission from the administrative department. Once approved, the mA-level beam commissioning work can be carried out soon.

	\subsection{Target ion source development}\label{sub:5-2}
The BRIF has delivered some RIBs such as Na, K, Rb, and Al for experiments in past years, however, the RIBs are still not able to fulfill the needs of experiments. The improvement will be made to deliver higher intensity RIBs. Such measures are under developing or consideration.

The target has been a major issue for generating RIBs at BRIF, because light element target such as SiC can only generate the adjacent nuclide of Si by proton induced nuclear reaction.  As has been used in other ISOL facility, UC$_x$ target can be used to produce a lot of nuclides by proton induced fission. The target has been manufactured in laboratory. The process to produce UC$_x$ target is as follow: 71.2$\%$ UO, 28.2$\%$ graphite and 2$\%$ sugar are mixed and grinded into fine and homogeneous powder, then the powder is pressed into pellets under 600~MPa for one hour. The pellets are put into evacuated oven to sinter, the sintering procedure is very important to obtain good micro-structure target. Then pellets is heated to 300~ $^\text{o}$C for 8~h to allow the sugar to decomposed and micro-hole to be formed. Then the pellets temperature will be hold at 1600~ $^\text{o}$C for 24 hours so that UO can react with graphite fully to form UC$_{x}$. The examination shows the UC$_{x}$ composed of UC$_{2}$ and UC mainly. The grain size of the is about 1~$\mu$m, the porosity is 51.3$\%$. 
The porous UC$_{x}$ target produced and its SEM photo are shown in Fig~\ref{fig:5-2-1}.  The micro-structure is similar to the UC$_{x}$ targets fabricated in the Istituto Nazionale di Fisica Nucleare (INFN) with uranium dioxide, graphite and phenolic resin, its density of the targets is approximately 3.9~g/cm$^{3}$, with a porosity of approximately $56\%$. The grain size of UC$_{x}$ targets in INFN is approximately 10~mm, while targets in BRIF is around 1~mm.

\begin{figure}[!htbp]
    \centering
    \includegraphics[width=0.8\linewidth]{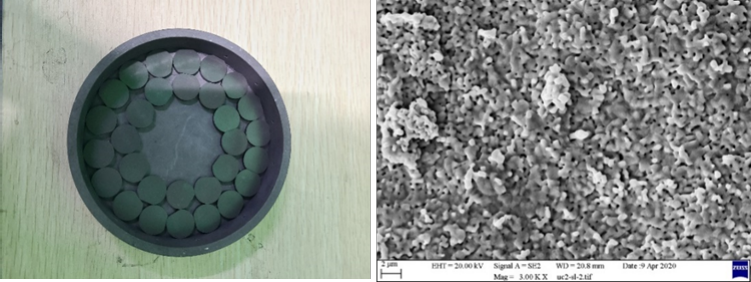}
    \caption{Left: the  UC$\_{x}$ target. Right: The SEM photo of UC$\_{x}$ target.}
    \label{fig:5-2-1}
\end{figure}

Primary test has been performed online on BRIF; Several RIBs have been generated. The Rb isotopes generated on line with UC$_{x}$ target are shown in Fig~\ref{fig:5-2-2}. 

So far only a few micro-amperes proton beam has been incident on UC$_{x}$ target, further development is scheduled to increase the proton beam current. 
The proton power is expected to be increased to 10~kW from currently 2~kW by improving the target radiation structure.

\begin{figure}[!htbp]
    \centering
    \includegraphics[width=0.6\linewidth]{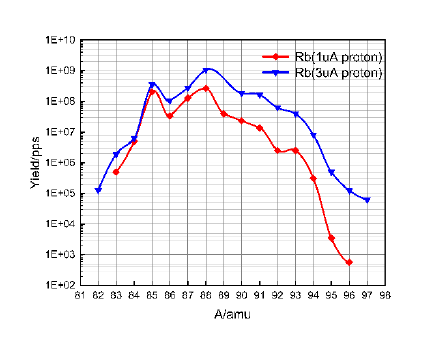}
    \caption{The yields of Rb isotopes with UC$_{x}$ target.}
    \label{fig:5-2-2}
\end{figure}

There are two type of ion sources can be chosen at BRIF, one is surface ion source for low ionization potential elements, the other is electron impact ion source for gaseous and high ionization potential elements. For high ionization efficiency and good selectivity, a laser ion source has been developed off-line. The ion source is based on a surface ion source with three all-solid state tunable Ti:Sapphire lasers. Each Ti: Sapphire laser is individually pumped by a Nd:YAG laser with the maximum power of 20~W. Sn$^+$ ion beam has been produced by the three-step resonance ionization scheme of 286.415, 811.63 and 823.704~nm. Fig~\ref{fig:5-2-3} shows the extracted mass spectrum of the ion beam, it is clear that when the lasers are turned on, the Sn$^{+}$ peak appears. The laser ion source is expected to be installed on BRIF soon. More RIBs will be available and unwanted nuclides can be better contained.

\begin{figure}[!htb]
    \centering
    \includegraphics[width=0.8\linewidth]{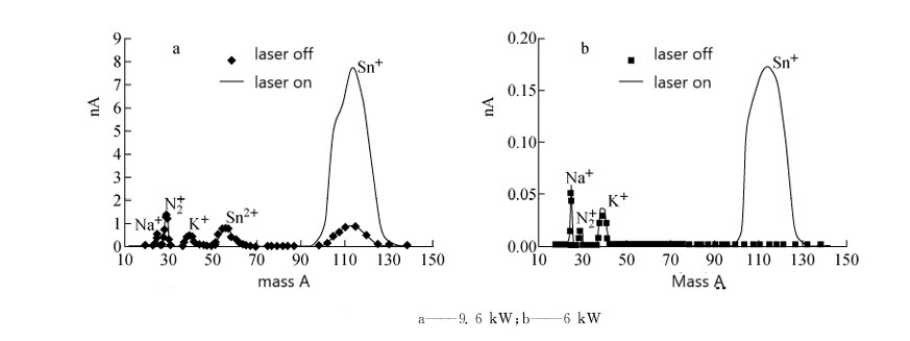}
    \caption{The mass spectrum of extracted ion beam from laser ion source at 
different target power.}
    \label{fig:5-2-3}
\end{figure}

	\subsection{New post-accelerator}\label{sub:5-3}
	
 The tandem accelerator requires the injection of negative ions, in comparison to positive ions, typically beam current is over two orders of magnitude lower. Additionally, the beam is stripped at the high voltage terminal of the tandem accelerator, which further reduces the beam current by approximately one order of magnitude. This significantly decreases the efficiency of BRIF facilities. To address this issue, BRIF plans to add an injector based on a linear accelerator to enhance the beam intensity of radioactive nuclear beams and some stable beams, operating in parallel with the tandem accelerator and complementing each other. The layout of this injector can be seen in Fig.~\ref{fig:5.3.1}.
\begin{figure}[!htb]
    \centering
    \includegraphics[width=0.5\linewidth]{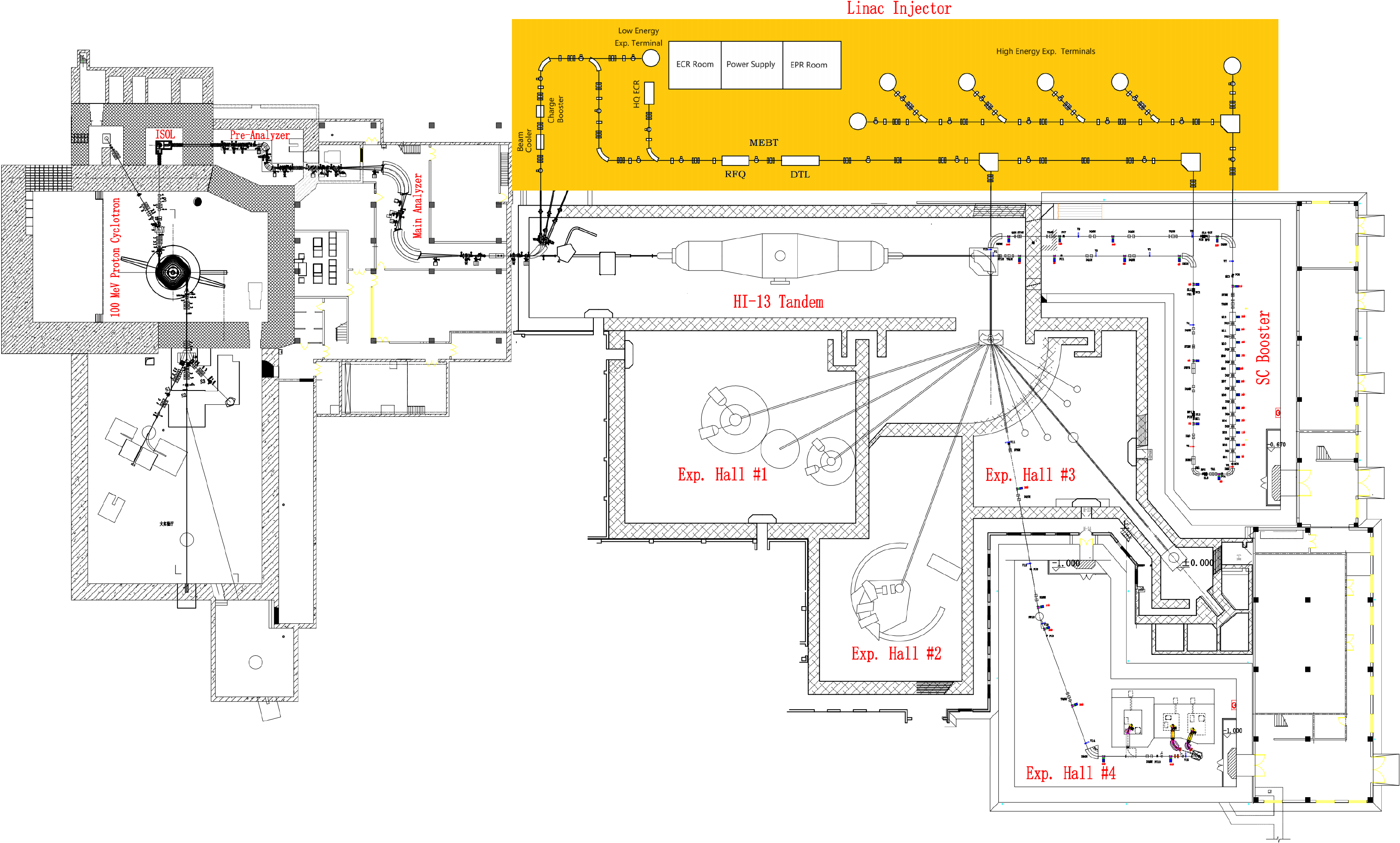}
    \caption{Layout of new injector(yellow area) composed of a RFQ and a DTL accelerator.}
    \label{fig:5.3.1}
\end{figure}
 
A new low-energy beam transfer line and a linear injector are being constructed next to the HI-13 tandem accelerator. The transfer line includes a 30~keV charge booster  to increase the charge state of radioactive nuclides,  enabling efficient acceleration for radio-frequency electric field. The linear injector consists of several components: an RFQ acceleration section, a medium energy beam transport (MEBT) section, a drift tube linac (DTL) acceleration section, and a bending section.
The RFQ acceleration section adopts a four-vane structure with an operating frequency of 75.2~MHz. It has an injection energy of 3~keV/u and an exit energy of 300~keV/u, with a maximum mass-to-charge ratio of 7. The total length is around 4 meters, and the total power loss is less than 40~kW.

\begin{figure}[!htbp]
    \centering
    \includegraphics[width=0.5\linewidth]{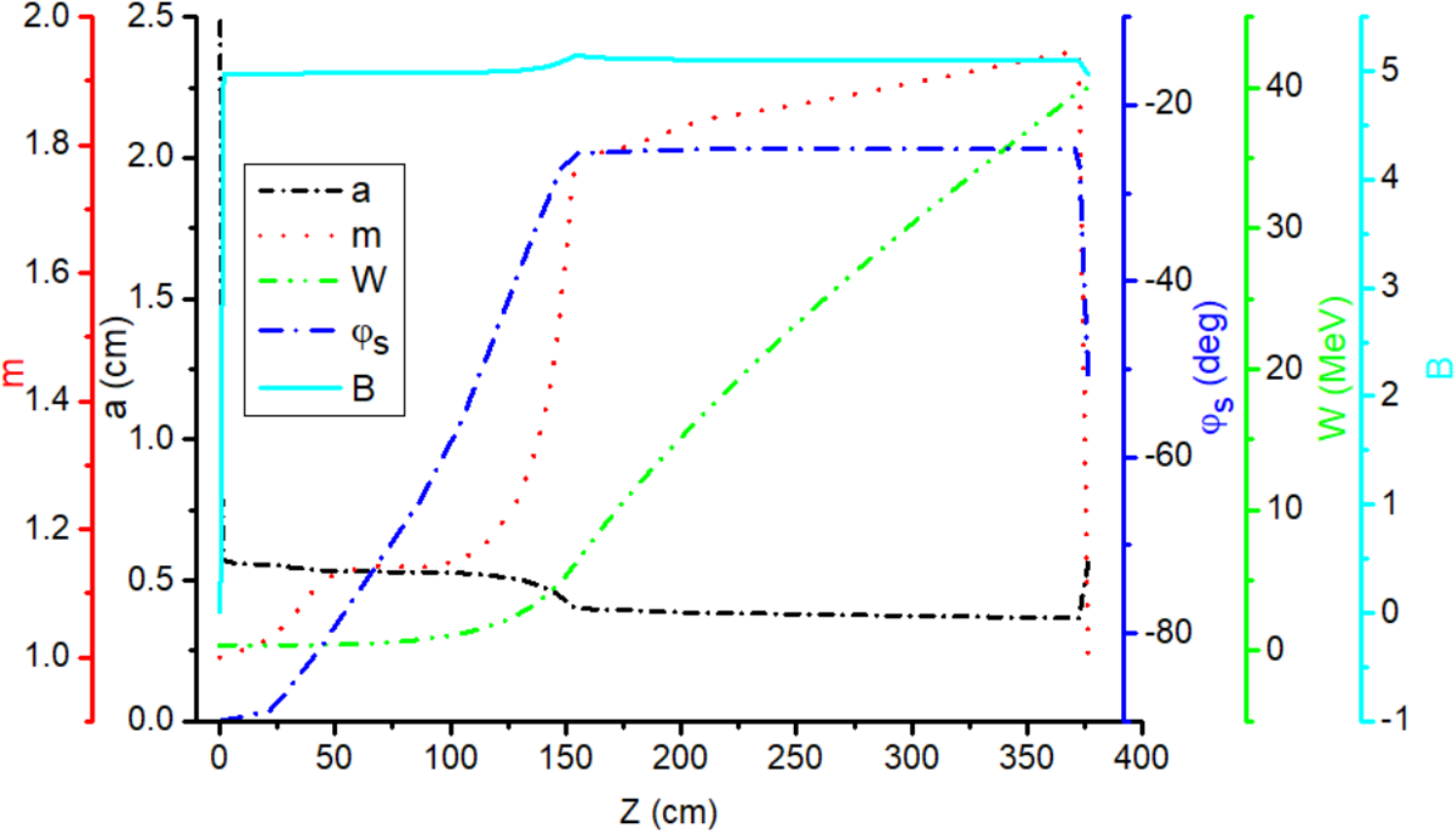}
    \caption{The key issues of RFQ modified along RFQ electrode's length with reference particle $^{132}Sn^{22+}$.}
    \label{fig:5.3.2}
\end{figure}

The MEBT section uses quadrupole magnets and room temperature resonators to match the beam from the RFQ to the DTL. The layout of the MEBT is shown in Fig.~\ref{fig:5.3.3}; the buncher operates at a frequency of 150.4~MHz, with a maximum effective voltage of 60~kV, and employs a quarter-wave double-gap structure with $\beta$ = 0.022.
\begin{figure}[!htbp]
    \centering
    \includegraphics[width=0.5\linewidth]{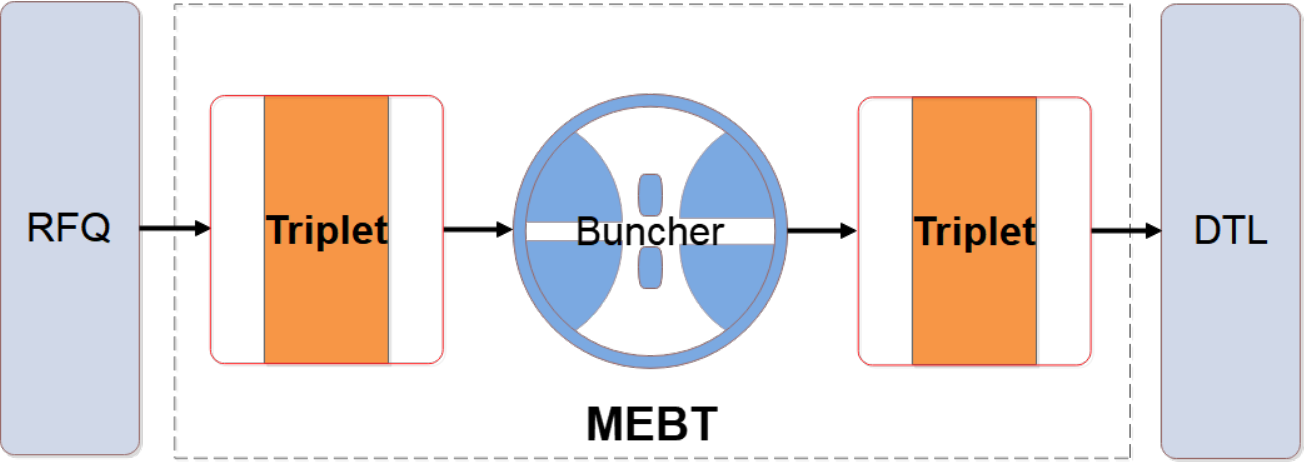}
    \caption{Schematic Diagram of MEBT.}
    \label{fig:5.3.3}
\end{figure}

The DTL acceleration uses an IH structure with an operating frequency of 150.4~MHz, boosting the beam energy from 300~keV/u to 1.5~MeV/u. It utilizes a KONUS dynamic structure with three cavities and four KONUS periods, as depicted in Fig.~\ref{fig:5.3.4}, with main parameters detailed in Tab.~\ref{tab:5.3.1}.
\begin{figure}[!htbp]
    \centering
    \includegraphics[width=0.5\linewidth]{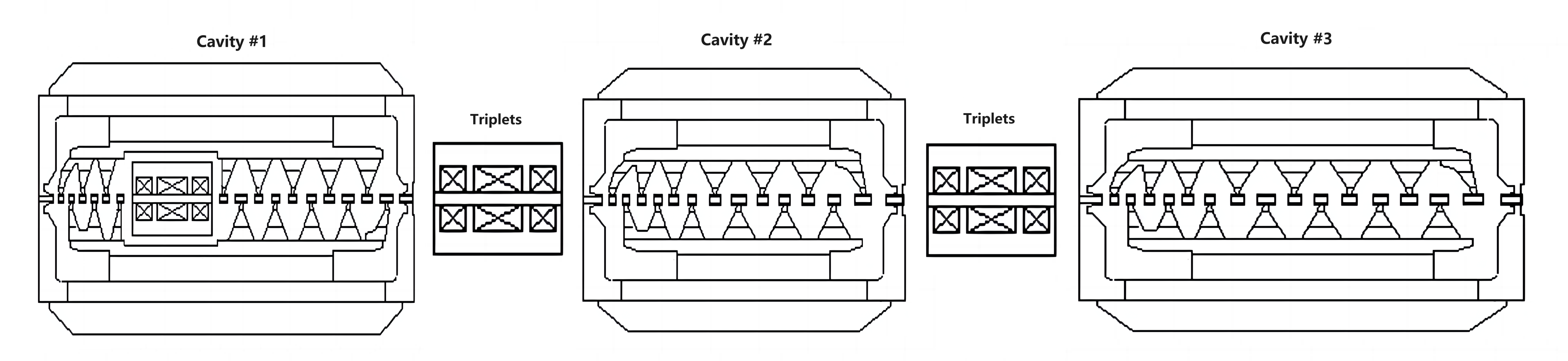}
    \caption{Structural Diagram of IH-DTL cavities.}
    \label{fig:5.3.4}
\end{figure}

\begin{table}[!htbp]
     \caption{The key parameters of IH-DTL Cavities.} 
     \centering
    \begin{tabular}{cccc}
    \hline
         &  Cavity \#1& Cavity \#2 & Cavity \#3\\
         \hline
      Injection Energy (MeV/u)   & 0.3 & 0.74 & 1.19\\
        Exit Energy (MeV/u) & 0.74 & 1.19 & 1.51\\
       Bunching Gaps  & 7 & 4 & 4\\
        Accelerating Gaps & 11 & 10 & 12\\
        Length of electrodes (m) & 1.63 & 1.55 & 1.94\\
       Power Consumption (kW)  & 61 & 55 & 68\\
       \hline
    \end{tabular}

    \label{tab:5.3.1}
\end{table}

The bending section employs an achromatic structure to deflect the beam into the original tandem beamlines or the superconducting booster for further acceleration.

To further increase facility utilization, it is proposed to add a high charge state ECR ion source before the RFQ, enhancing the variety and intensity of beams available at the experimental terminal. The intensity of stable nuclide beams at the experimental terminal is expected to increase by more than one order of magnitude. The initially accelerated beam from the injector can also be injected into the existing beam line behind the tandem accelerator via a double 45$^\circ$ magnet achromatic system for medium-energy nuclear physics experiments at the experimental terminal.

A new beam line is being built in the main analysis section of the original BRISOL, using two small bending magnets to reduce beam loss instead of the high-resolution main analyzing magnets. 
\begin{figure}[!htbp]
    \centering
    \includegraphics[width=0.5\linewidth]{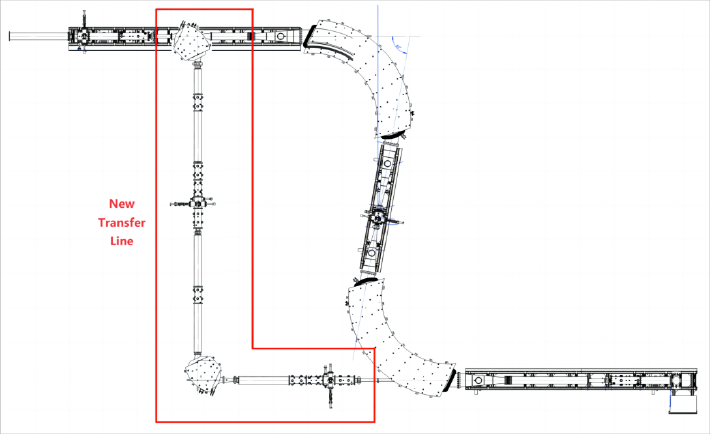}
    \caption{New transfer line to reduce beam loss instead of the high-resolution main analyzing magnets.}
    \label{fig:5.3.5}
\end{figure}
\newpage
	\subsection{High-sensitivity collinear resonance ionization spectroscopy}\label{sub:5-4}

As discussed in Section~\ref{sub:3-1}, a collinear laser spectroscopy setup has been installed and commissioned at BRIF to measure HFS spectra by detecting laser-induced fluorescence (LIF) photons. However, even with the implementation of the RFQ, collinear laser spectroscopy experiments are generally used to study unstable nuclei with production rates around 10$^{3\sim4}$ particle per second (pps). This limitation is primarily due to the significant photon-detection background caused by scattered laser light. In contrast, in-source resonance ionization spectroscopy (RIS) experiment~\cite{RILIS} has demonstrated high sensitivity, enabling the study of unstable nuclei with production rates as low as 1~pps~\cite{Hg-radii2018}. Nonetheless, the spectral resolution achievable with this technique is constrained by Doppler and/or pressure broadening effects, as the measurements are conducted within a hot-cavity (over 2000~$^\text{o}$C) laser ion source. 

\begin{figure*}[!htbp]
\centering
\includegraphics[width=0.99\linewidth]{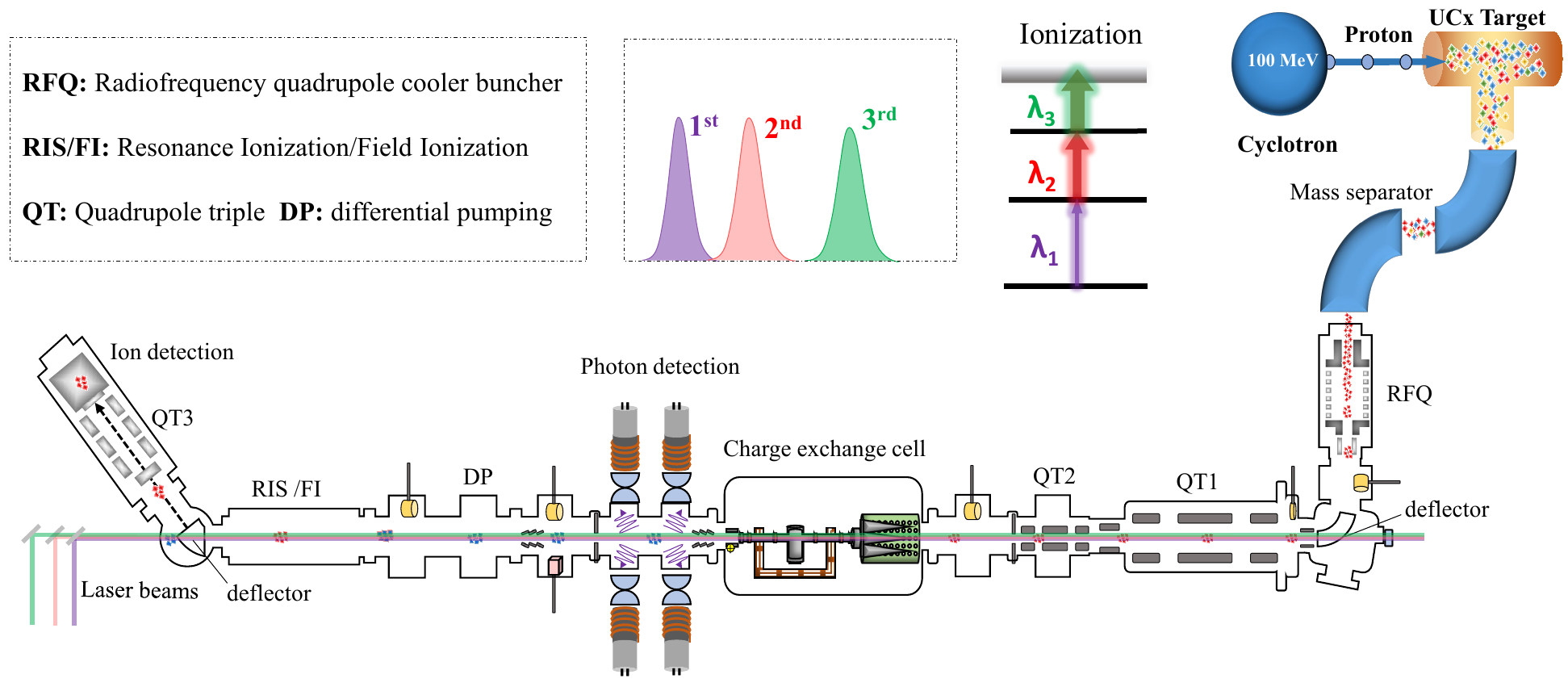}
\caption{Collinear resonance ionization laser spectroscopy setup planned at BRIF~\cite{CRIS-NIMB-PKU,PKU-CRIS-2024}. A radio frequency quadrupole cooler buncher~\cite{PKU-RFQ-2024} is also included to provide bunched radioactive ion beam that needs to be synchronized with the laser pulses.}
\label{figcris}
\end{figure*} 

To further overcome the issue of photon-detection background, an upgrade to the current collinear laser spectroscopy setup is underway at the BRIF facility to implement collinear resonance ionization laser spectroscopy~\cite{CRIS-NIMB-PKU,PKU-CRIS-2024}, in addition to the installation of an RFQ. Collinear resonance ionization laser spectroscopy technique combines the high-resolution capabilities of conventional collinear laser spectroscopy, which uses LIF detection, with the high-sensitivity advantages of in-source laser spectroscopy based on the resonance ionization detection. This method enables the measurement of HFS spectra with RIS by overlapping the multiple laser beams collinearly or anti-collinearly along the radioactive ion beam.

The first Collinear Resonance Ionization Spectroscopy (CRIS) setup was established at ISOLDE-CERN~\cite{CRIS-NIM2020} and achieved its highest sensitivity to study exotic $^{78}$Cu isotope at a production rate of approximately 20~pps while maintaining spectral resolution comparable to that of standard collinear laser spectroscopy~\cite{Cu-moment2017}. Further advancements include the integration of $\beta$-decay detection into the CRIS setup, which has allowed the study of isotopes with low production rates despite significant contamination from stable or long-lived isobar~\cite{k-radii2021}. More recently, this CRIS method has been successfully applied to study RaF radioactive molecules, providing new opportunities for research in fundamental symmetries and astrophysics~\cite{CRIS-molecule1}. The success of the CRIS experiments has motivated ongoing development of the technique at facilities such as Rare Isotope Beams (FRIB) at MSU, USA~\cite{FRIB}, the Ion Guide Isotope Separation On-Line (IGISOL) facility at Jyv\"askyl\"a, Finland~\cite{IGISOL2014} and the BRIF facility in China.

The layout of the planned collinear resonance ionization laser spectroscopy setup at BRIF is shown in Fig.~\ref{figcris}. Radioactive isotopes will be produced at BRIF by impinging 100-MeV protons onto a thick UC$_x$ target. The resulting radioactive isotopes will be ionized by a surface ion source before being extracted and accelerated to 30~keV. The ion beams will then be separated by the mass separator, and cooled and trapped in the RFQ~\cite{PKU-RFQ-2024}, which will deliver a bunched ion beam with a temporal width less than 5 $\mu$s to collinear resonance ionization setup~\cite{CRIS-NIMB-PKU,PKU-CRIS-2024}. In the setup, the bunched ion beam will first be neutralized via a charge exchange cell (CEC) filled with sodium vapor. Non-neutralized ions will be removed using deflector plates located downstream of the CEC. The neutralized beam bunch will then be transported to the ultra-high vacuum laser-atom interaction region (resonance ionization region). In this region, the atom bunch will be overlapped anti-collinearly in space and synchronized in time with three pulsed lasers for resonant excitation and ionization, as shown in the inset of the Fig.~\ref{figcris}. 

The resonantly ionized ions will be deflected and delivered to an ion detector where they will be recorded as a function of laser frequency to obtain the HFS spectrum. To suppress the ion background from stable or long-lived contaminants, the resonantly ionized ions can alternatively be injected into a decay station. There, the decay particles of the ions can be collected as functions of the laser frequency to reconstruct the HFS spectrum. The first planned online resonance ionization laser spectroscopy experiment will focus on neutron-rich rubidium isotopes and is currently under preparation.

        %

 	\subsection{Advanced superconducting solenoid spectrometer}\label{sub:5-5}

Light-ion induced direct reactions provide a sensitive probe for various aspects of nuclei, including the direct single- and multi-nucleon transfer, elastic, and inelastic scattering reactions. Recognizing the crucial role of direct reactions in uncovering atomic nuclear properties, substantial efforts have been devoted to developing instruments capable of measuring these reactions in inverse kinematics using radioactive ion beams. The challenges in this endeavor are significant: radioactive ion beams typically exhibit low intensities, often reduced by over six orders of magnitude compared to stable beams. Additionally, the detection of outgoing ions at forward center-of-mass angles is difficult due to their relatively low energy. The Q-value spectrum, if recorded as a function of the laboratory angle, becomes kinematically compressed, so it is difficult to achieve a good resolution of the Q-value spectrum with the regular silicon array setup. The superconducting solenoidal spectrometer is designed to fully exploit the potential of radioactive beams through direct nuclear reactions. 

Coupling this unique instrumentation system with ISOL-type beams enables exploration of nuclear structure, nuclear astrophysics, fundamental symmetries, and various nuclear applications. 

The superconducting solenoidal spectrometer can be used to benefit the physics program of BRIF by integrating two proven technologies into one instrument: a large-bore superconducting solenoid operating in vacuum with an on-axis silicon array, and a gas-filled detection mode functioning as an active-target time-projection chamber. These two operational modes, the silicon-array mode and the active-target mode, will offer a wide range of scientific opportunities. Pioneering work with devices like HELIOS~\cite{HELIOS, HELIOS_NIMA, Schiffer_1999}, ISS~\cite{ISS} and SOLARIS~\cite{Kay} has demonstrated the effectiveness of such spectrometers for direct reactions. This advances multiple research themes at BRIF, including modifications in nuclear shells, emerging collective phenomena in nuclei, nuclear reactions relevant to nucleosynthesis, and verification of critical nuclear data. 

In the silicon-array mode of this device, detection is achieved by transporting the outgoing ions through a solenoidal field to the on-axis detector. In this configuration, the beam enters the solenoidal field along the axis, in the same detection as the field axis, and interacts with the target. The outgoing ions return to the axis one cyclotron period later based on their mass-to-charge ratio. A silicon array surrounding the magnetic axis records the energy, flight time, and distance from the target. The timing signature identifies the outgoing ions according to the different B$\rho$ without relying on telescope arrangements. This operation mode unfolds the kinematic compression experienced in direct reactions conducted in inverse kinematics and results in a straightforward linear relationship between the spectrum of outgoing ions in the laboratory frame. This optimizes the resolution of the resulting Q-value spectrum (see examples in Ref.~\cite{CHEN, CHEN2024}). The successful implementation of solenoidal spectrometers worldwide has inspired their development at BRIF. Usually, the beam intensity requirement is more than $10^4$ particles per second. For weaker beams or more complex reactions involving many-body final states, other appealing solutions exist, such as the gas-filled time projection chamber (TPC), exemplified by the AT-TPC developed at NSCL~\cite{ATTPC} and MATE developed by IMP~\cite{Lxb2024,Zhang2021}. 

In a gas-filled TPC, the nuclei of the gas molecules serve as the target isotopes. This setup results in an effective luminosity potentially up to 100 times greater than what can be achieved in conventional solid-target experiments. The luminosity is adjustable by varying the gas pressure and type. This large luminosity makes it possible to conduct direct reactions with beams as weak as hundreds of particles per second, which are a few orders of magnitude weaker than those required for conventional methods. The reaction vertex within the gas volume is determined for each event, allowing the energy of the beam at the time of the reaction to be deduced. This is a significant improvement over the solid targets, where the beam energy must be averaged over the energy loss of the entire target thickness, often limiting achievable energy resolution. Consequently, the resolution with active targets is usually comparable to that with passive targets but also independent of the target thickness. Additionally, particles with low recoil energies can be detected with very low thresholds as long as their tracks can be recorded outside the vertex region.

The combination of the cylindrical TPC with the solenoid spectrometer is especially appealing because longer trajectories can be recorded and more information can be obtained, including the energy loss in the gas, $B\rho$ values, energies and angles. This allows for particle identifications even if punching through the detector. Therefore, a nearly 4$\pi$ coverage is achievable for the direct reactions of interest.

The development of a solenoidal spectrometer at BRIF is particularly interesting in several cases listed as below. One-neutron adding reactions may be carried out with ISOL beams of $^{20\sim25}$Na and $^{37\sim46}$K, to investigate the corresponding single-particle structures, which serve to enhance our understanding of the mechanism driven behind the $N=20$ and $28$ shell closure. These measurements also could be used to deduce the ($n, \gamma$) reaction rate used to constrain the heavy element synthesis. In addition, $\gamma$-ray branching ratios could be obtained for the unbound states with a set of recoil detectors, for both operation modes.

	\newpage
	\section{Summary and outlook}\label{sec:sum}
	Present and planned radioactive ion beam accelerators offer opportunities to explore the exotic structure of unstable nuclei and explosive nuclear burning in astrophysical environments. The BRIF, which is the only ISOL-type facility that has been commissioned in Asia, is a high-precision facility for investigation of nuclear reaction dynamics of unstable nuclei as it can deliver almost pure unstable ion beams, complementary to PF-type technique. It consists of a 100-MeV 200-$\mu$A compact proton cyclotron serving as the driving accelerator, a two-stage ISOL system with mass resolution of 20,000, a 13-MV tandem accelerator for post-acceleration, a superconducting linac for further boosting beam energies and various fundamental research terminals including a collinear laser spectroscopy system, total absorption gamma spectroscopy, Q3D magnetic spectrograph and a heavy-ion time-of-flight spectrometer. Here we summarize the experiments performed at the BRIF so far on the decay properties of $^{20}$Na and $^{88}$Rb and reaction dynamics of $^{21,22}$Na. In the future more heavy neutron-rich isotopes will be produced due to the use of UC$_x$ target for investigation of their exotic structure and basic properties.

    Although the present facility has started contributing to the investigation of nuclear structure physics, major improvements are desirable in the future. Firstly, we will try to deliver more intense proton beam than the present 200~$\mu$A. In fact, we have commissioned mA-level beam on the internal target of our proton cyclotron, and will try to commission mA-level beam on the external target soon. Secondly, we will develop UC$_x$ target which has better performance to endure radiation from more intense proton beam. Thirdly, we will construct a linear accelerator instead of the tandem accelerator to enhance the intensity of radioactive ion beams. This could increase the beam intensity by over three orders of magnitude because it is no longer necessary for the rare positive RIBs to be changed into negative ions required by the tandem accelerator. In addition, we have started installing a high-sensitivity collinear resonance ionization spectroscopy and plan to develop an advanced superconducting solenoid spectrometer. We welcome interested users to participate in the design and construction of experimental setups and conduct research at the BRIF.
	
	
	\newpage
	\section*{Acknowledgements}
	It is a great pleasure to thank our colleagues of the BRIF collaboration, without whom the results reviewed in this article would not have been achieved. This work is supported by the National Key Research and Development Project, China under Grant No. 2022YFA1602301, the National Natural Science Foundation of China under Grants No. 12125509, 12222514 and 12435010, the National Key Research and Development Project under Grant No. 2023YFA1607000, the National Natural Science Foundation of China under Grants No. 10125524, 10775185, 12027809, U2167204 and U2267205, and the Continuous-Support Basic Scientific Research Project.\\

	\bibliography{brif}
	


	
	\newpage
	\appendix
	\renewcommand*{\thesection}{\Alph{section}}
	

\end{document}